\documentclass[aps,groupedaddress,superscriptaddress]{revtex4-1}

\usepackage{graphicx}
\usepackage{amsmath}%
\usepackage{subfig}
\usepackage{psfrag}
\usepackage[usenames,dvipsnames]{xcolor}
\usepackage{float}
\usepackage{natbib}
\usepackage{pifont}
\usepackage{wasysym}
\usepackage{amssymb}

\DeclareGraphicsExtensions{.pdf,.png,.jpg}

\newcommand\Rey{\mbox{\textit{Re}}}  
\newcommand\Pran{\mbox{\textit{Pr}}} 
\newcommand\St{\mbox{St}} 

\def\bulloverlay{\mathrel{%
    \mathchoice{\BULLOVERLAY}{\BULLOVERLAY}{\scriptsize\BULLOVERLAY}{\tiny\BULLOVERLAY}%
}}
\def\BULLOVERLAY{{%
    \setbox0\hbox{---$\!$---}%
    \rlap{\hbox to \wd0{\hss$\bullet$\hss}}\box0
}}

\def\circleoverlay{\mathrel{%
    \mathchoice{\CIRCLEOVERLAY}{\CIRCLEOVERLAY}{\scriptsize\CIRCLEOVERLAY}{\tiny\CIRCLEOVERLAY}%
}}
\def\CIRCLEOVERLAY{{%
    \setbox0\hbox{---$\!$---}%
    \rlap{\hbox to \wd0{\hss$\circ$\hss}}\box0
}}

\def\squareoverlay{\mathrel{%
    \mathchoice{\SQUAREOVERLAY}{\SQUAREOVERLAY}{\scriptsize\SQUAREOVERLAY}{\tiny\SQUAREOVERLAY}%
}}
\def\SQUAREOVERLAY{{%
    \setbox0\hbox{---$\!$---}%
   \rlap{\hbox to \wd0{\hss{\scriptsize $\square$}\hss}}\box0
}}

\def\blacksquareoverlay{\mathrel{%
    \mathchoice{\BLACKSQUAREOVERLAY}{\BLACKSQUAREOVERLAY}{\scriptsize\BLACKSQUAREOVERLAY}{\tiny\BLACKSQUAREOVERLAY}%
}}
\def\BLACKSQUAREOVERLAY{{%
    \setbox0\hbox{---$\!$---}%
    \rlap{\hbox to \wd0{\hss{\scriptsize $\blacksquare$}\hss}}\box0
}}

\newsavebox{\astrutbox}
\sbox{\astrutbox}{\rule[-5pt]{0pt}{20pt}}

\begin{document}

\title{Heat transfer and wall temperature effects in shock wave turbulent boundary layer interactions}

\author{M. Bernardini}
\email[]{matteo.bernardini@uniroma1.it}
\affiliation{Dipartimento di Ingegneria Meccanica e Aerospaziale, Universit\`a di Roma `La Sapienza' Via Eudossiana 18, 00184 Roma, Italia}
\author{J. Larsson}
\affiliation{Dept. of Mechanical Engineering, University of Maryland, College Park, MD 20742}
\author{S. Pirozzoli}
\affiliation{Dipartimento di Ingegneria Meccanica e Aerospaziale, Universit\`a di Roma `La Sapienza' Via Eudossiana 18, 00184 Roma, Italia}
\author{F. Grasso}
\affiliation{Cnam-Laboratoire DynFluid, 151 Boulevard de L'Hopital, 75013 Paris}

\date{\today}

\begin{abstract}
Direct numerical simulations are carried out to investigate the effect of the wall temperature
on the behavior of oblique shock-wave/turbulent boundary layer interactions
at freestream Mach number $2.28$ and shock angle of the wedge generator $\varphi = 8^{\circ}$.
Five values of the wall-to-recovery-temperature ratio ($T_w/T_r$) are considered, corresponding
to cold, adiabatic and hot wall thermal conditions. We show that the main effect of cooling is
to decrease the characteristic scales of the interaction in terms of upstream influence and
extent of the separation bubble. The opposite behavior is observed in the case of heating,
that produces a marked dilatation of the interaction region.
The distribution of the Stanton number shows that a strong amplification of the heat transfer occurs
across the interaction, and the maximum values of thermal and dynamic loads are found in the case
of cold wall. The analysis reveals that the fluctuating heat flux exhibits a strong intermittent
behavior, characterized by scattered spots with extremely high values compared to the mean. Furthermore,
the analogy between momentum and heat transfer, typical of compressible, wall-bounded, equilibrium turbulent
flows does not apply for most part of the interaction domain.
The pre-multiplied spectra of the wall heat flux do not show any evidence of the influence of the low-frequency
shock motion, and the primary mechanism for the generation of peak heating is found to be linked
with the turbulence amplification in the interaction region.
\end{abstract}

\maketitle

\section{Introduction}

In a wide range of high-speed applications in the aerospace industry shock-wave turbulent boundary layer
interactions (SBLI) are crucial for an efficient aerodynamic and thermodynamic design, SBLI being 
responsible for increased internal machine losses, thermal and structural fatigue
due to increased heat transfer rates and substantial modification
of the wall-pressure signature, flow unsteadiness, shock/vortex interaction and broadband noise
emission. Improving the understanding of these critical features is essential to enhance the capability
to predict important quantities like the location and magnitude of peak heating,
as well as for the development of effective flow control methods~\citep{dolling01}.

Most of prior scientific work on SBLI, of both experimental~\citep{dolling83,delery86,dupont_06,piponniau_09,humble09,souverein_10_b} and
numerical nature~\citep{Adams2000,wu_martin07,touber_09,pirozzoli_11_3,grilli12,aubard13,morgan13}, has been aimed at the case 
of adiabatic wall condition and many efforts have been invested in
the last decade to characterize the large-scale, low-frequency unsteadiness typically
found in the interaction region. This phenomenon can be particularly severe when the shock
is strong enough to produce separation of the incoming boundary layer~\citep{clemens14}.

The influence of wall thermal conditions on the characteristics of SBLI can be considerable and wall cooling is often
advocated as a possible candidate for flow control, strong cooling being capable of~\citep{delery85}: i) shifting the
laminar-turbulent boundary layer transition toward higher Reynolds numbers; ii) producing a fuller incoming boundary layer
velocity profile; and iii) reducing the thickness of the subsonic layer by decreasing the local speed of sound.
Unfortunately, only a few experimental studies have been conducted on this topic, all based on the analysis of mean flow properties.

The effects of heat transfer in turbulent interactions over a compression ramp have been
investigated by~\citet{spaid72}, who performed experiments at freestream Mach number $M_{\infty} = 2.9$,
by considering a cold (wall-to-recovery-temperature ratio $T_w/T_r = .47$) and a nearly adiabatic wall
($T_w/T_r = 1.05$).  Their results showed that the effect of wall cooling, relative to the adiabatic condition, is to increase the incipient separation angle and to decrease the separation distance.
Similar conclusions were later reported by~\citet{back76},
who considered an oblique shock-wave impinging on a turbulent boundary layer at $M_{\infty} = 3.5$ with surface cooling ($T_w/T_r = .44$).

An in depth experimental analysis of a shock reflection over a strongly heated wall ($T_w/T_r = 2$) was carried out
by~\citet{delery92}, who considered a two-dimensional test arrangement for an upstream Mach number $M_{\infty} = 2.4$ and
two incident shock wave intensities. The experimental measurements showed that heating the surface greatly increases the extent
of the interaction zone and the separation point moves much farther upstream than under adiabatic conditions.
More recently, an investigation of the impact of wall temperature on a $M_{\infty} = 2.3$ shock-induced
boundary layer separation has been carried out by~\citet{jaunet14} for shock deflection angles ranging from $3.5^o$ to $9.5^o$
under adiabatic ($T_w/T_r =1$) and wall heating conditions ($T_w/T_r = 1.4, 1.9$).
Their extensive experimental analysis based on Schlieren visualizations, particle image velocimetry (PIV)
and time-resolved hot-wire measurements highlighted that a hot wall leads to an increase of the interaction length-scales,
which is mainly associated with changes of the wall incoming conditions. A slight influence was also observed on
the onset of separation, shifted to smaller flow deviations in the heated case.
This scale change due to wall thermal conditions has also an effect on the flow unsteadiness,
the lower frequencies becoming more and more important by heating the wall.

Measurements of heat transfer in SBLI were first reported by~\citet{hayashi84},
who considered a $M_{\infty} = 4$ boundary layer
developing over an isothermal cold wall ($T_w/T_r \approx 0.6$) interacting with
an oblique shock at various incident angles.
They observed a complex spatial variation of the heat transfer coefficient, characterized by 
a rapid increase near the separation point, followed by a sharp reduction within the separation bubble and a further
increase in the proximity of the reattachment point. Combined measurements of skin friction and heat transfer
have been recently reported by~\citet{schulein06} who considered an impinging shock at $M_{\infty}=5$ and three
values of the incident angle. Their results show a strong increase of the heat flux in the separation zone,
characterized by a complex non-equilibrium behavior, in which the Reynolds analogy between momentum
and heat flux is not valid.

A relatively large number of direct numerical and large-eddy simulations (DNS/LES) of both compression ramp
and impinging shock interactions have appeared over the last decade~\citep{Adams2000,touber_09,pirozzoli_11_3,hadjadj12,Nichols2016}.
However, all these studies addressed the case of adiabatic wall conditions and to our knowledge,
no high-fidelity simulations have been carried out to explore the effect of neither wall heating nor cooling
in SBLI.
The main objective of the present work is to fill this gap by providing a numerical study on the influence
of wall thermal conditions on the behavior of oblique SBLI. The analysis is based on direct numerical simulations 
to explore the effect of different wall-to-recovery-temperature ratios.
This can be beneficial for the improvement of current turbulence modeling for SBLI, in particular for the computation
of the heat transfer, which is the most challenging aspect of these flows,
and it is well known that numerical predictions based on the solution of Reynolds Averaged Navier-Stokes equations
are rather poor~\citep{fedorova2001experimental,knight2003advances}, with significant differences (up to 100\%)
among different turbulence models.
A careful characterization of how the separation bubble, the skin friction and heat transfer are affected by the wall thermal conditions
is a core objective of this work, and it represents the key stepping-stone towards harnessing wall-cooling to stabilize and control SBLI.

The paper is organized as follows. After the introduction,
the numerical strategy and the flow conditions of the simulations are described in Section 2.
The main results are presented in Section 3, where we also provide a comparison
with experimental data. The discussion first focuses on the modifications induced by the wall temperature
on the structure of SBLI, both in terms of lenghtscales and turbulence amplification. Emphasis is then put on the
heat transfer behavior across the interaction and for the first time fluctuating heat flux data are reported and analyzed.
Conclusions are finally provided in Section 4.

\section{Computational setup} \label{sec:computational}

\subsection{Flow solver}

We solve the three-dimensional Navier-Stokes equations for a perfect compressible gas with Fourier heat law
and Newtonian viscous terms. The molecular viscosity $\mu$ is assumed to depend on temperature $T$ through Sutherland's law,
and the thermal conductivity is computed as $k = c_p \mu / \Pran$, the molecular Prandtl number being set to $\Pran=0.72$.

The Navier-Stokes equations are discretized on a Cartesian mesh and solved by means of
an in-house finite-difference flow solver, extensively validated for 
wall-bounded flows and shock boundary layer interactions in the transonic and supersonic regime~\citep{pirozzoli_10_2,PirozzoliJFM2011}.
The solver incorporates state-of-the-art numerical algorithms, specifically designed to cope with
the challenging problems associated with the solution of high-speed turbulent flows, i.e. the need to
accurately resolve a wide spectrum of turbulent scales and to capture steep gradients without undesirable numerical oscillations.
In the current version of the code the convective terms are discretized by means of a hybrid conservative sixth-order central/fifth-order WENO scheme,
with a switch based on the Ducros sensor~\citep{ducrosetal99}.
To improve numerical stability, the triple splitting of the convective terms~\citep{kennedy_08}
is used in a locally conservative implementation~\citep{pirozzoli_10}.
The viscous terms are approximated with sixth-order central differences,
after being expanded to Laplacian form to guarantee physical dissipation
at the smallest scales resolved by the computational mesh.
Time advancement is performed by means of a third-order, low-storage, explicit
Runge-Kutta algorithm~\citep{bernardini09}.

\subsection{Flow conditions and computational arrangement}

\begin{figure}
 \centering
 \psfrag{W}[][][.7]{$x_{rec}$}
 \psfrag{J}[][][.7]{$x_0$}
 \psfrag{Z}[][][.7]{$x_T$}
 \psfrag{Q}[][][.7]{$x_{sh}$}
 \psfrag{F}[][][.5]{$(T_w/T_r = 1)$}
 \psfrag{G}[][][.5]{$(T_w/T_r = s)$}
 \includegraphics[width=12cm,clip]{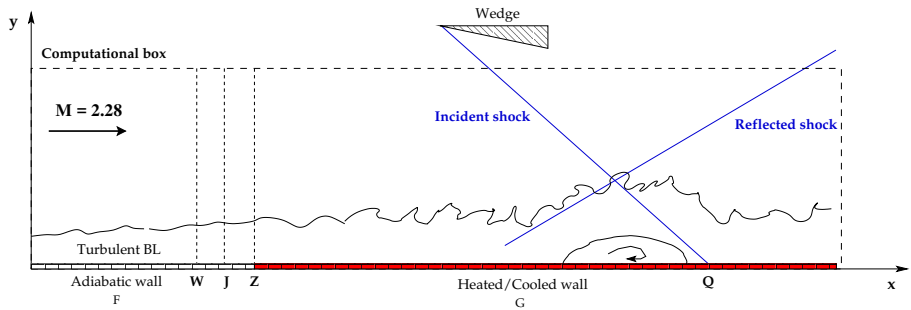} \vskip 0.5em
 \caption{Sketch of the flow configuration under investigation.}
 \label{fig:configuration}
\end{figure}
\begin{table}
 \begin{center}
  \begin{tabular*}{1.\textwidth}{@{\extracolsep{\fill}}cccccccccc}
   \hline
   Test case & Line-style & $M_{\infty}$ & ${\Rey_{\theta}}_0$ & $\varphi$ & $s$ & $T_w/T_{\infty}$ & $\textrm{T} u_{\infty}/\delta_0$ & $L/\delta_0$ & $L_{sep}/\delta_0$ \\
   \hline
   BL-s0.5    & \textcolor{blue}{-$\!$- -$\!$- -$\!$-} & 2.28 & 2500 & $8^{\circ}$ & $0.5$  & $0.96$ & 218.9  &  --  &  --  \\
   BL-s1.9    & \textcolor{red}{-$\!$- -$\!$- -$\!$-}  & 2.28 & 2500 & $8^{\circ}$ & $1.9$  & $3.66$ & 187.2  &  --  &  --  \\
   SBLI-s0.5  & \textcolor{blue}{$\bulloverlay$}       & 2.28 & 2500 & $8^{\circ}$ & $0.5$  & $0.96$ & 598.7  & 2.90 & 0.55 \\
   SBLI-s0.75 & \textcolor{cyan}{$\circleoverlay$}     & 2.28 & 2500 & $8^{\circ}$ & $0.75$ & $1.44$ & 455.5  & 3.31 & 1.67 \\
   SBLI-s1.0  & \textcolor{black}{---$\!$---}          & 2.28 & 2500 & $8^{\circ}$ & $1.0$  & $1.93$ & 840.1  & 3.74 & 2.11 \\
   SBLI-s1.4  & \textcolor{orange}{$\squareoverlay$}   & 2.28 & 2500 & $8^{\circ}$ & $1.4$  & $2.70$ & 669.6  & 4.32 & 2.89 \\
   SBLI-s1.9  & \textcolor{red}{$\blacksquareoverlay$} & 2.28 & 2500 & $8^{\circ}$ & $1.9$  & $3.66$ & 1002.6 & 4.97 & 3.98 \\
   \hline
  \end{tabular*}
 \end{center}
 \caption{Flow parameters for DNS simulations.
 $\varphi$ is the incidence angle of shock generator, $s = T_w/T_r$ the wall-to-recovery temperature ratio in the interaction zone,
 $L$ is the interaction lengthscale, and $L_{\mathrm{sep}}$ is the length of the recirculation bubble.
 The subscript $0$ refers to properties taken upstream of the temperature step change at $x_0 = 50 \delta_{in}$.
 $\textrm{T}$ is the time span used for the computation of the flow statistics.}
 \label{tab:testcases}
\end{table}

A schematic view of the flow configuration investigated is shown in figure~\ref{fig:configuration}.
A turbulent boundary layer developing over a flat plate 
is made to interact with an impinging shock.
The computational domain extends for $L_x \times L_y \times L_z = 96 \, \delta_{in} \times 11.7 \, \delta_{in} \times 5.5 \, \delta_{in}$,
in the streamwise ($x$), wall-normal ($y$) and spanwise ($z$) directions,
$\delta_{in}$ being the inflow boundary layer thickness.
The oblique shock is introduced in the simulation by locally imposing 
the inviscid Rankine-Hugoniot jump conditions at the top boundary so as to
mimic the effect of the shock generator and the nominal shock impingement point
is $x_{sh} = 69.5 \delta_{in}$,
Non-reflecting boundary conditions are enforced
at the outflow and at the top boundary, away from the incoming shock.
A recycling/rescaling procedure is used for turbulence generation
at the inflow plane, whereby staggering in the spanwise direction is used
to minimize spurious flow periodicity~\citep{pirozzoli_10_2}.
The recycling station is placed at $x_{rec} = 48 \delta_{in}$, sufficiently distant from the inflow to guarantee proper
streamwise decorrelation of the boundary layer statistics~\citep{simens_09} and to prevent
any spurious low-frequency dynamics associated with the recycling procedure.
A characteristic wave decomposition is used at the no-slip wall, where perfect
reflection of acoustic waves is enforced, and the wall temperature is held fixed.
The turbulent boundary layer develops under nominal adiabatic conditions
up to $x_T = 54 \delta_{in}$ (the wall temperature $T_w$ being equal to the
recovery temperature $T_r$) and local cooling/heating is applied for $x > x_T$ by specifying 
the wall-to-recovery-temperature ratio $s = T_w/T_r$ to the desired value.
To avoid a discontinuity in the wall temperature distribution a smoothed
step change is prescribed according to
\begin{equation}
 \nonumber
 T_w (x) = T_r \left [1+\frac{s-1}{2} \left (1+\tanh \frac{2 (x - x_T)}{\delta_{in}} \right) \right].
\end{equation}

Five DNS have been carried out at various values of the wall-to-recovery-temperature
ratio, spanning cold ($s=0.5, 0.75$), adiabatic ($s=1.0$) and hot ($s=1.4,1.9$) walls.
These cases are labelled as SBLI-s0.5, SBLI-s0.75, SBLI-s1.0, SBLI-s1.4, SBLI-s1.9, respectively.
The flow conditions for the various runs are reported in table~\ref{tab:testcases}.
For all cases, the free-stream Mach number is $M_{\infty}=2.28$, the deflection angle of the wedge
shock generator is $\varphi = 8^{\circ}$ and the Reynolds number of the incoming boundary layer based on the momentum
thickness, evaluated at a reference station upstream the temperature step change ($x_0 = 50 \delta_{in}$)
is $\Rey_{\theta_0} \approx 2500$.
For reference purposes, two additional simulations have been also carried out, corresponding to DNS of
spatially evolving boundary layers (in the absence of impinging shock)
subjected to the same temperature step change as in SBLI-s0.5 and SBLI-s1.9.
These two cases are denoted as BL-s0.5 and BL-s1.9, respectively.

The domain is discretized with a mesh consisting of $6144 \times 448 \times 448$ grid nodes,
that are uniformly distributed in the spanwise direction.
In the streamwise and wall-normal directions stretching functions are employed to better
resolve the interaction region and to cluster grid nodes towards the wall.
In particular, a hyperbolic sine mapping is applied from the wall
$y = 0$ up to $y = 3.5 \delta_{in}$. A uniform mesh spacing is then used above this location and
an abrupt variation of the metrics is avoided by a suitable smoothing of the connection zone.
In terms of wall units (based on the friction velocity $u_{\tau}$ and viscous length-scale $\delta_v$)
evaluated in the undisturbed turbulent boundary layer at $x_0$, the streamwise and spanwise spacings are $\Delta x^+ = 5.9$, $\Delta z^+ = 3.1$;
in the wall-normal direction the spacing ranges from $\Delta y^+ = 0.49$ at the wall
to $\Delta y^+ = 6.7$ at the edge of the boundary layer. We point out that such mesh spacings
are significantly smaller than those usually employed for DNS of SBLI under adiabatic conditions.
The motivation is dictated by the need of maintaining adequate resolution even when strong cooling is applied,
which is the most challenging case in terms of spacing requirements, due to the drastic reduction of
the viscous length-scale.

The simulations have been run on a parallel cluster using 4096 cores, for a total of 7 Mio CPU hours.
The time span over which the flow statistics have been computed is reported in table~\ref{tab:testcases}.
In the following, the boundary layer thickness in the undisturbed boundary layer at station $x_0$ is assumed as
reference length for all flow cases ($\delta_0 = 1.45 \, \delta_{in}$). The results are reported using scaled
interaction coordinates $x^* = (x-x_{sh})/\delta_0$, $y^* = y/\delta_0$.
For the sake of notational clarity, the streamwise, wall-normal and spanwise velocity components will be hereafter
denoted as $u$, $v$, $w$, respectively, and either the Reynolds ($\varphi = \overline{\varphi} + \varphi'$)
or the mass-weighted ($\varphi = \widetilde{\varphi} + \varphi'', \widetilde{\varphi} =
\overline{\rho \, \varphi}/\overline{\rho}$) decomposition will be used for the generic variable $\varphi$.

It is worth pointing out that the flow conditions of case SBLI-s1.0 are essentially identical to that of our previous DNS,
reported in~\citet{pirozzoli_11_3}, based on the experiment by~\citet{piponniau_09}.
The extensive comparison available in that paper (not repeated here) showed that the global structure of the flow
(mean velocities and turbulence velocity fluctuations) predicted by DNS is in very good agreement with that observed
in the experiment, provided that the differences in the overall size of the interaction zone are suitably compensated.
Indeed, the size of the separation bubble found in the computation is approximately $30 \%$ smaller than the
experimental one. As later shown by~\citet{bermejo14} this difference can be ascribed to
the assumption of spanwise periodicity applied in the numerical simulation,
which avoids confinement effects from lateral walls that are inevitable in the experiment
and are known to cause substantial increase of the separation bubble size.

\section{DNS Results} \label{sec:results}

\begin{table}
 \centering
 \begin{tabular*}{1.\textwidth}{@{\extracolsep{\fill}}ccccccc}
  Run & $\Rey_{\theta}$ & $\Rey_{\delta_2}$ & $\Rey_{\tau}$ & $H$ & $H_i$ & $C_f (\cdot 10^3)$ \\
  \hline
  ALL & 2410 & 1509 & 450 & 3.64 & 1.41 & 2.45 \\
 \end{tabular*}
 \caption{Global properties of the incoming turbulent boundary layer at $x_0 = 50 \delta_{in}$. $\theta$ denotes the
          momentum thickness. The Reynolds number are defined as $\Rey_{\theta} = \rho_{\infty} \, u_{\infty} \theta / \mu_{\infty}$;
          $\Rey_{{\delta}_2} = \rho_{\infty} \, u_{\infty} \theta / \overline{\mu}_w$; $\Rey_{\tau} = \overline{\rho}_w \, u_{\tau} \delta / \overline{\mu}_w$;
          $H$ and $H_i$ are the compressible and incompressible shape factor, respectively, computed with mean velocity $\overline{u}$. $C_f$ is the skin
          friction coefficient.}
 \label{tab:blprop}
\end{table}

\begin{figure}
 \centering
 \psfrag{x}[t][][1.0]{$y^+$}
 \psfrag{y}[][][1.0]{$U_{VD}^+$}
 \psfrag{p}[a][][0.8]{$(a)$}
 \includegraphics[width=4.1cm,angle=270,clip]{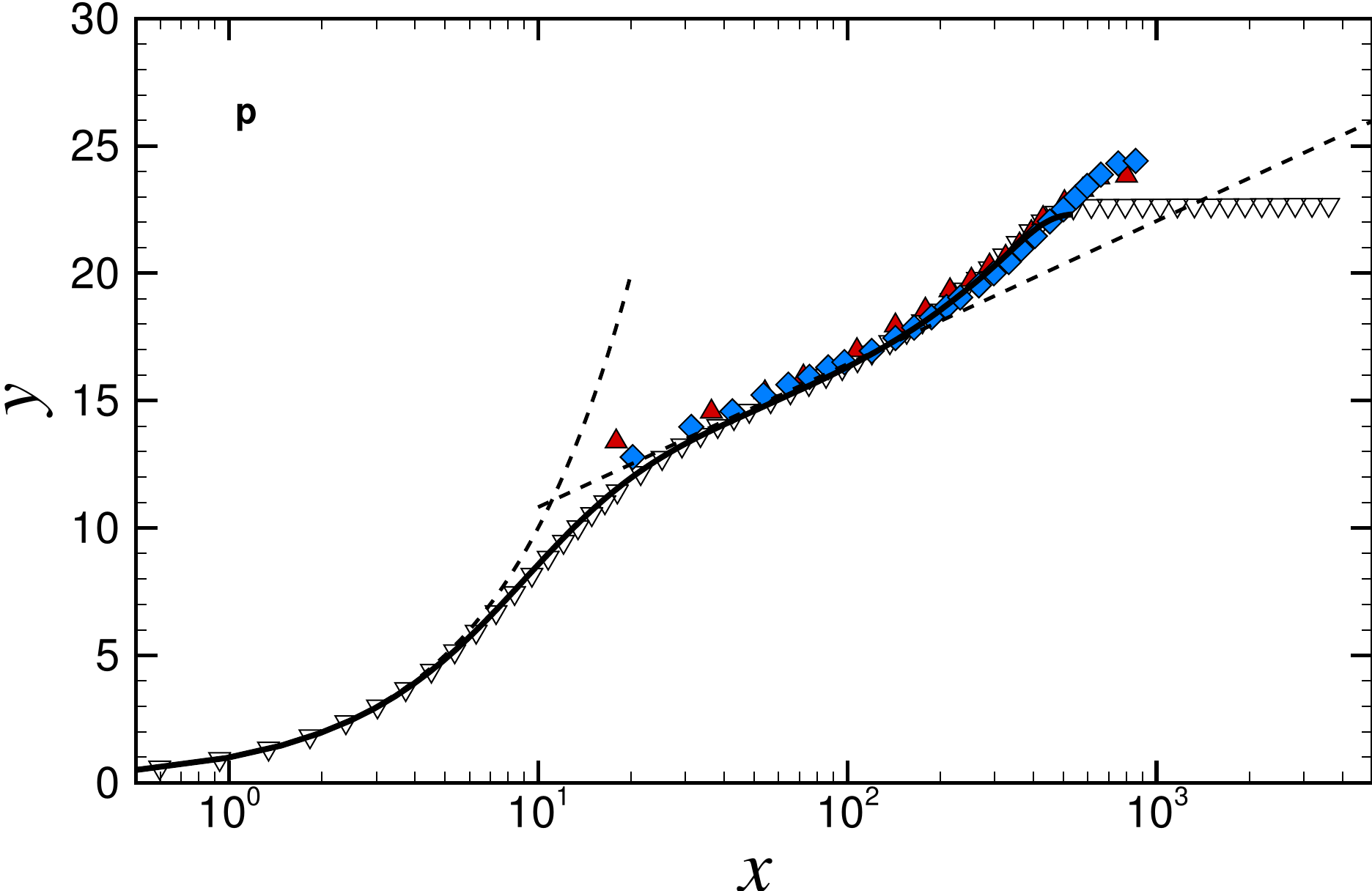} \hskip 1.0em
 \psfrag{x}[t][][1.0]{$y/\delta_0$}
 \psfrag{y}[][][1.0]{$\overline{\rho}/\overline{\rho}_w \, \widetilde{u_i^{\prime \prime} u_j^{\prime \prime}}^+$}
 \psfrag{p}[a][][0.8]{$(b)$}
 \includegraphics[width=4.1cm,angle=270,clip]{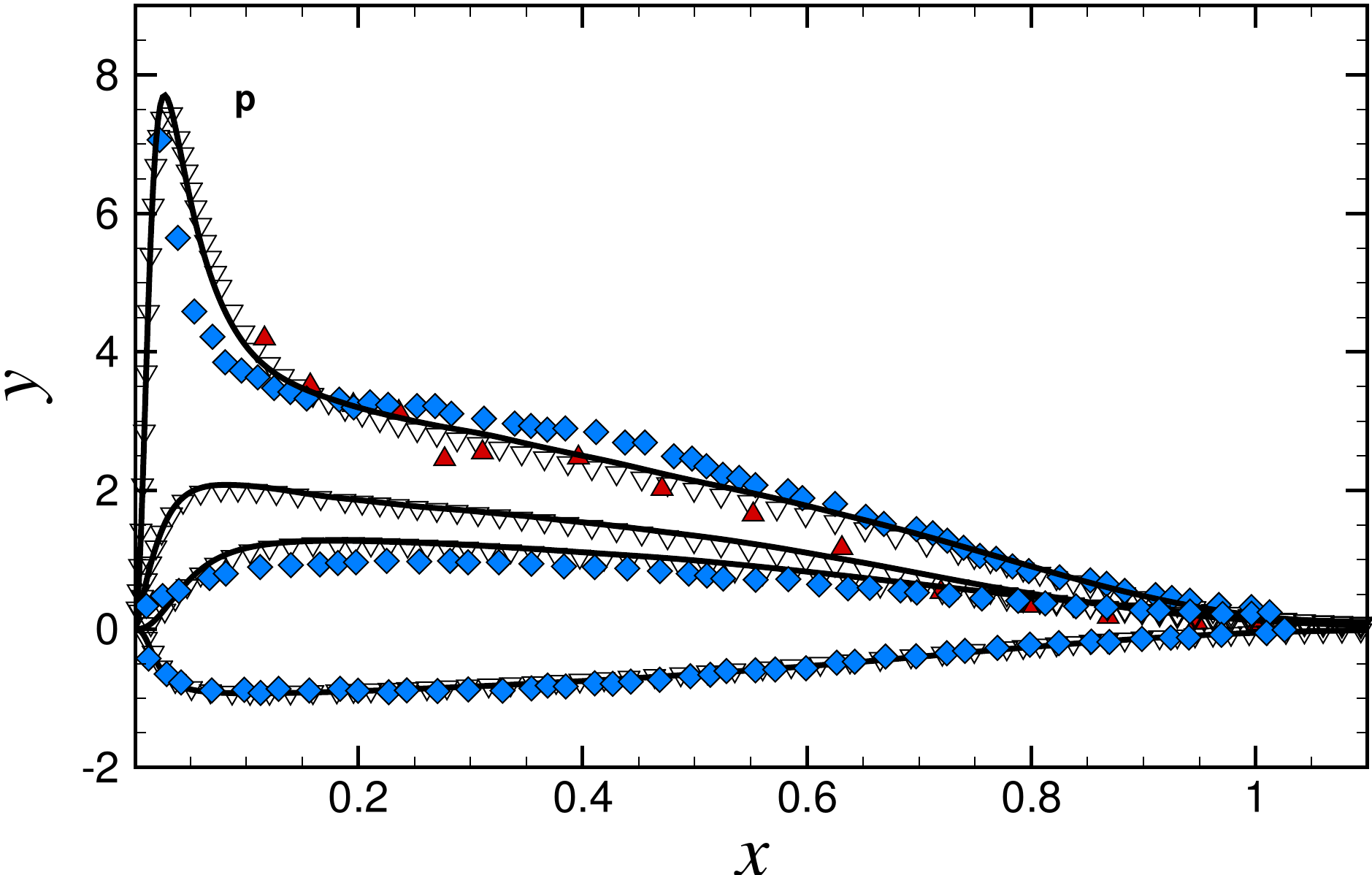} \vskip 0.5em
 \caption{Comparison of (a) van Driest-transformed mean velocity profile and (b) density-scaled Reynolds stress components
         at the adiabatic station $x_0$ with reference numerical and experimental data. Symbols denote experiments
         by~\citet{elena_88} (triangles, $M_{\infty} = 2.32$, $\Rey_{\theta} = 4700$), \citet{piponniau_09},
         (diamonds, $M_{\infty} = 2.28$, $\Rey_{\theta} = 5100$) and the incompressible DNS data by~\citet{schlatter_10b}
         (gradients, $\Rey_{\theta} = 1410$).}
 \label{fig:bl_incoming}
\end{figure}
\begin{figure}
 \centering
 \psfrag{x}[t][][1.2]{$\overline{u}/u_{\infty}$}
 \psfrag{y}[b][][1.2]{$\overline{T}/T_{\infty}$}
 \psfrag{b}[ ][][1.2]{$x$}
 \psfrag{c}[ ][][1.2]{$x$}
 \includegraphics[width=5.5cm,angle=270,clip]{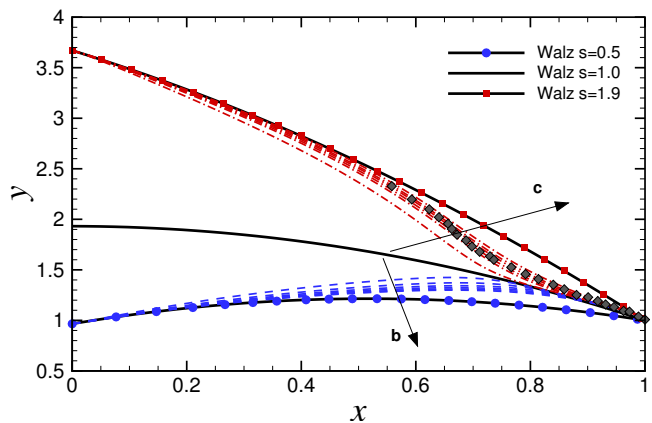} \vskip 0.5em
 \caption{Distribution of temperature-velocity relationship at various stations for BL-s0.5 and BL-s1.9.
         Refer to table~\ref{tab:testcases} for nomenclature of the DNS data.
         The solid lines indicate the equilibrium solution~\ref{eq:walz} for cold, adiabatic and hot walls.
         The black arrows indicate the direction of increasing $x$.
         The grey diamonds denote reference experiments with $s=2$ by~\citet{debieve97}.}
 \label{fig:t-u}
\end{figure}

\begin{figure}
 \centering
 \psfrag{x}[t][][1.2]{$\overline{T}_0/T_{0 \infty}$}
 \psfrag{y}[b][][1.2]{$y/\delta_{x_T}$}
 \psfrag{a}[][][1.2]{$x$}
 \psfrag{b}[][][1.2]{$x$}
 \includegraphics[width=5.5cm,angle=270,clip]{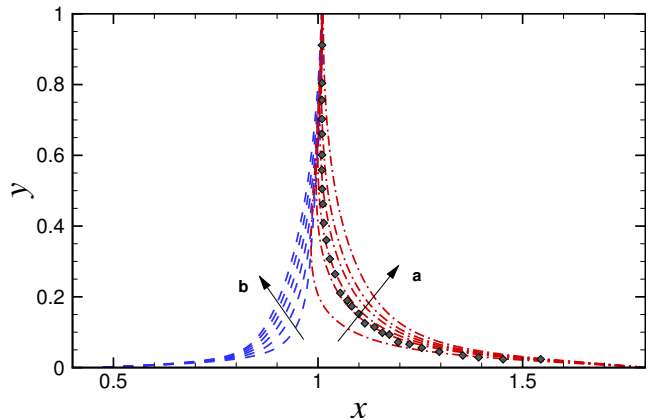} \vskip 0.5em
 \caption{Distribution of total temperature profiles at various streamwise stations for BL-s0.5 and BL-s1.9.
          The wall-normal coordinate is normalized through the boundary layer thickness at $x_T$ ($\delta_{x_T}$).
          Refer to table~\ref{tab:testcases} for nomenclature of the DNS data.
          The black arrows indicate the direction of increasing $x$.
          The grey diamonds denote reference experiments with $s=2$ by~\citet{debieve97}.}
 \label{fig:debieve}
\end{figure}

\subsection{Characterization of the incoming flow}

A comparison of the basic velocity statistics of the incoming turbulent boundary layer
with reference experiments and numerical simulations is shown in figure~\ref{fig:bl_incoming}.
The DNS data are taken at the reference station $x_0 = 50 \delta_{in}$,
which is still in the adiabatic portion of the wall, and where
the friction Reynolds number (ratio between the boundary
layer thickness and the viscous length-scale) is $\Rey_{\tau} \approx 450$.
The global properties of the boundary layer at this location are summarized
in Table~\ref{tab:blprop}.

As expected, when the van Driest-transformation
$\mathrm{d}U_{VD} = (\overline{\rho}/\overline{\rho}_w)^{1/2} \mathrm{d}\overline{u}$
is applied to take into account for the variation of the thermodynamic properties through the boundary layer,
a collapse with reference low-speed data at comparable $\Rey_{\tau}$~\citep{schlatter_10b} is observed,
and the mean velocity profile exhibits the onset of a small region with a nearly logarithmic behavior.
The density-scaled Reynolds stresses, reported in~\ref{fig:bl_incoming}b, highlight close
similarities with the incompressible distributions and a very good agreement is also obtained
with reference compressible experiments, except for the wall-normal velocity variance, which
is typically underestimated by measurements.

\begin{figure}
 \centering
 \psfrag{x}[t][][1.2]{$(x-x_T)/\delta_{x_T}$}
 \psfrag{y}[b][][1.2]{$C_h (\cdot 10^3)$}
 \includegraphics[width=5.5cm,angle=270,clip]{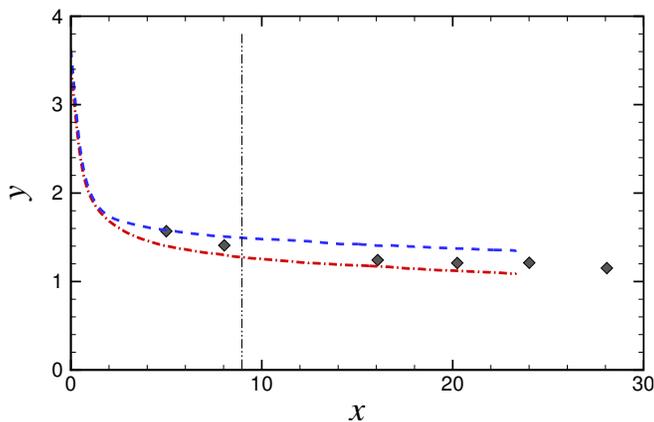} \vskip 0.5em
 \caption{Distribution of the Stanton number as a function of the streamwise distance from the temperature step change,
          normalized by the boundary layer thickness at $x_T$ ($\delta_{x_T}$).
          Refer to table~\ref{tab:testcases} for nomenclature of the DNS data.
          The grey diamonds denote reference experiments with $s=2$ by~\citet{debieve97}.
          The vertical line denotes the impingment shock location for SBLI simulations.}
 \label{fig:ch_comparison}
\end{figure}
The main effect of the temperature step change on the incoming flow can be understood by looking at figure~\ref{fig:t-u}
where the temperature-velocity relationship in the boundary layer
is reported for simulations BL-s0.5 and BL-s1.9 at various stations along the streamwise direction,
from $x_0$ to the end of the computational domain.
This representation is very suited to describe the adaptation process
of the boundary layer to the new thermal conditions at the wall. The shape of the profiles at the various
x-stations suggests that the outer region of the boundary layer significantly deviates from the
equilibrium Walz solution,
\begin{equation}
 \label{eq:walz}
 \frac{\overline{T}}{T_{\infty}} = \frac{T_w}{T_{\infty}} +
                                   \frac{T_r-T_w}{T_{\infty}} \frac{\overline{u}}{u_{\infty}} +
                                   \frac{T_{\infty}-T_r}{T_{\infty}} \left ( \frac{\overline{u}}{u_{\infty}} \right)^2 \quad
 T_r = T_{\infty} + r \frac{u_{\infty}^2}{2 \, C_p}, r = (Pr)^{1/3}
\end{equation}
and even at the end of the computational domain
the recovery process is not yet completed for both the cold and hot wall cases.
A similar conclusion was also reported by~\citet{debieve97}, who investigated the effect of heating
by considering a step change in the wall temperature distribution of a spatially evolving supersonic
turbulent boundary layer at freestream Mach number $M_{\infty}=2.3$, wall-to-recovery
temperature ratio $s=2$ and Reynolds number based on the momentum thickness at the temperature step change
$\Rey_{\theta} = 4100$. 
Their data, taken 8 boundary layer thicknesses downstream the beginning of the heated wall,
are also included in figure~\ref{fig:t-u}. The close agreement between the experimental measurements,
and the DNS profile at the corresponding location provides a confirmation of the quality
of the present simulations with non-adiabatic wall conditions.

A further comparison is shown in figure~\ref{fig:debieve}, where the distribution of the total temperature
in the boundary layer is shown. The figure allows to appreciate the rapid growth of the thermal boundary layer
starting from the step change position and again highlights a remarkable agreement between the experimental measurements
and DNS data, despite the slightly different nominal conditions in the wall temperature and Reynolds number.

To highlight the effect of heating/cooling on the heat transfer rate, the spatial distribution of the Stanton number 
\begin{equation}
 \nonumber
 C_h = \frac{q_w}{\rho_{\infty} \, u_{\infty} C_p \, \left ( T_w - T_r \right )}, \qquad q_w = -k \left . \frac{\mathrm{d}\overline{T}}{\mathrm{d}y}\right|_w
\end{equation}
is shown in figure~\ref{fig:ch_comparison}, where the origin of the streamwise coordinate is located at the beginning of the step change ($x_T$).
For both cooling and heating, the simulation predicts a rapid decay of the heat transfer coefficient towards values typical of an equilibrium
boundary layer, and in agreement with recent DNS data~\citep{hadjadj15}, $C_h$ is found to increase when $s$ decreases.
In this case the agreement with the experimental data (available for the hot wall) is reasonably good, the computed values
being approximately $8 \%$ lower than the measurements. These differences might be explained recalling that in the experiment $C_h$ was
computed through an iterative procedure based on the theoretical Walz's temperature-velocity relationship,
which is far from being valid past the step change location, as previously seen in figure~\ref{fig:debieve}.

\subsection{Effect of wall temperature on SBLI flow fields}

\begin{figure}
 \centering
 \includegraphics[width=1.8cm,angle=270,clip]{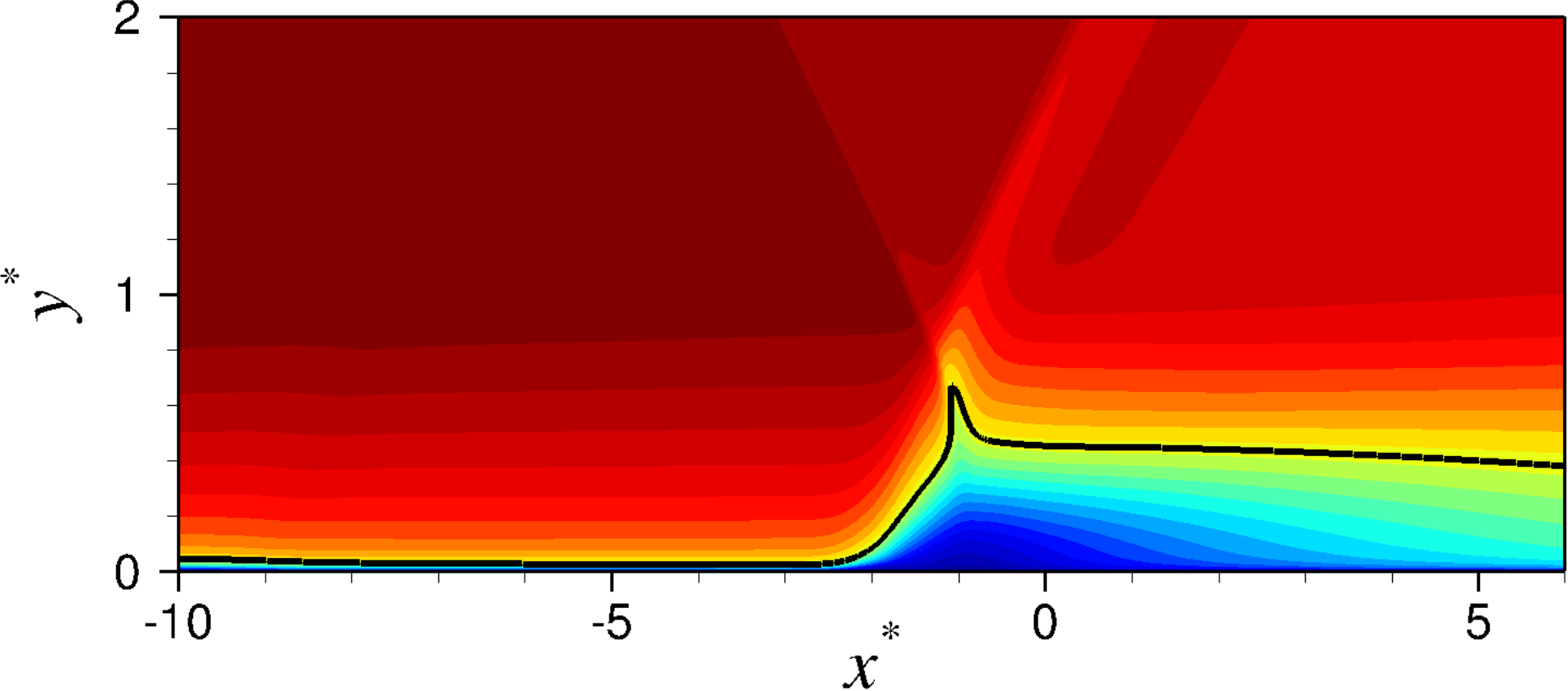} \hskip 0.5em
 \includegraphics[width=1.8cm,angle=270,clip]{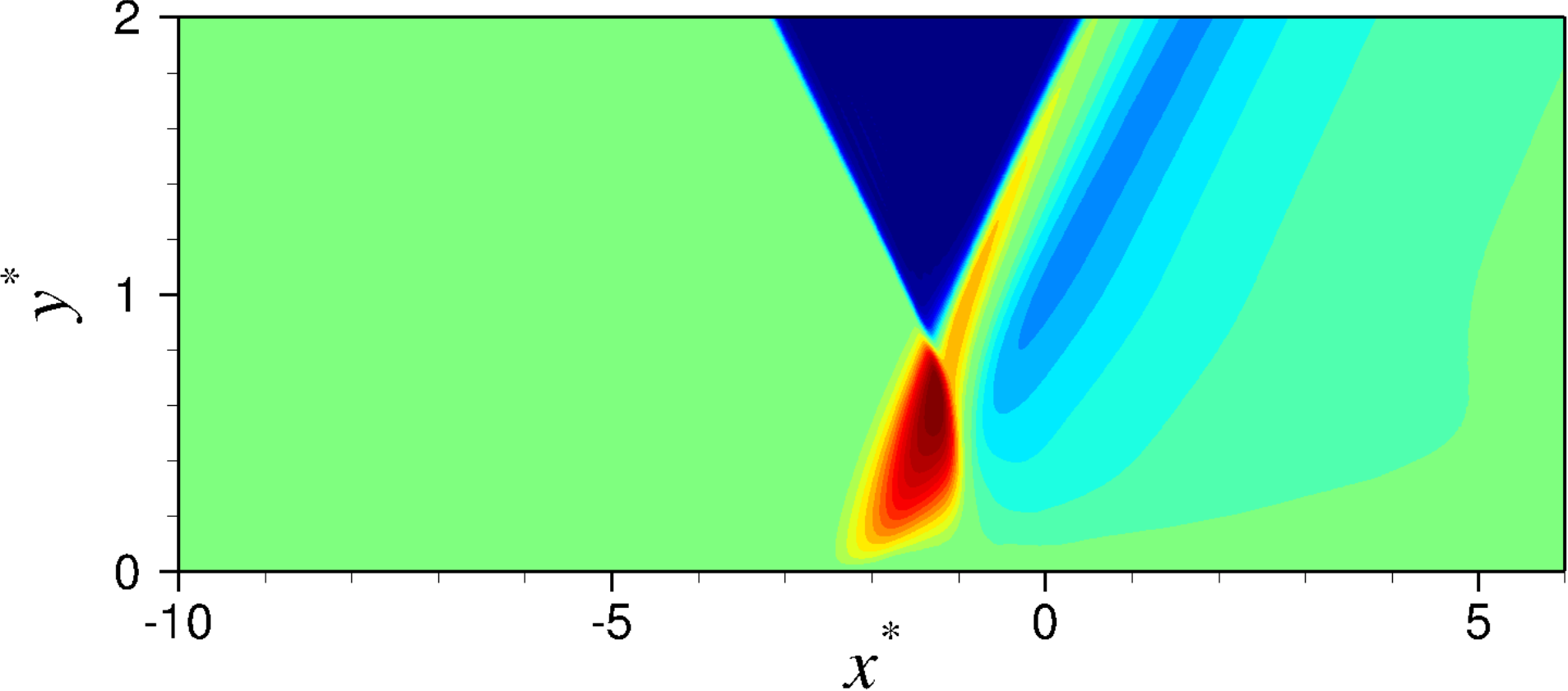} \hskip 0.5em
 \includegraphics[width=1.8cm,angle=270,clip]{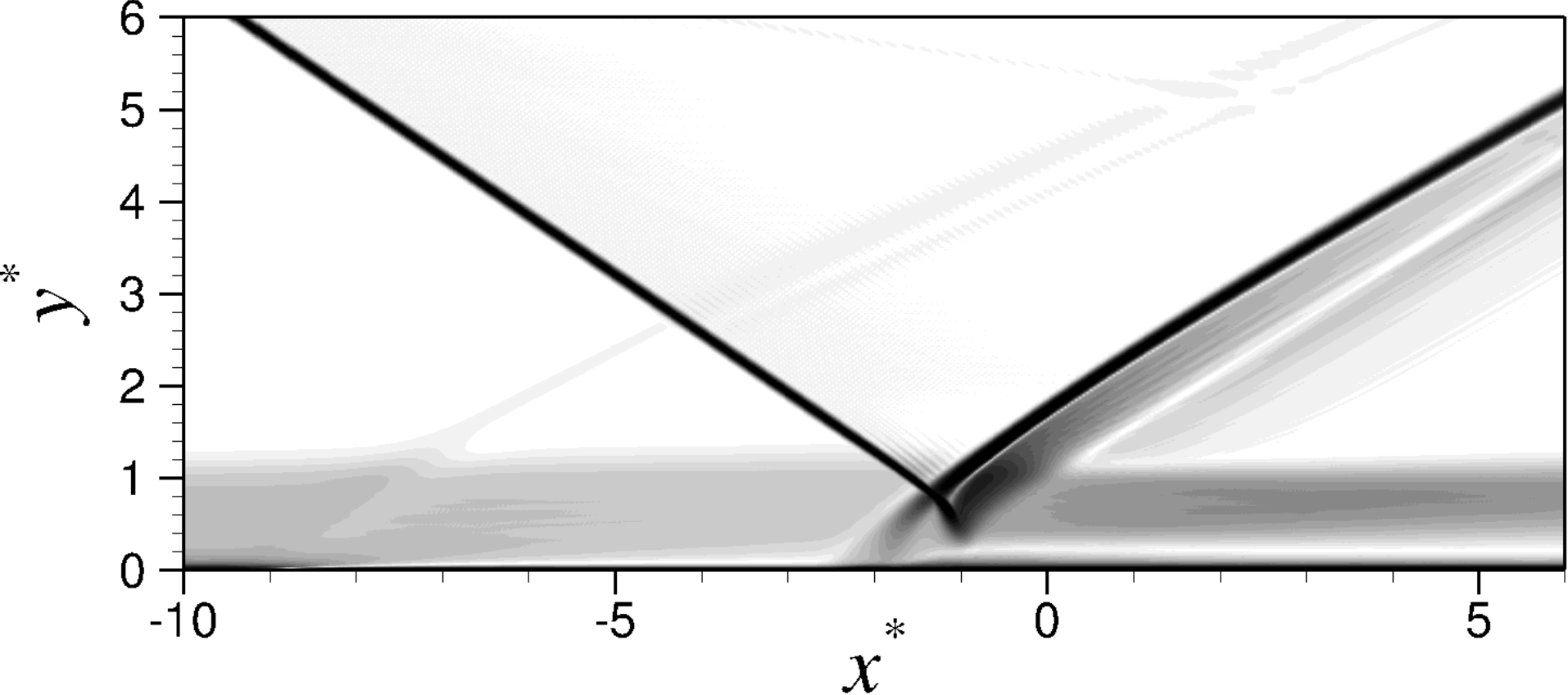} \vskip 0.5em
 \includegraphics[width=1.8cm,angle=270,clip]{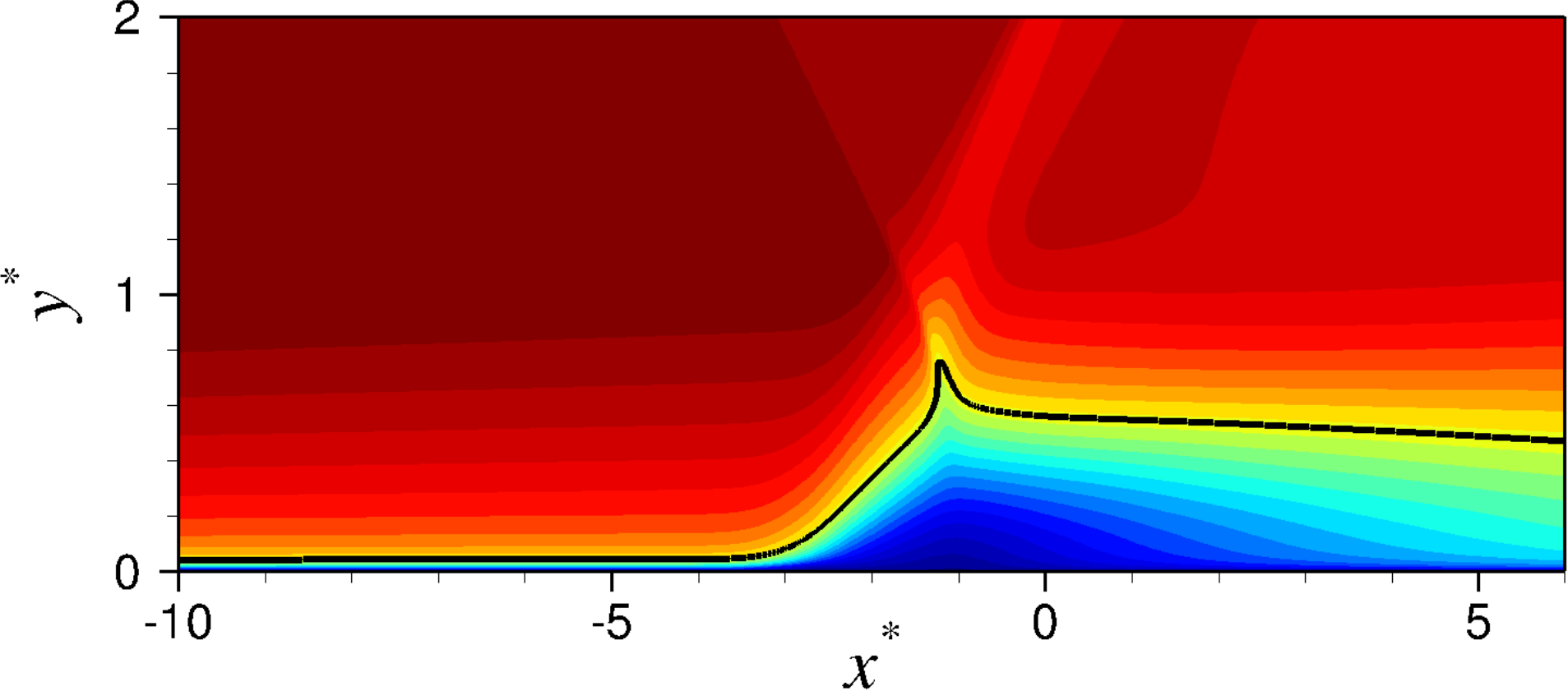} \hskip 0.5em
 \includegraphics[width=1.8cm,angle=270,clip]{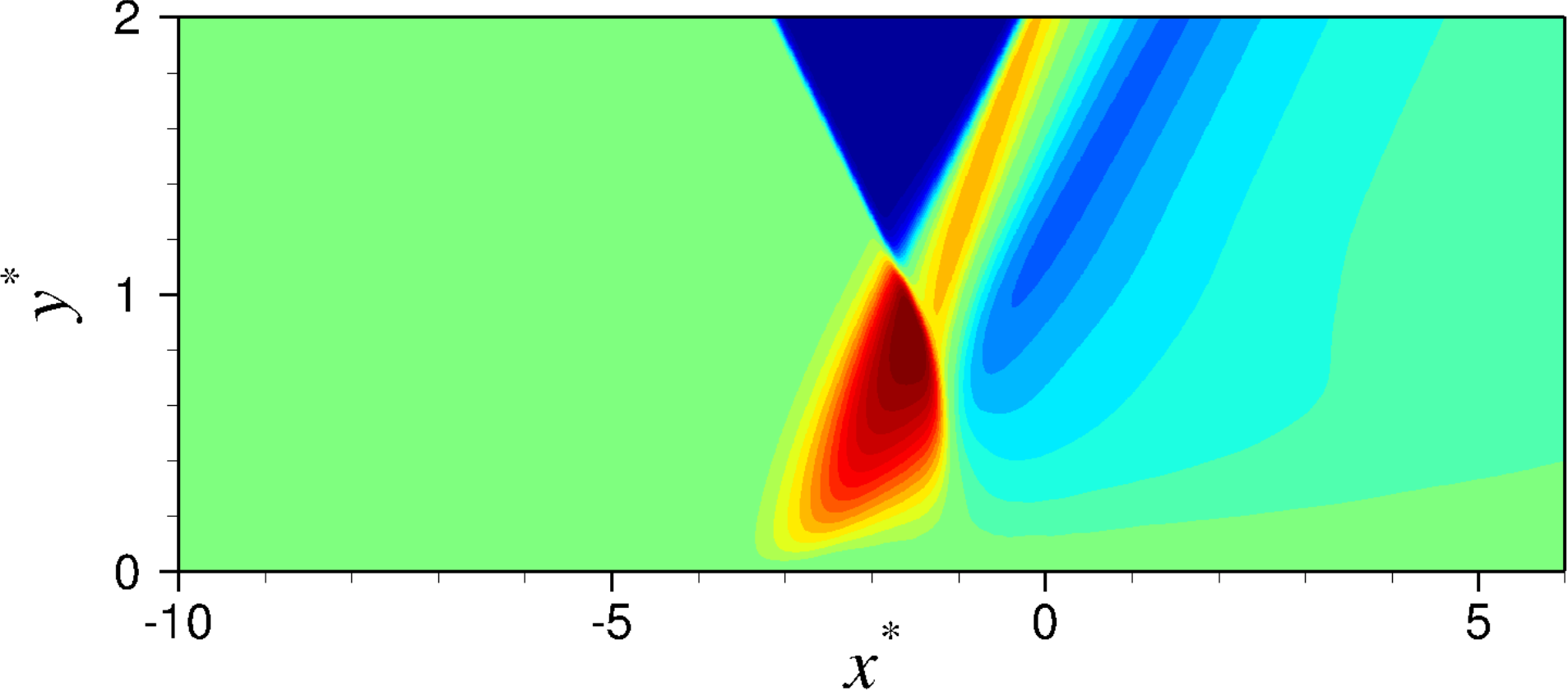} \hskip 0.5em
 \includegraphics[width=1.8cm,angle=270,clip]{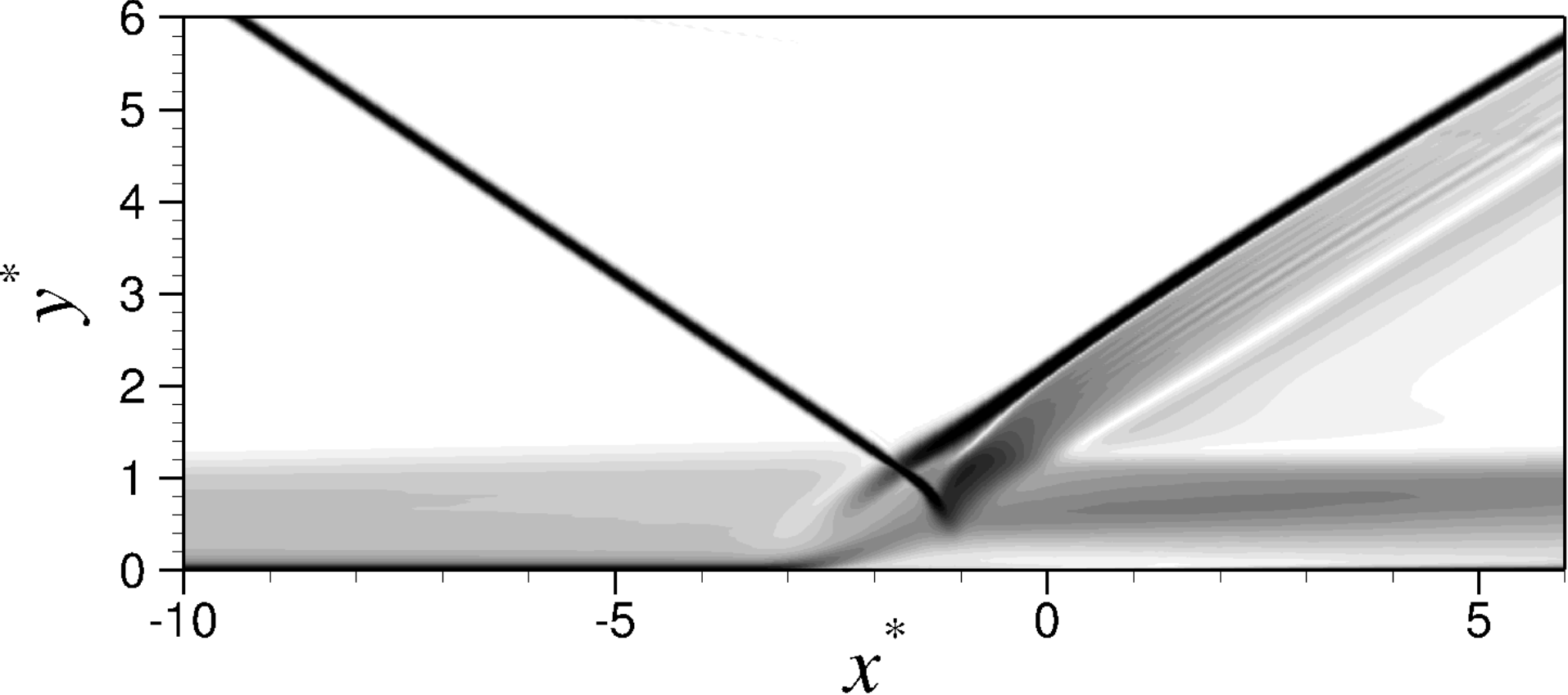} \vskip 0.5em
 \includegraphics[width=1.8cm,angle=270,clip]{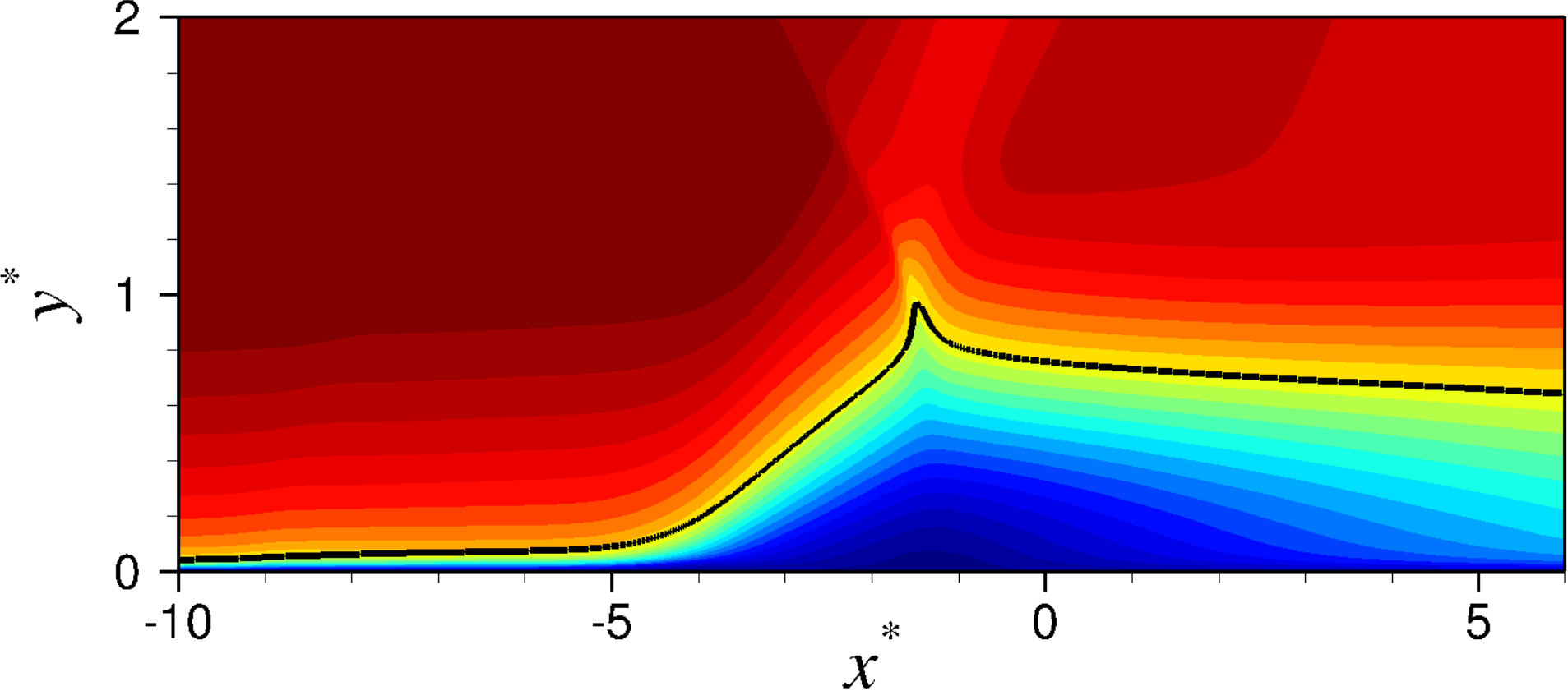} \hskip 0.5em
 \includegraphics[width=1.8cm,angle=270,clip]{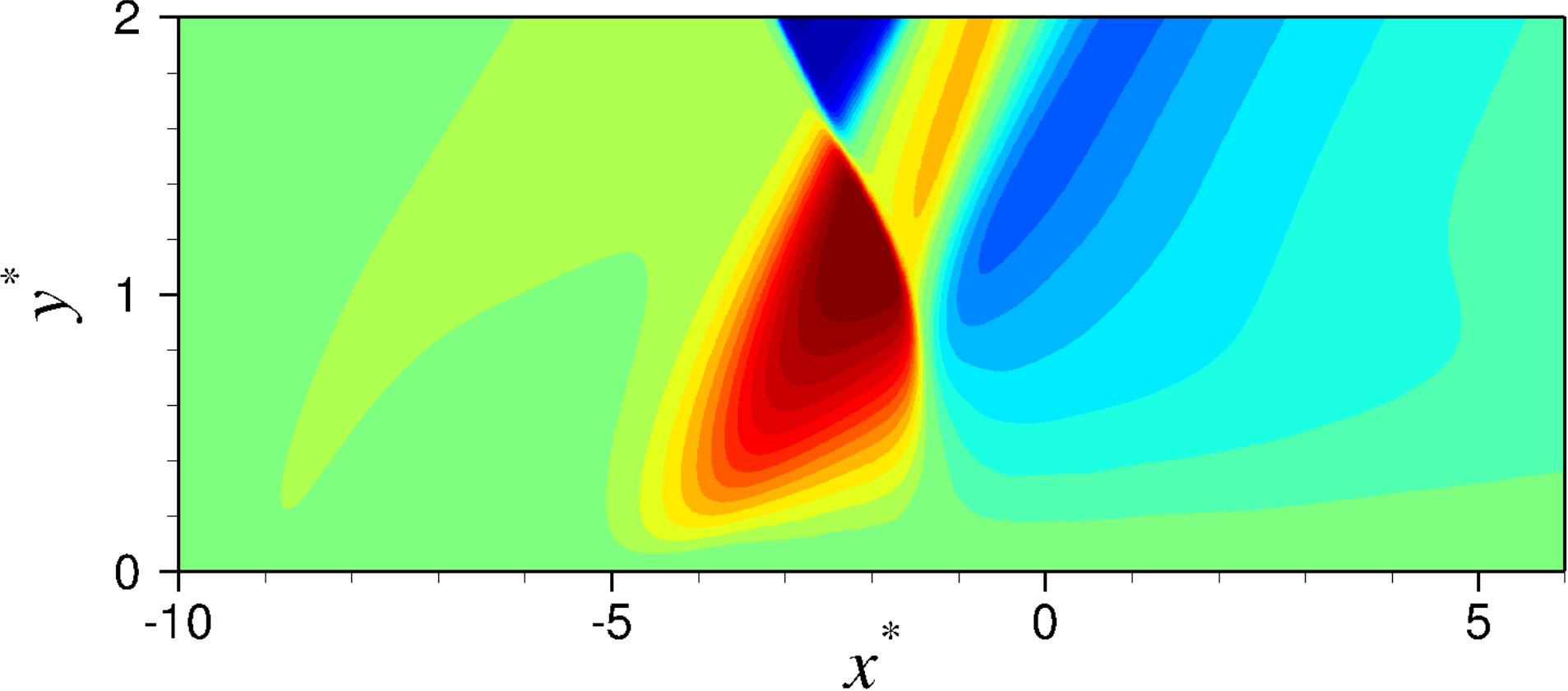} \hskip 0.5em
 \includegraphics[width=1.8cm,angle=270,clip]{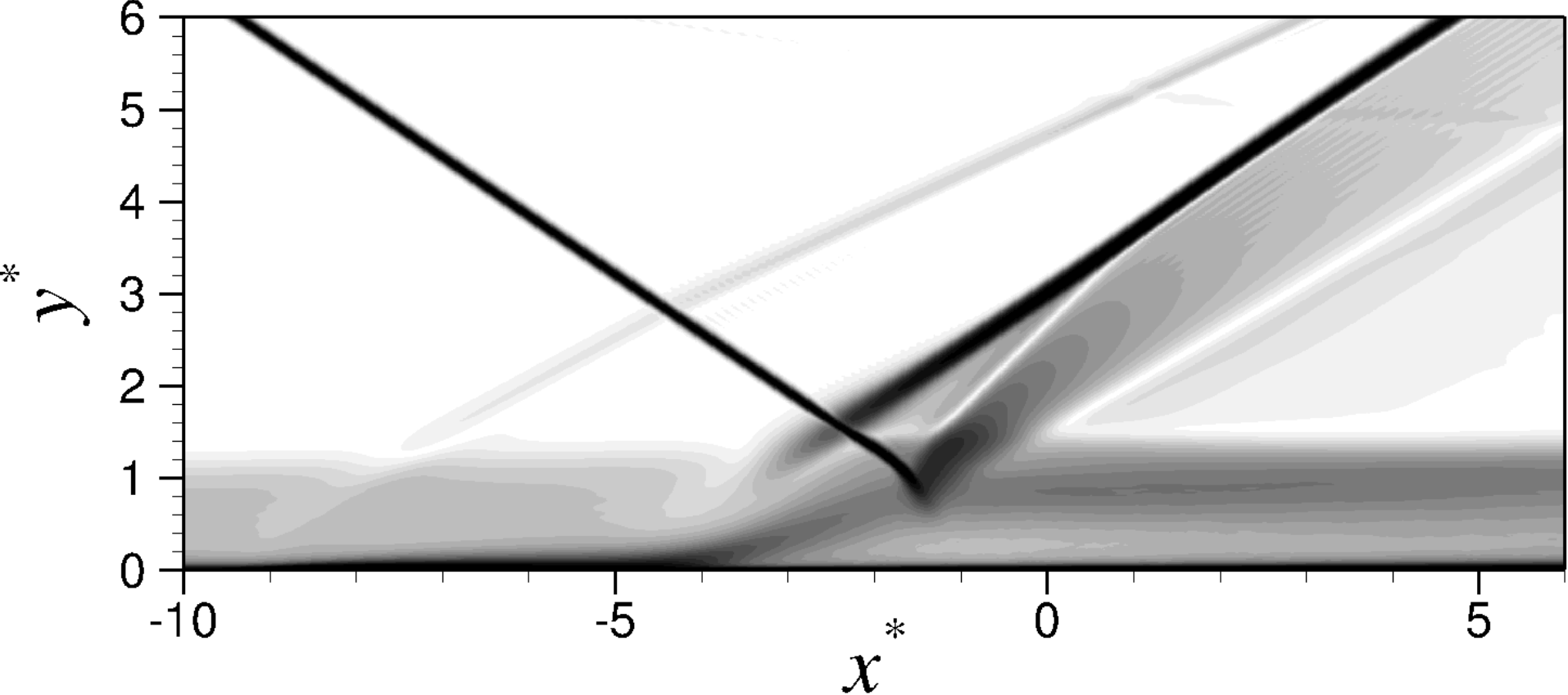} \vskip 0.5em
 \caption{Contours of mean streamwise velocity (left panels), vertical velocity (middle panels)
          and magnitude of the mean density gradient (right panels) at various wall-to-recovery-temperature ratios, increasing
          from top to bottom ($s = 0.5,1,1.9$). The black line denotes the sonic line.
          Twenty-four contour levels are shown in the range: $0 < \overline{u}/u_{\infty} < 1$;
          $-0.13 < \overline{v}/u_{\infty} < 0.13$; $0 < e^{- \left |\nabla \overline{\rho} \right |/\rho_{\infty}} < 1$.}
 \label{fig:uvpmean}
\end{figure}

To provide an overview of the flow organization and a qualitative perception of the
influence of the wall thermal conditions
we report in figure~\ref{fig:uvpmean} contours of mean velocity components and of mean density gradient magnitude
for some representative values of $s$ (0.5, 1 and 1.9).
The typical topology of SBLI is observed for all flow cases, independently of the wall temperature:
i) the incoming turbulent boundary layer thickens within the interaction region and
relaxes to a new equilibrium state further downstream;
ii) a compression fan develops near the separation point well upstream of the nominal impinging location;
iii) away from the wall the compression waves coalesce to form the principal reflected shock; and
iv) the flow turns through an expansion fan towards the wall and reattaches.

\begin{figure}
 \centering
 \includegraphics[width=1.8cm,angle=270,clip]{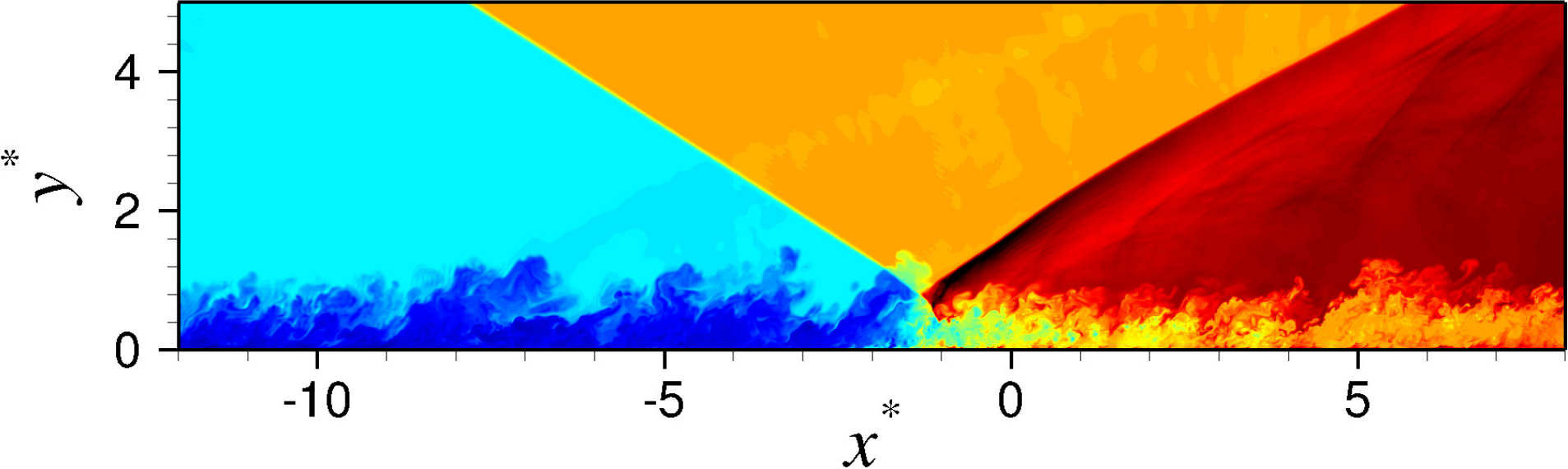} \hskip 0.5em
 \includegraphics[width=1.8cm,angle=270,clip]{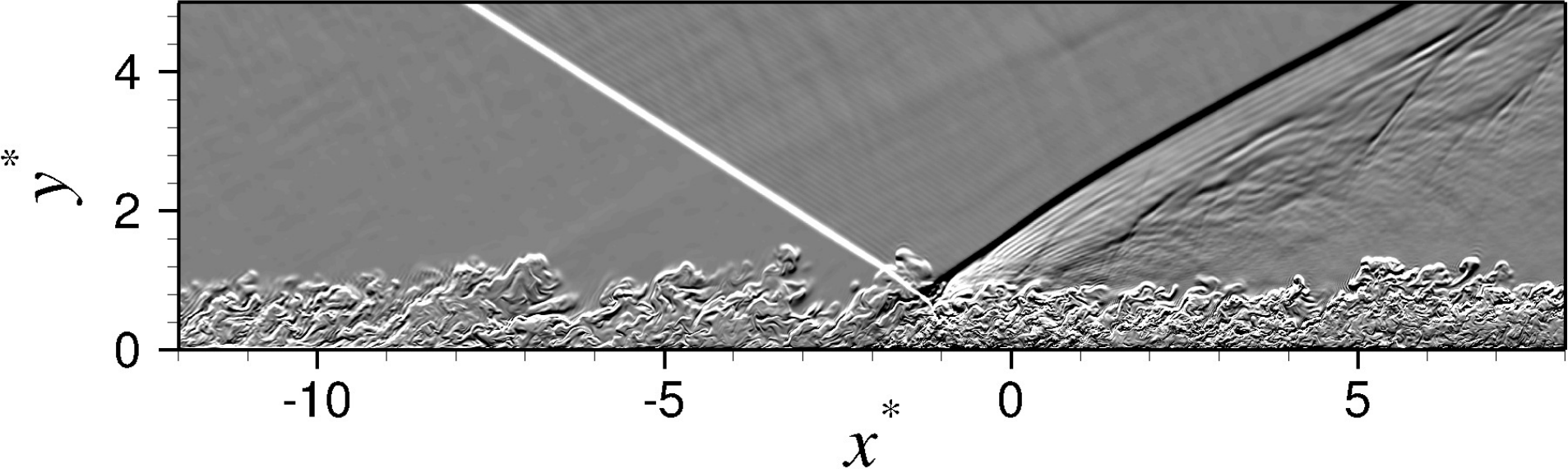} \vskip 0.5em
 \includegraphics[width=1.8cm,angle=270,clip]{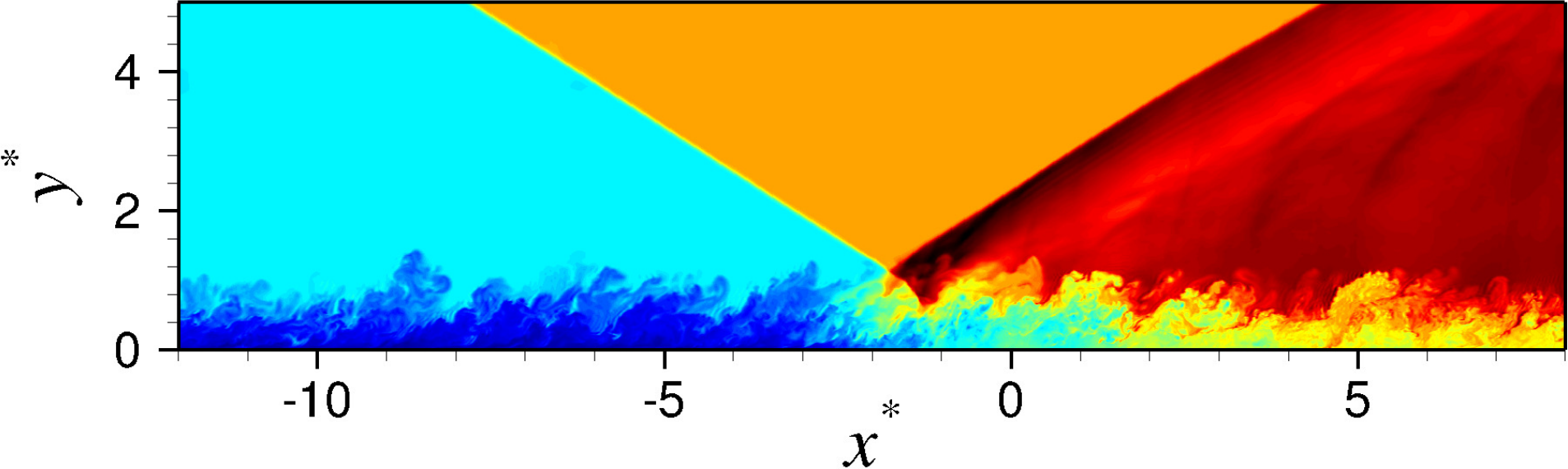} \hskip 0.5em
 \includegraphics[width=1.8cm,angle=270,clip]{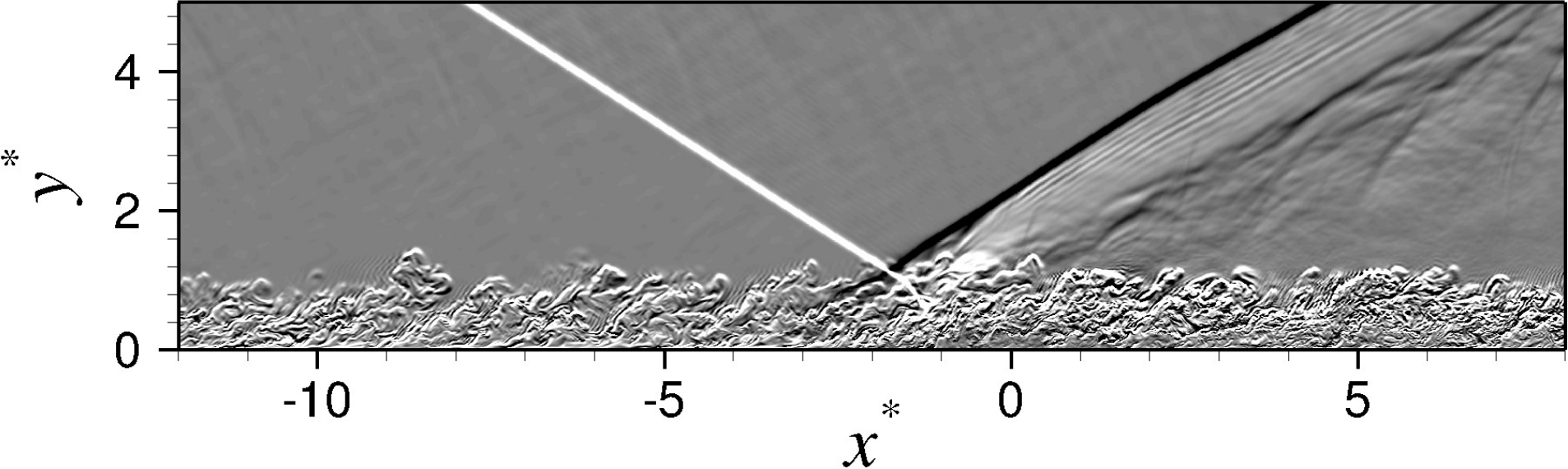} \vskip 0.5em
 \includegraphics[width=1.8cm,angle=270,clip]{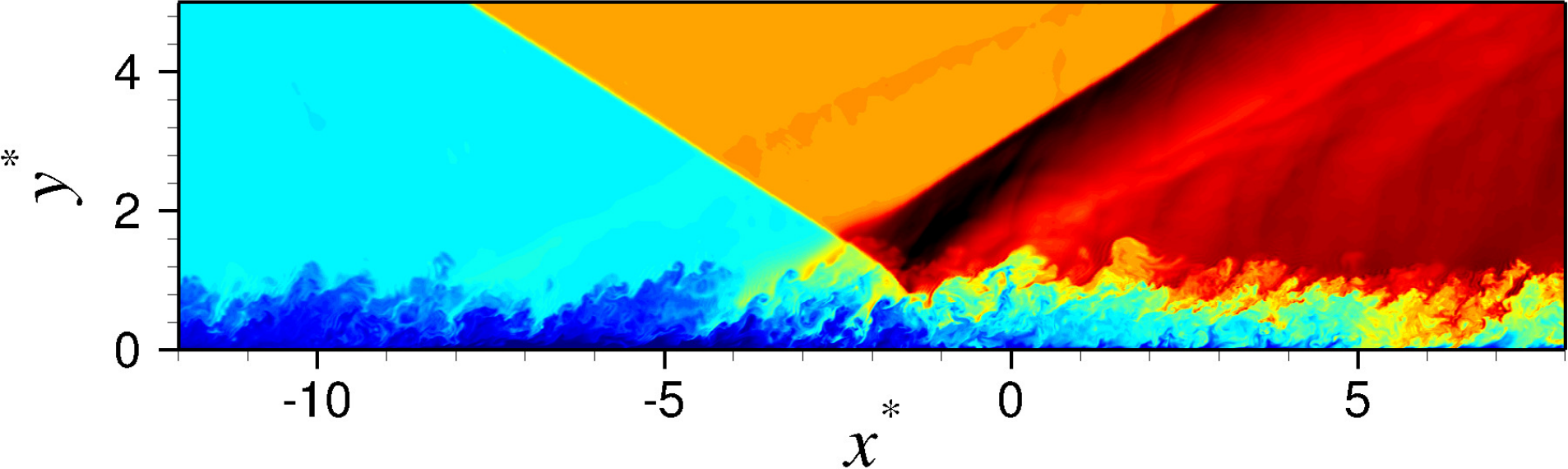} \hskip 0.5em
 \includegraphics[width=1.8cm,angle=270,clip]{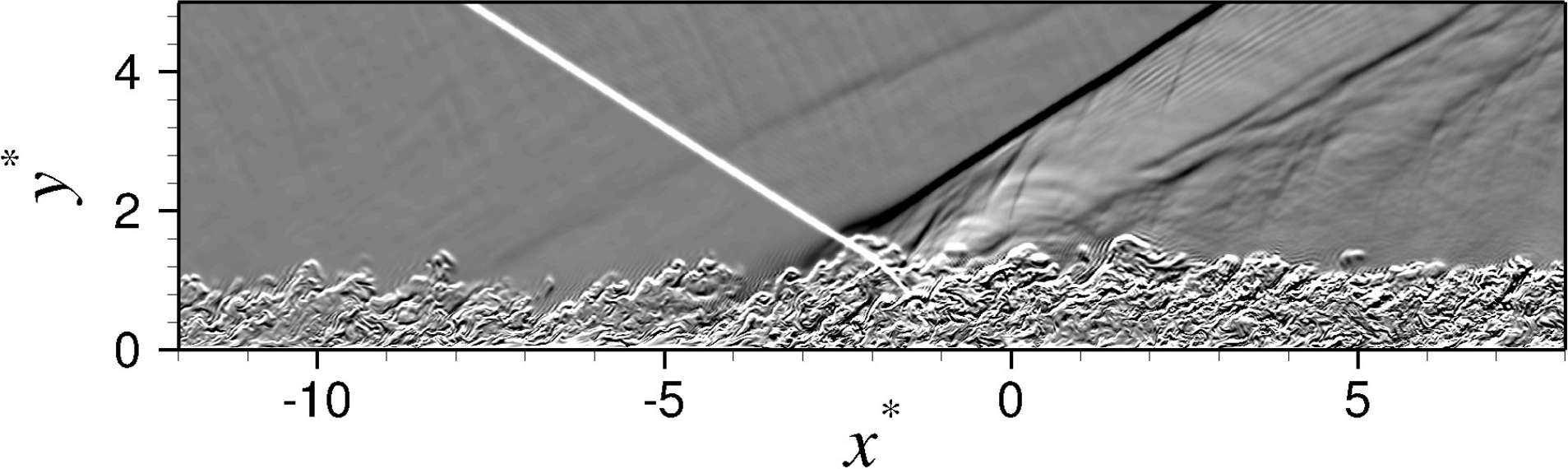} \vskip 0.5em
 \caption{Contours of instantaneous density (left panels) and wall-normal density gradient (right panels)
          in the longitudinal mid-plane at various wall-to-recovery-temperature ratios, increasing
          from top to bottom ($s = 0.5,1,1.9$). Sixty-four contour levels are shown in the range:
          $0.48 < \rho/\rho_{\infty} < 2.12$;
          $-1.5 < \mathrm{d}\rho/\mathrm{d}y/\rho_{\infty} < 1.5$.}
 \label{fig:rhoxy}
\end{figure}
Snapshots of the instantaneous density field and of its
wall-normal derivative (numerical schlieren) in the longitudinal mid-plane
are reported in figure~\ref{fig:rhoxy}.
These visualizations bring to light the convoluted structures
of the turbulent boundary layer and allow to appreciate the complex pattern of waves originating
from the interaction with the impinging shock. The step change imposed in the wall temperature
distribution is also revealed in figure~\ref{fig:rhoxy} by the formation of a weak disturbance
originated at $x^* \approx -9$, also visible in the mean density gradient of figure~\ref{fig:uvpmean}.
The main effect of the wall thermal condition is a change in the interaction scales,
well highlighted by the mean and instantaneous visualizations, that clearly shows that the impinging shock
penetrates deeper in the incoming turbulent boundary layer when the wall temperature is reduced
This effect is mainly associated with the displacement of the sonic line (displayed in~\ref{fig:uvpmean})
towards to (away from) the wall with wall cooling (heating).
The interaction length-scale $L$ (see table~\ref{tab:testcases}),
defined as the distance between the nominal incoming shock impingement point and the
apparent origin of the reflected shock, is strongly affected by $s$.
Compared to the adiabatic case, $L$ decreases (increases)
significantly with wall cooling (heating), in agreement with previous experimental findings
for impinging shock and compression ramp configurations~\citep{spaid72,jaunet14}.

\begin{figure}
 \centering
 \includegraphics[width=2.4cm,angle=270,clip]{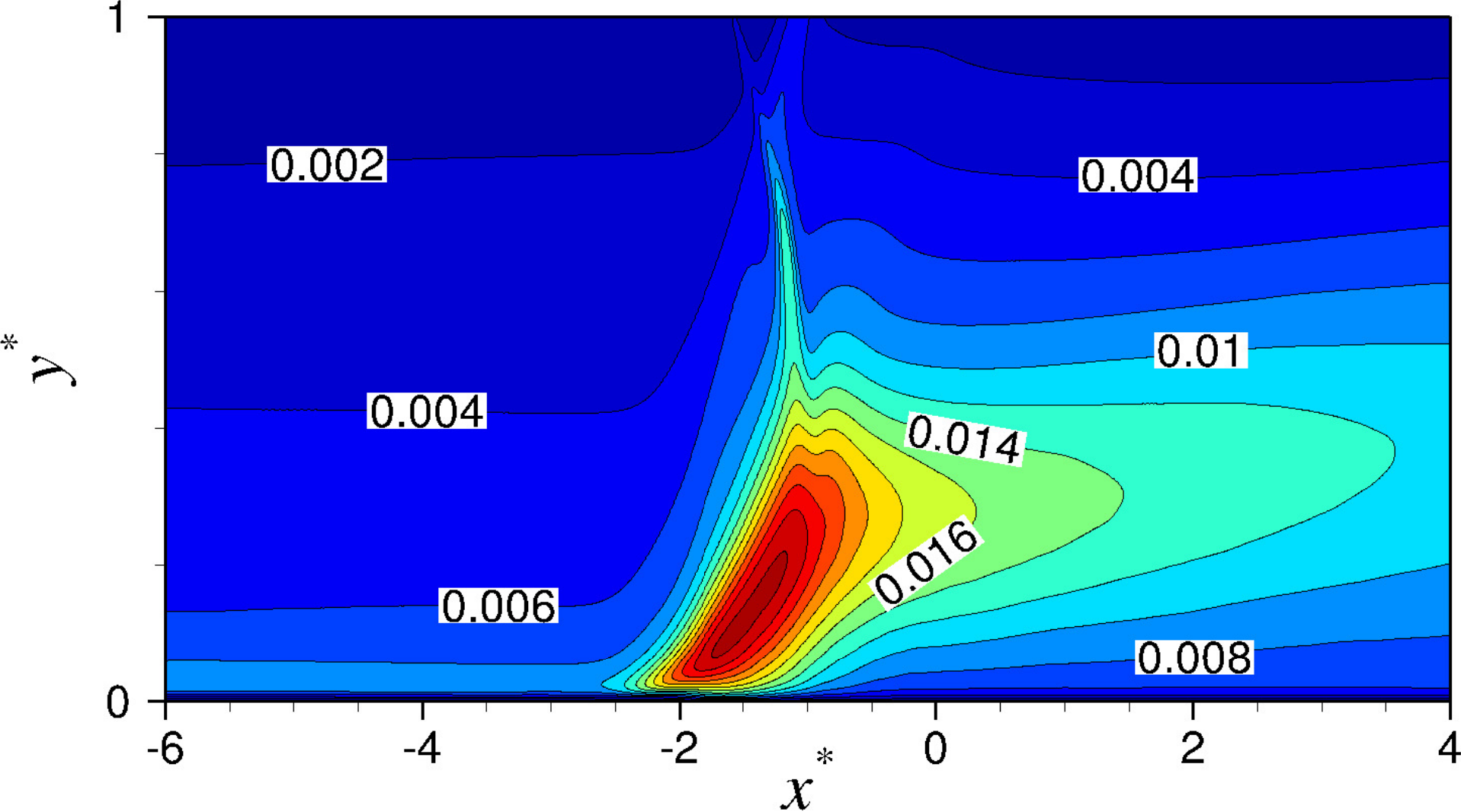} \hskip 0.5em
 \includegraphics[width=2.4cm,angle=270,clip]{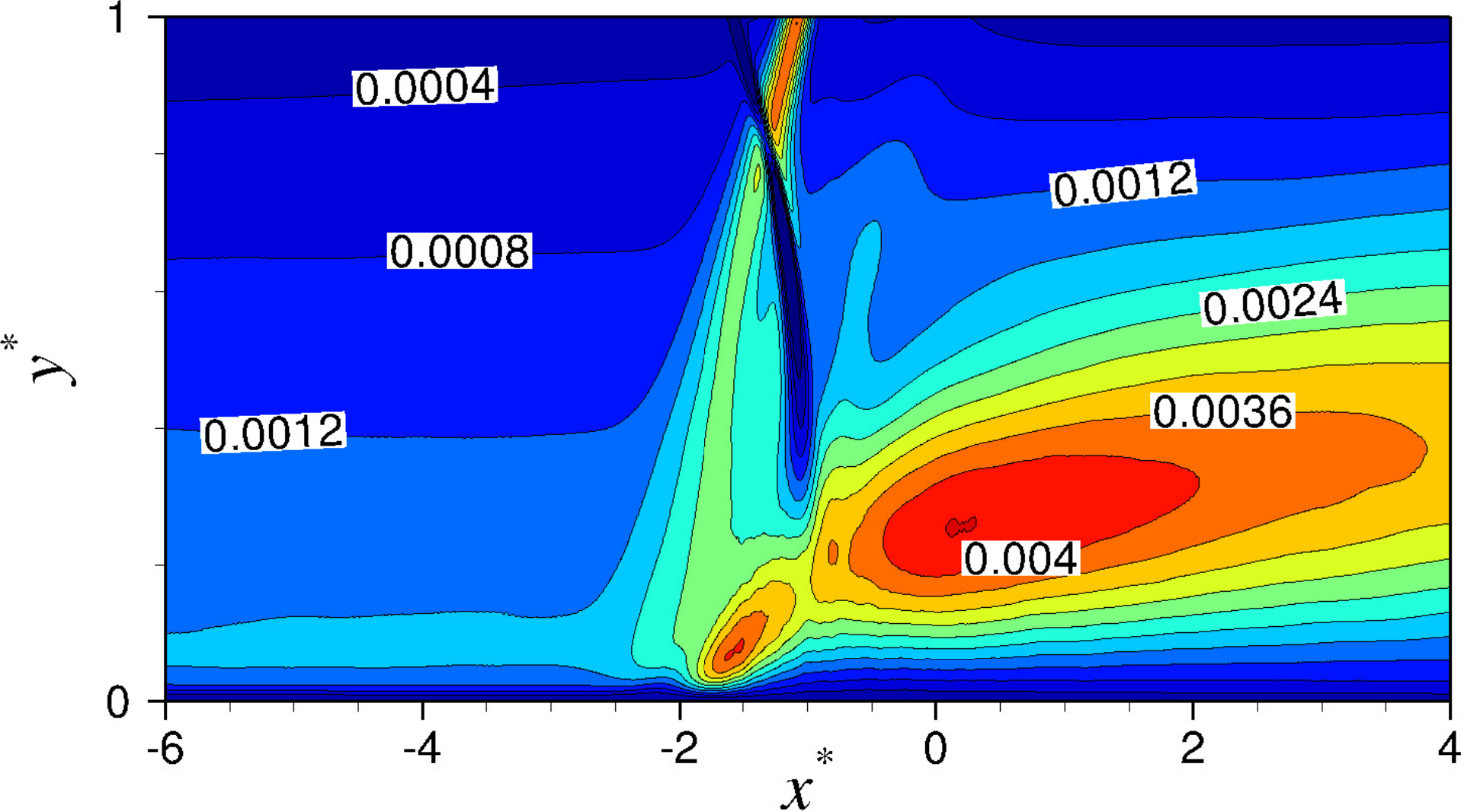} \hskip 0.5em
 \includegraphics[width=2.4cm,angle=270,clip]{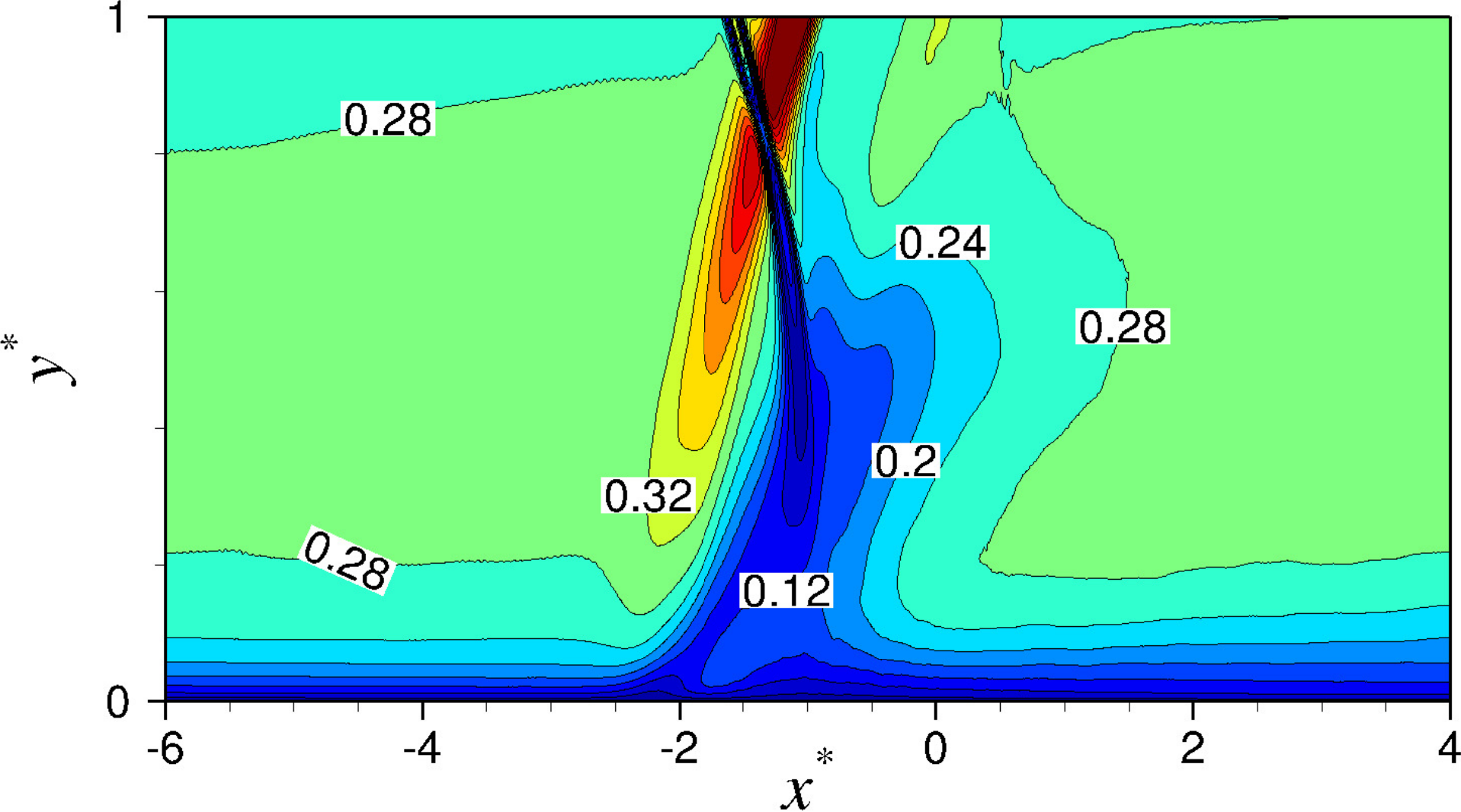} \vskip 0.5em
 \includegraphics[width=2.4cm,angle=270,clip]{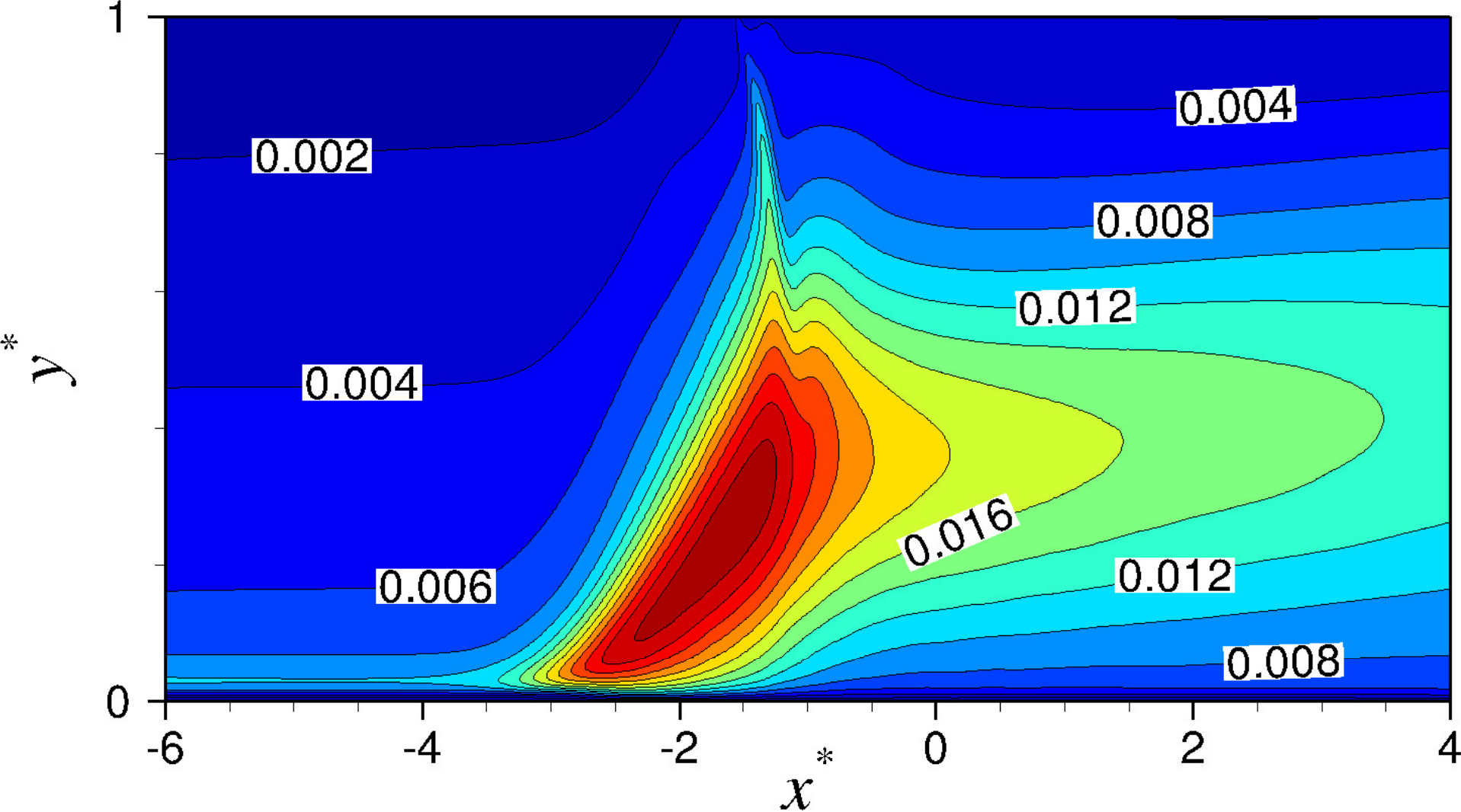} \hskip 0.5em
 \includegraphics[width=2.4cm,angle=270,clip]{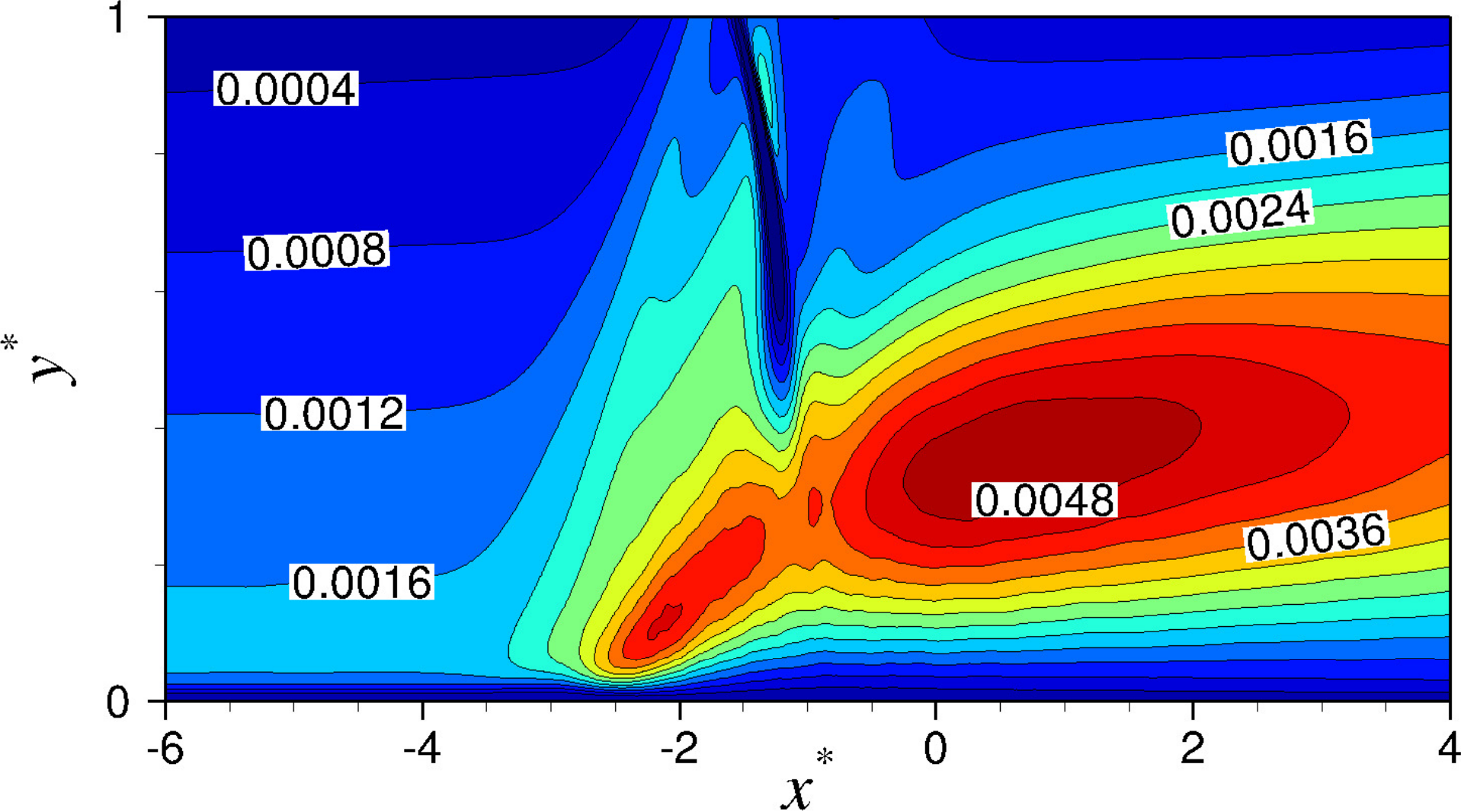} \hskip 0.5em
 \includegraphics[width=2.4cm,angle=270,clip]{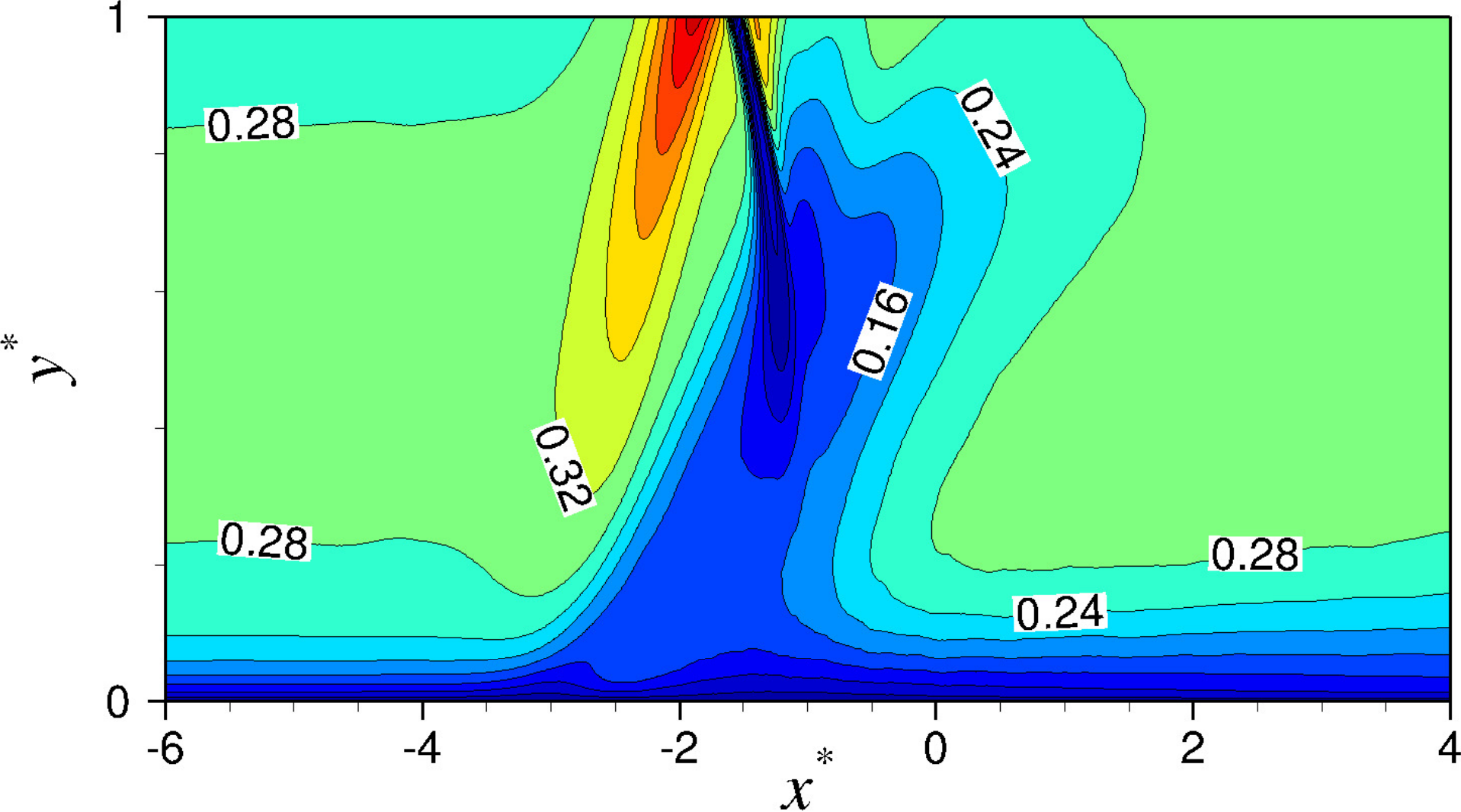} \vskip 0.5em
 \includegraphics[width=2.4cm,angle=270,clip]{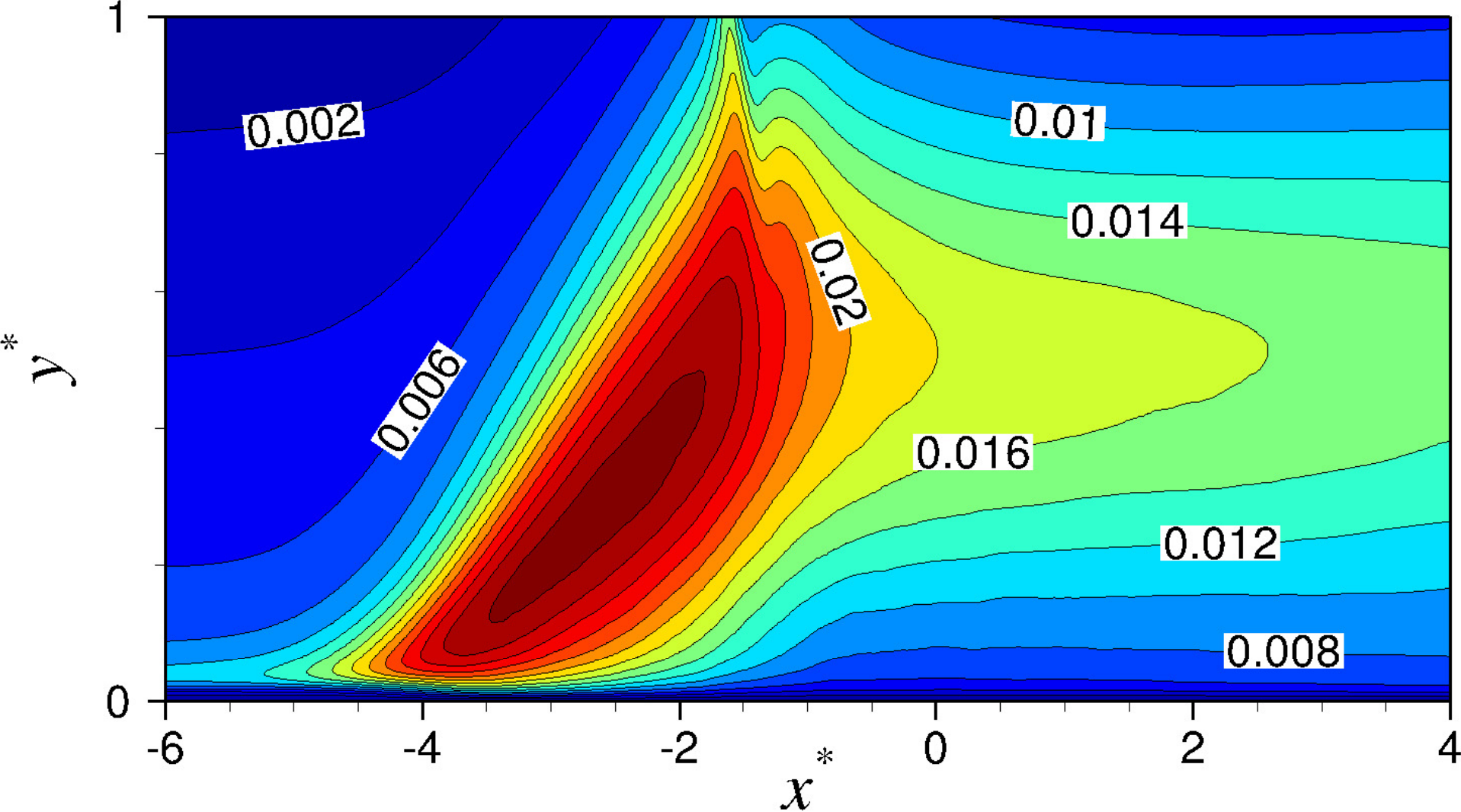} \hskip 0.5em
 \includegraphics[width=2.4cm,angle=270,clip]{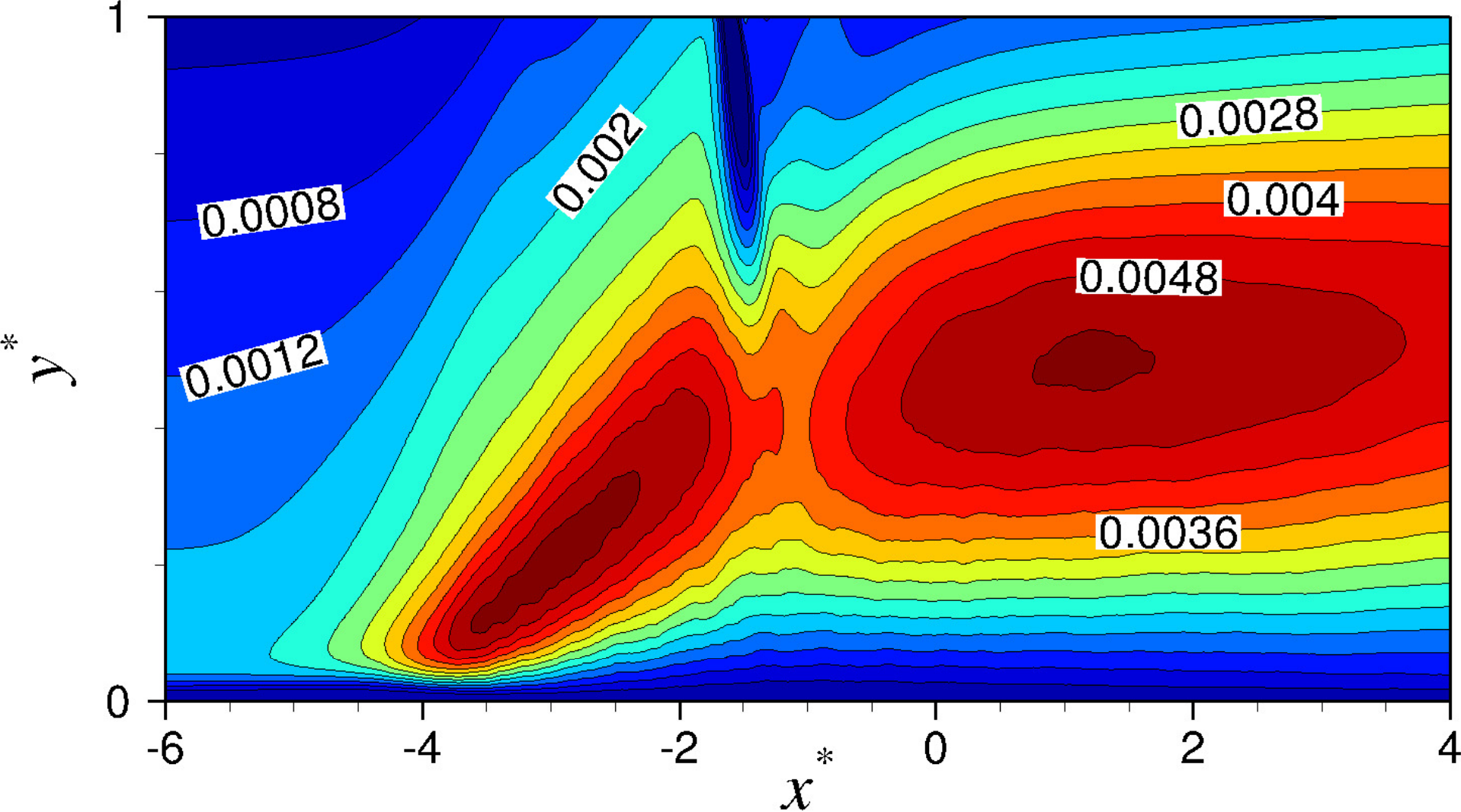} \hskip 0.5em
 \includegraphics[width=2.4cm,angle=270,clip]{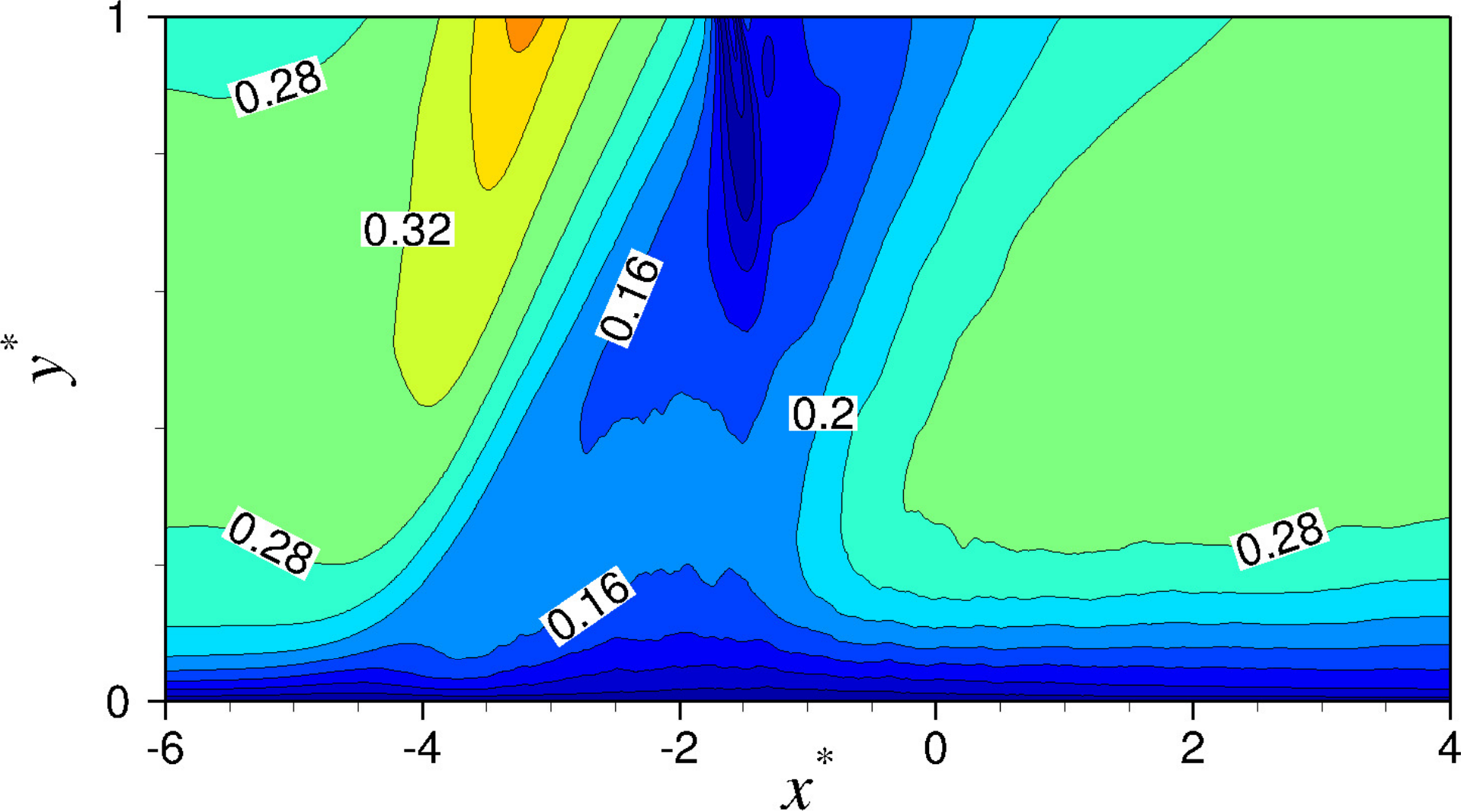} \vskip 0.5em
 \caption{Contours of turbulence kinetic energy $k/u_{\infty}^2$ (left panels),
          Reynolds shear stress $\widetilde{u^{\prime \prime} v^{\prime \prime}}/u_{\infty}^2$ (middle panels) and
          structure parameter $\Pi$ (right panels)
          at various wall-to-recovery-temperature ratios, increasing from top to bottom ($s = 0.5,1,1.9$).}
 \label{fig:tke_uv}
\end{figure}
\begin{figure}
 \centering
 \psfrag{x}[t][][1.0]{$x^*$}
 \psfrag{y}[b][][1.0]{$\widetilde{u^{\prime \prime}u^{\prime \prime}}/u_{\infty}^2$}
 \includegraphics[width=3.4cm,angle=270,clip]{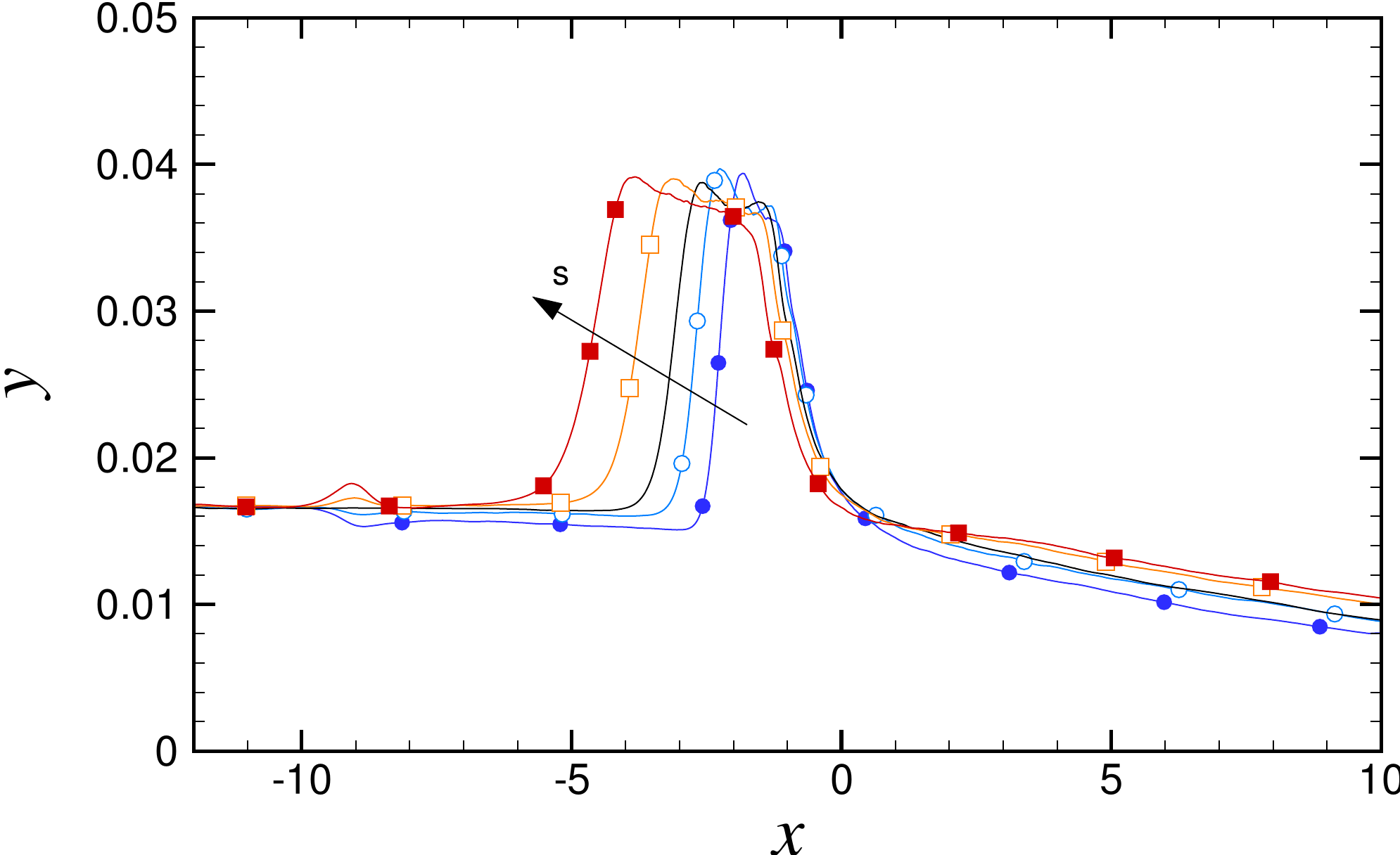} \hskip 1em
 \psfrag{y}[b][][1.0]{$\widetilde{v^{\prime \prime}v^{\prime \prime}}/u_{\infty}^2$}
 \includegraphics[width=3.4cm,angle=270,clip]{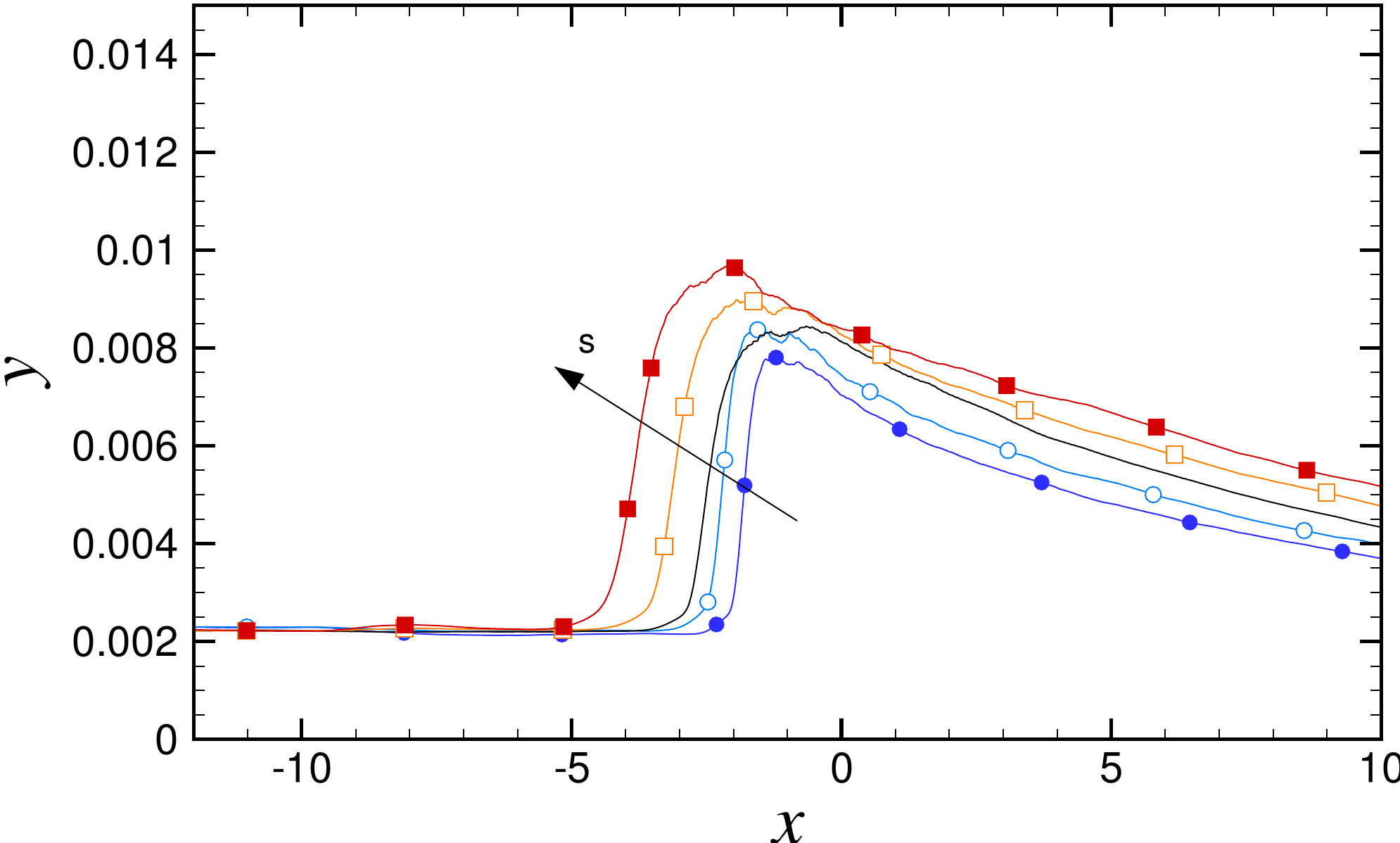} \vskip 0.5em
 \psfrag{y}[b][][1.0]{$\widetilde{w^{\prime \prime}w^{\prime \prime}}/u_{\infty}^2$}
 \includegraphics[width=3.4cm,angle=270,clip]{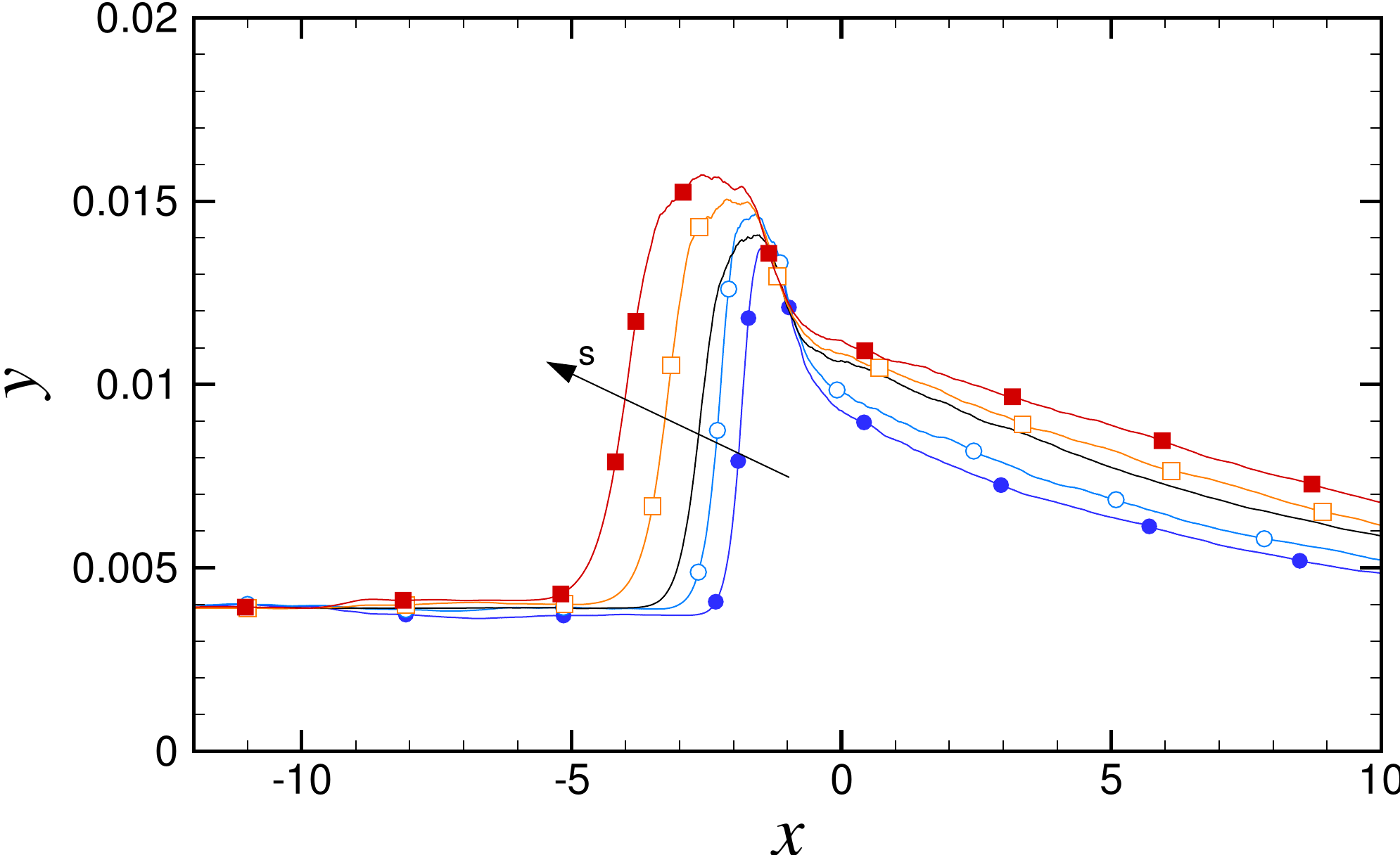} \hskip 1em
 \psfrag{y}[b][][1.0]{$-\widetilde{u^{\prime \prime}v^{\prime \prime}}/u_{\infty}^2$}
 \includegraphics[width=3.4cm,angle=270,clip]{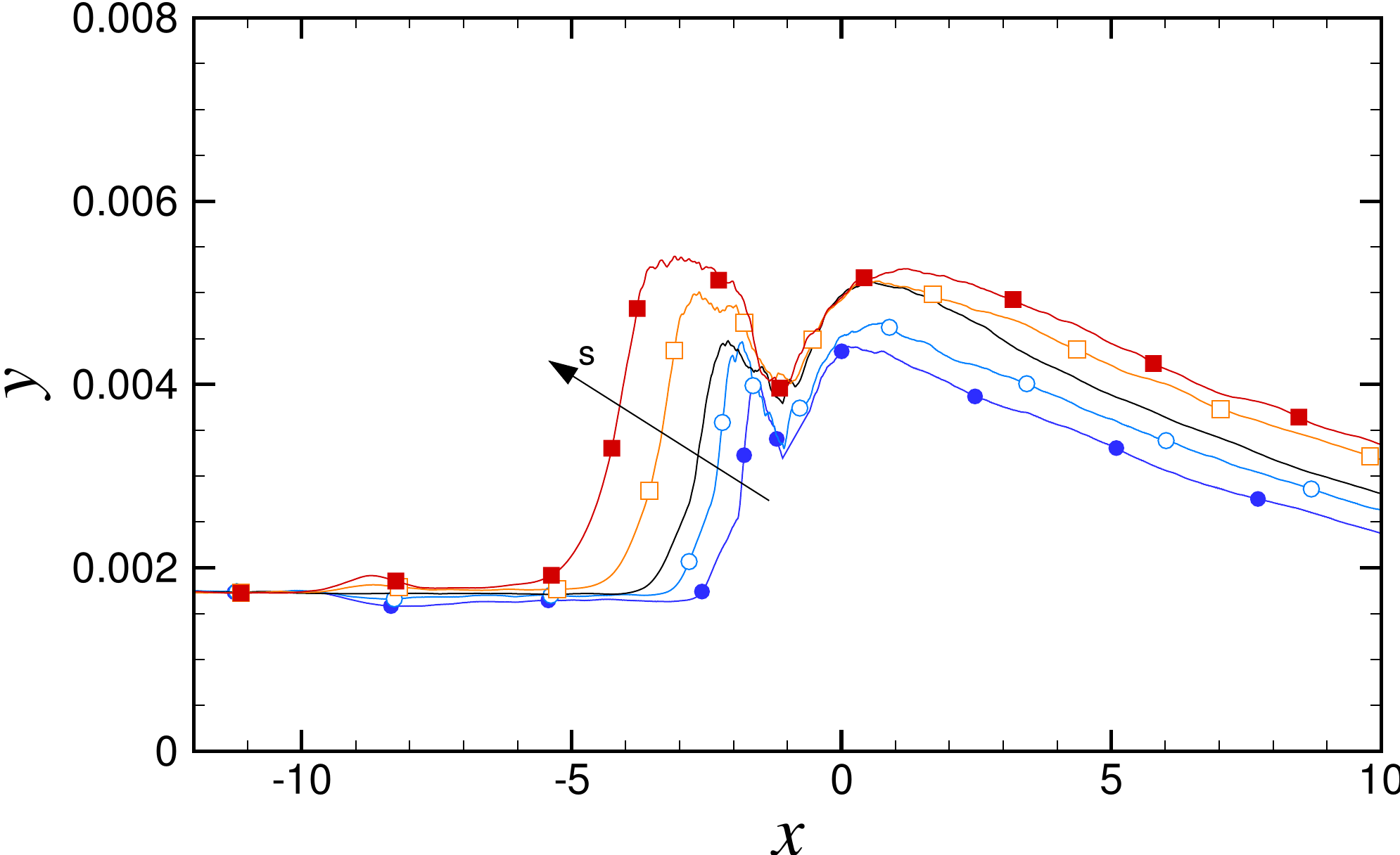} \vskip 0.5em
 \caption{Streamwise sistribution of the peaks of the Reynolds stress components for the various SBLI flow cases.
          Refer to table~\ref{tab:testcases} for nomenclature of the DNS data.}
 \label{fig:reymax}
\end{figure}
A strong amplification of turbulence kinetic energy $k = \widetilde{u_i^{\prime \prime} u_i^{\prime \prime}}/2$
and Reynolds shear stress $\widetilde{u^{\prime \prime} v^{\prime \prime}}$
is found across the interaction region, as revealed by figure~\ref{fig:tke_uv}. For all SBLI cases, a remarkable
growth is observed in the first part of the interaction and the maximum values of both
$k$ and $\widetilde{u^{\prime \prime} v^{\prime \prime}}$ are seen to gradually detach from the wall.
This behavior is associated with the development of a shear layer at the separation shock
and is consistent with previous numerical and experimental findings in supersonic~\citep{delery92,dupont_06} and
transonic interactions~\citep{pirozzoli_10_2}. To characterize the behavior of turbulence across the interaction,
the ratio between the absolute value of the shear stress and the turbulence kinetic energy,
known as structure parameter ($\Pi$), is also reported in figure~\ref{fig:tke_uv}. 
In the upstream region this quantity is approximately constant for all cases, assuming a value
typical of a turbulent boundary layer not too far from the equilibrium ($\Pi \approx 0.3$).
At the beginning of the interaction, independently of $s$, a rapid decrease is observed and 
$\Pi$ attains values in the range $0.1 \div 0.15$, before gradually recovering the original value.
The influence of the wall temperature on the behavior of the structure parameter is found to be marginal,
except for the previously mentioned shrinking/expansion effect of the interaction domain.

To better quantify the enhancement of turbulence across the interaction, we have computed 
at each $x$ station the peak values of the Reynolds stress components,
reported in figure~\ref{fig:reymax} as a function of the scaled streamwise coordinate.
The distributions are strongly influenced by the wall temperature, an increment of
$s$ implying an upstream shift of the turbulence amplification location. Furthermore, the intensity
of all the Reynolds stress components is seen to increase when the wall is heated, with the exception
of $\widetilde{u^{\prime \prime} u^{\prime \prime}}$, whose peak is
identical for the various SBLI cases. The maximum amplification (approximately a factor 4 with respect to the upstream level)
is attained by the wall-normal component $\widetilde{v^{\prime \prime} v^{\prime \prime}}$,
whose behavior is qualitatively similar to that of $\widetilde{w^{\prime \prime} w^{\prime \prime}}$,
whereas the shear stress displays a second maximum immediately past the nominal impingment location.

\begin{figure}
 \centering
 \includegraphics[width=2.6cm,angle=270,clip]{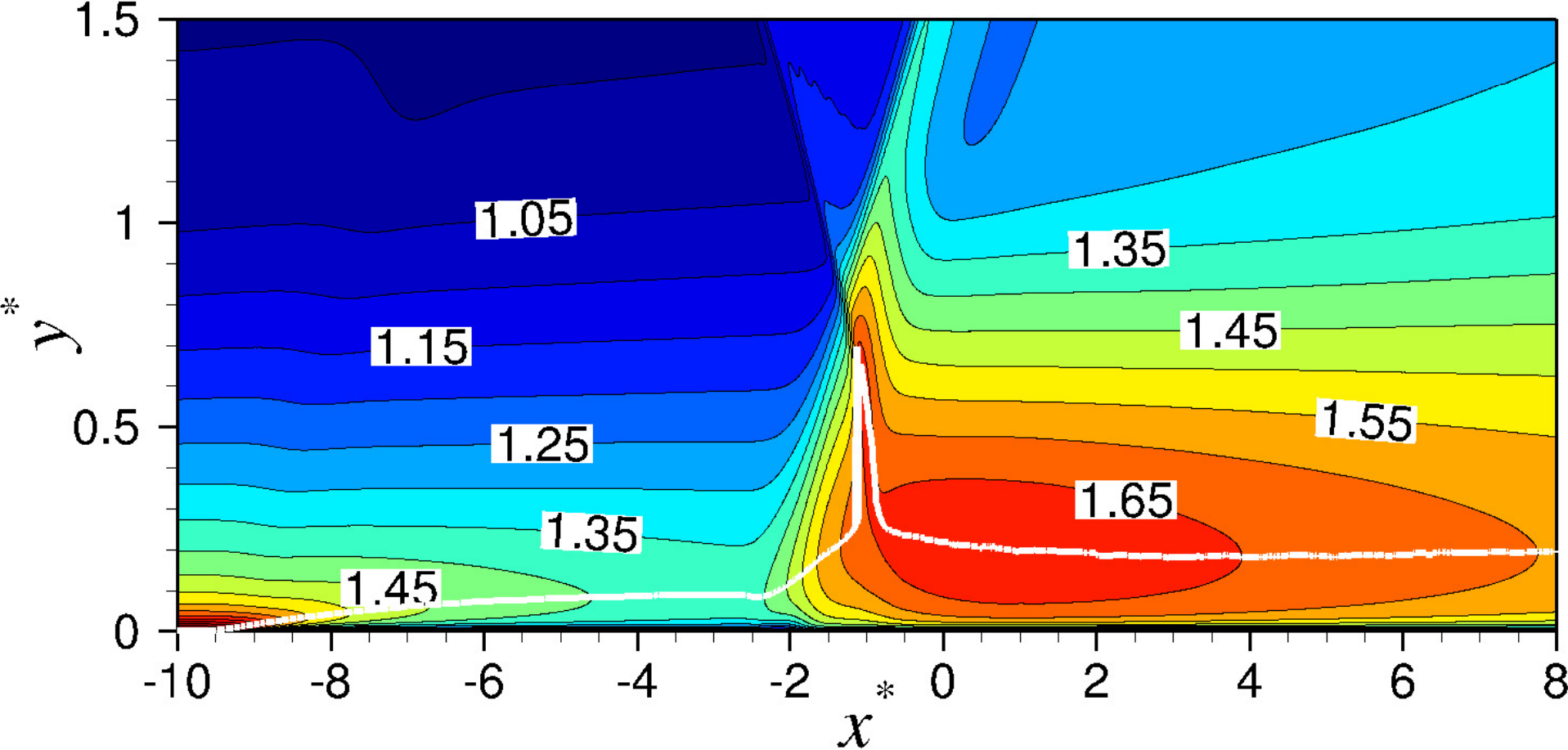} \hskip 1.0em
 \includegraphics[width=2.6cm,angle=270,clip]{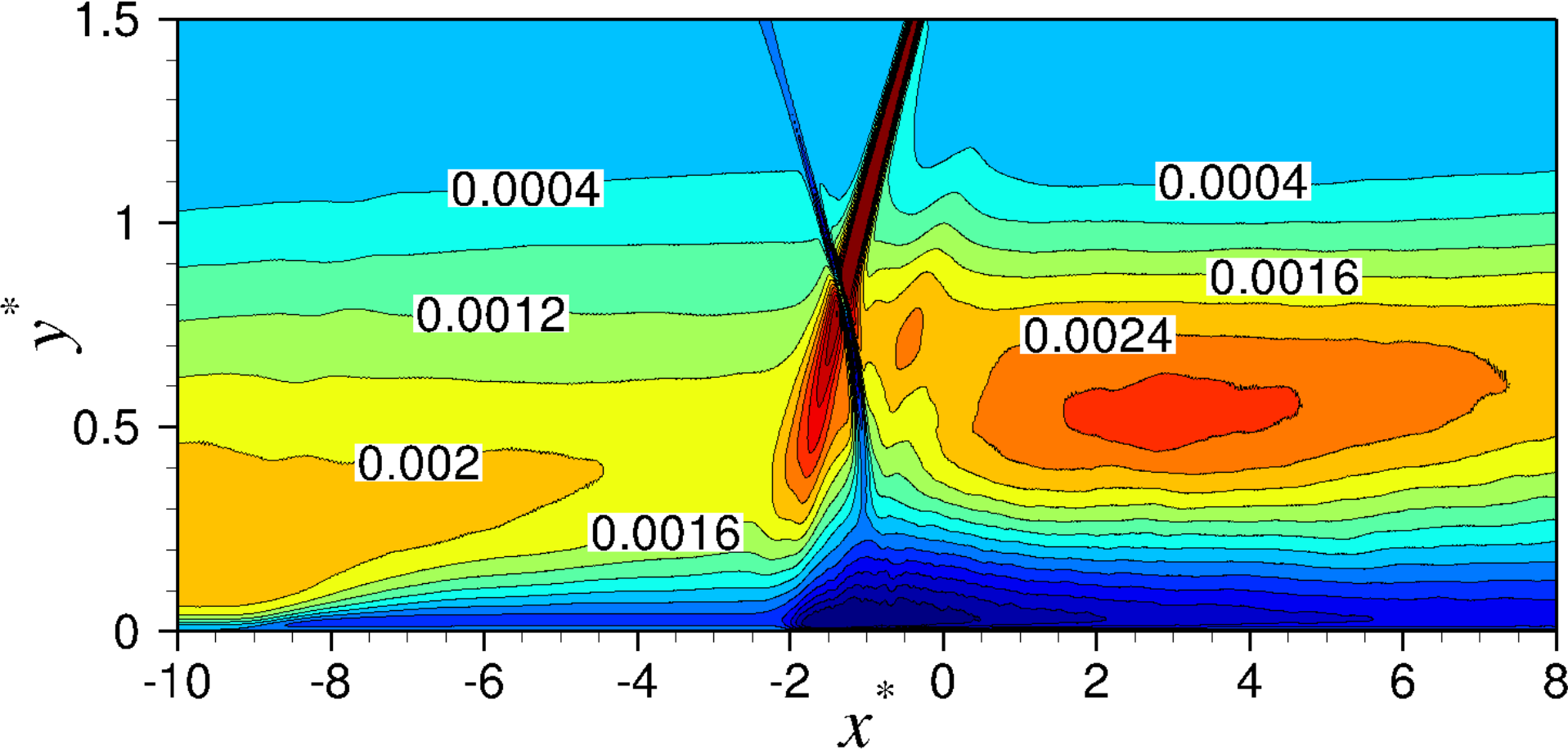} \vskip 0.5em
 \includegraphics[width=2.6cm,angle=270,clip]{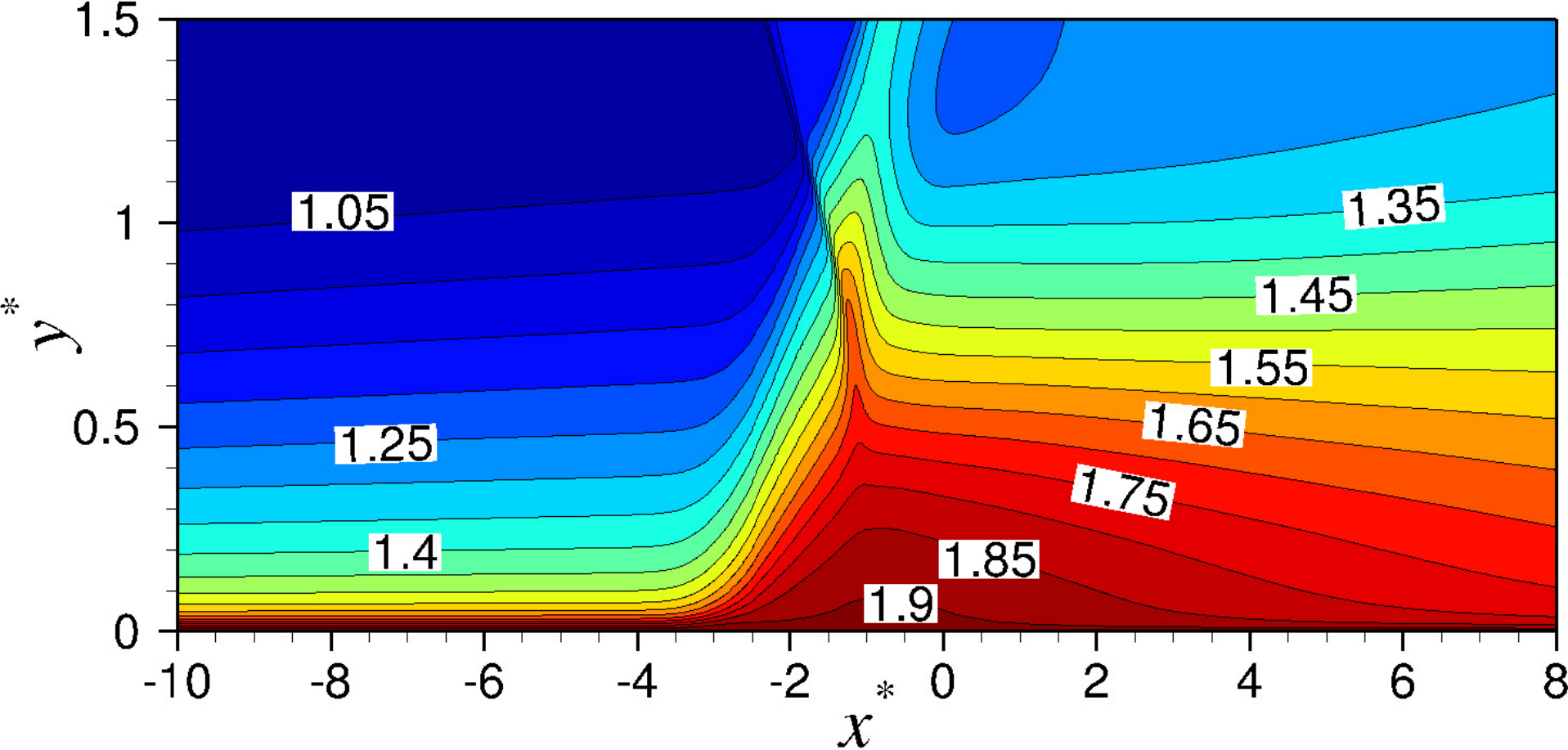} \hskip 1.0em
 \includegraphics[width=2.6cm,angle=270,clip]{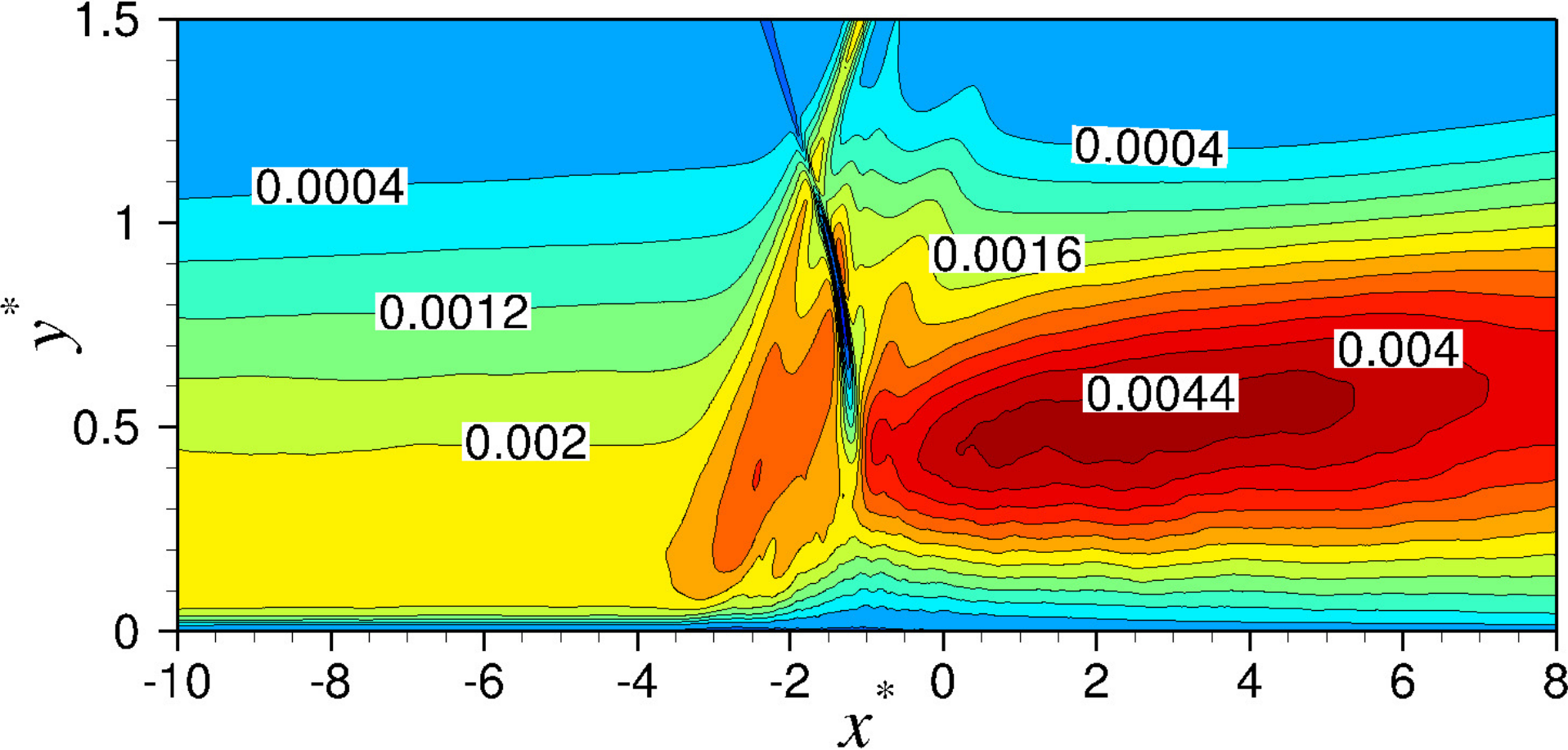} \vskip 0.5em
 \includegraphics[width=2.6cm,angle=270,clip]{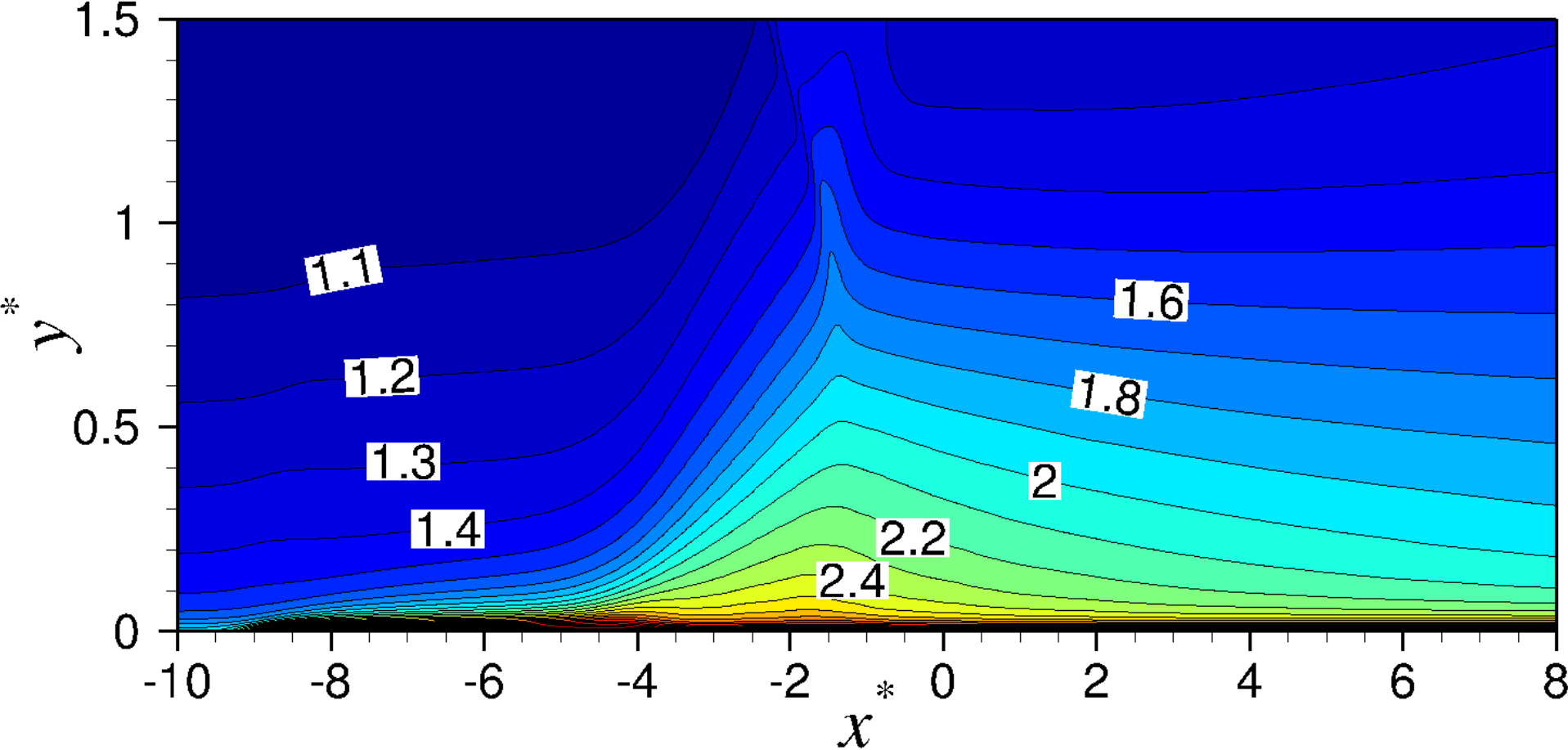} \hskip 1.0em
 \includegraphics[width=2.6cm,angle=270,clip]{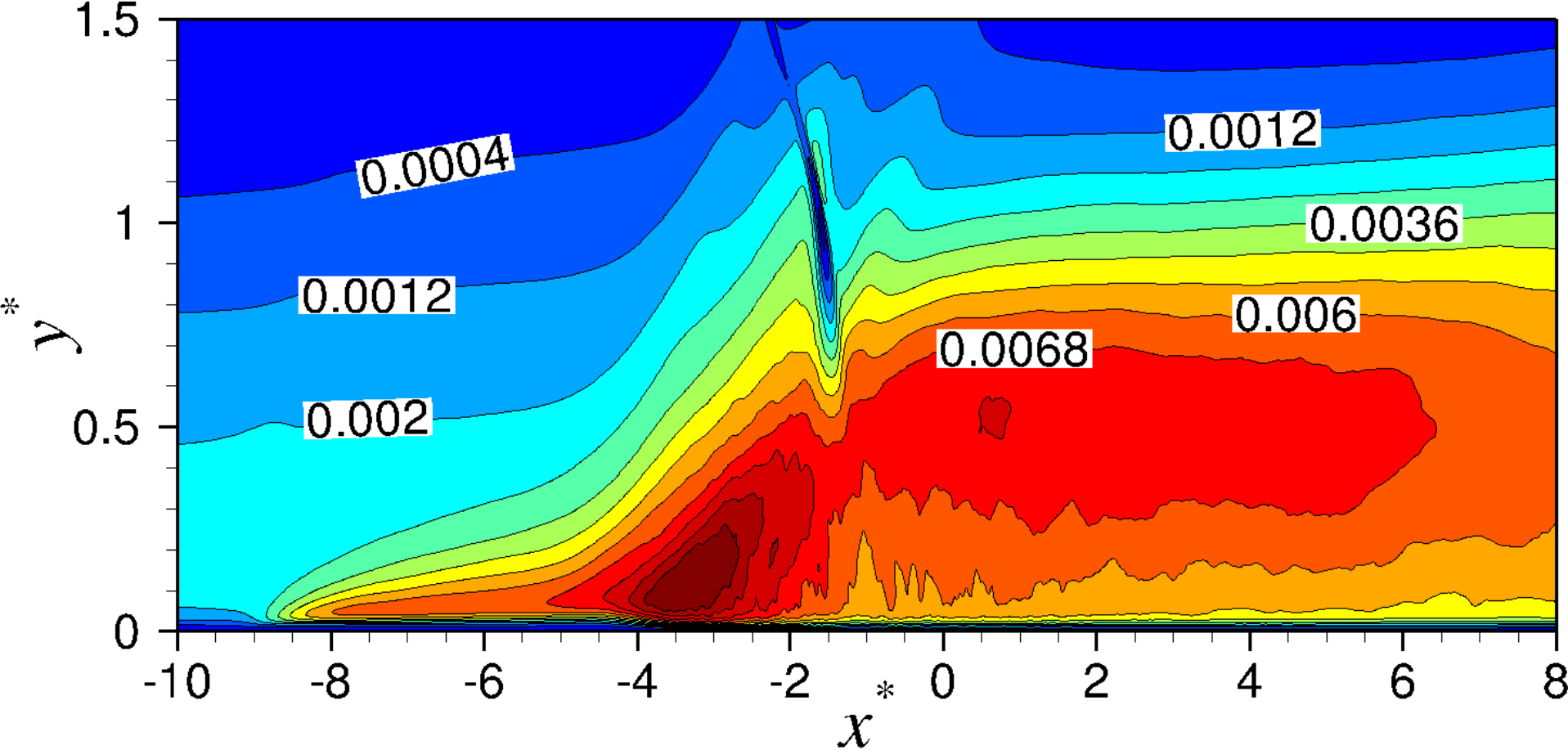} \vskip 0.5em
 \caption{Contours of mean temperature $\widetilde{T}/T_{\infty}$ (left panels),
          and wall-normal turbulent heat flux $\widetilde{v^{\prime \prime} T^{\prime \prime}}/u_{\infty} T_{\infty}$ (right panels)
          at various $s$, increasing from top to bottom ($s = 0.5,1,1.9$).}
 \label{fig:tfav_vsts}
\end{figure}
A major effect of cooling/heating is found in the fields of the mean temperature $\widetilde{T}$ and of
the wall-normal turbulent heat-flux $\widetilde{v^{\prime \prime} T^{\prime \prime}}$,
displayed in figure~\ref{fig:tke_uv}, where the y-axis has been magnified to better highlight
the near-wall behavior. 
The impinging shock greatly affects both $\widetilde{T}$ and $\widetilde{v^{\prime \prime} T^{\prime \prime}}$,
leading to a thickening of the thermal boundary layer and to a strong amplification of the
turbulent heat flux. However, the specific behavior of the flow significantly depends on
the wall thermal condition. In particular, in both the adiabatic and hot wall case
the mean temperature attains its maximum at the wall and a positive correlation is always found
between temperature and wall-normal velocity fluctuations across the interaction region.
On the other hand, when surface cooling is applied, a local maximum of the mean temperature within the boundary
layer starts to develop
(white solid line in figure~\ref{fig:tke_uv}a)
, which moves far away from the wall
at the beginning of the interaction process. In this case, a negative $\widetilde{v^{\prime \prime} T^{\prime \prime}}$
correlation is found close to the wall, and as observed for a cold spatially
evolving boundary layer~\citep{hadjadj15}, the crossover position ($\widetilde{v^{\prime \prime} T^{\prime \prime}} = 0)$
occurs close to the location of maximum mean temperature.

\subsection{Wall properties in adiabatic and non-adiabatic SBLI}

\begin{figure}
 \centering
 \psfrag{x}[t][][1.2]{$x^*$}
 \psfrag{y}[b][][1.2]{$C_f \cdot 10^3$}
 \psfrag{p}[][][0.8]{$(a)$}
 \includegraphics[width=4.0cm,angle=270,clip]{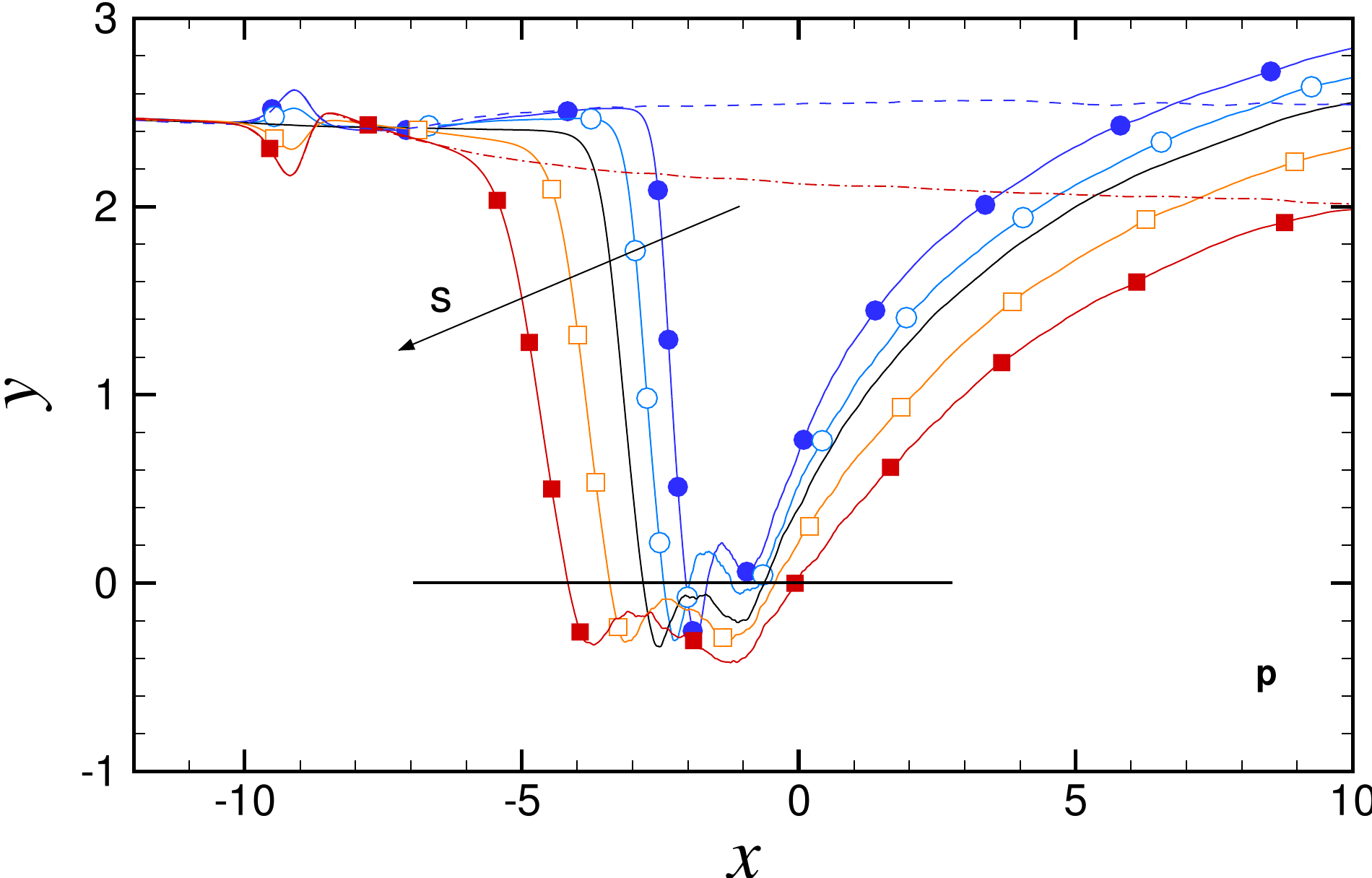} \hskip 1em
 \psfrag{y}[b][][1.2]{$p_w/p_{\infty}$}
 \psfrag{p}[][][0.8]{$(b)$}
 \includegraphics[width=4.0cm,angle=270,clip]{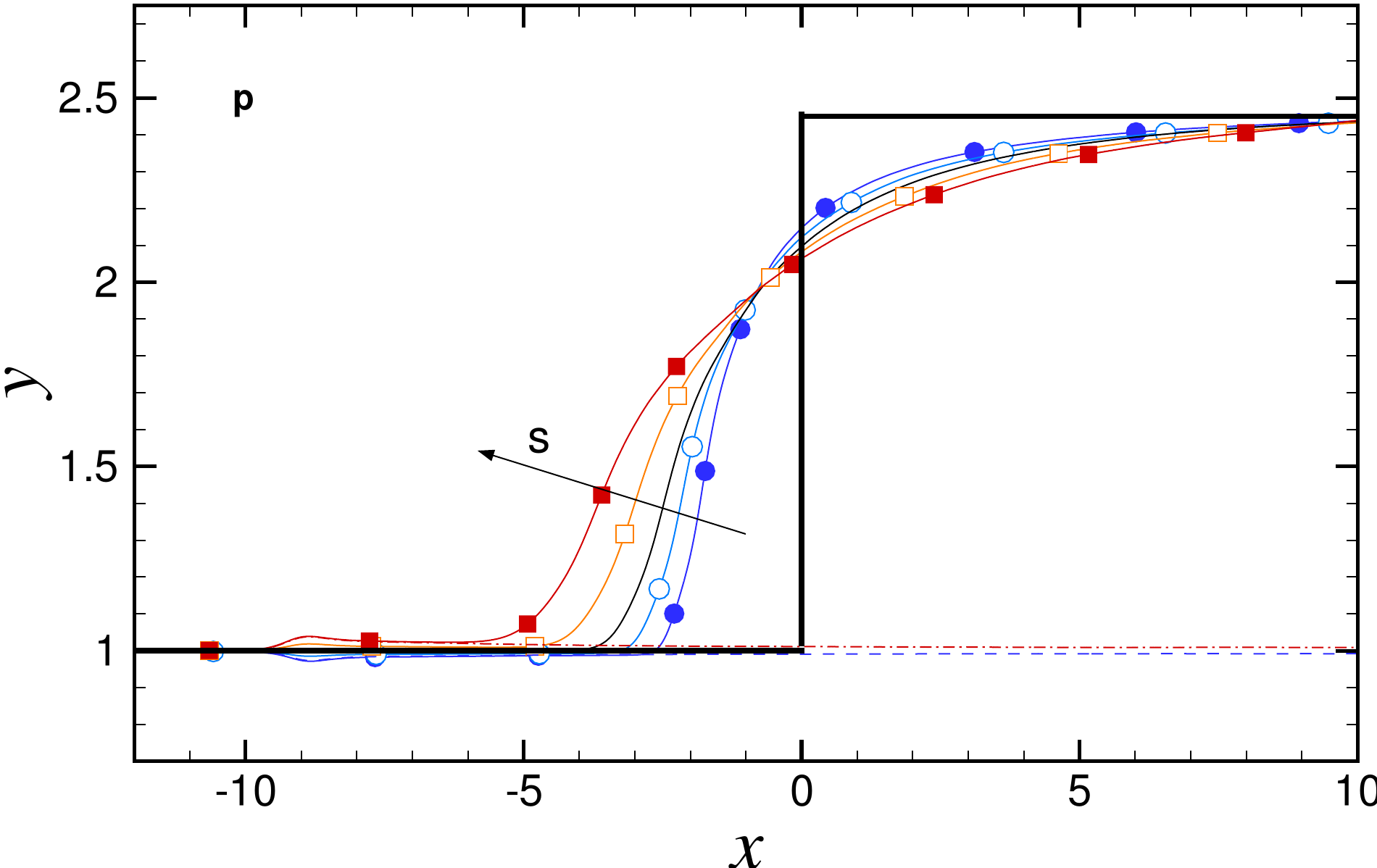} \vskip 0.5em
 \caption{Distribution of (a) skin friction coefficient and (b) mean wall pressure at various wall-to-recovery-temperature ratios.
          Refer to table~\ref{tab:testcases} for nomenclature of the DNS data.
          The dotted line denotes the pressure jump predicted by the inviscid theory.} 
 \label{fig:cf_pmean}
\end{figure}
 
\begin{figure}
 \centering
 \psfrag{x}[t][][1.2]{$s$}
 \psfrag{y}[b][][1.2]{$L_{sep}$}
 \psfrag{p}[][][0.8]{$(a)$}
 \includegraphics[width=4.0cm,angle=270,clip]{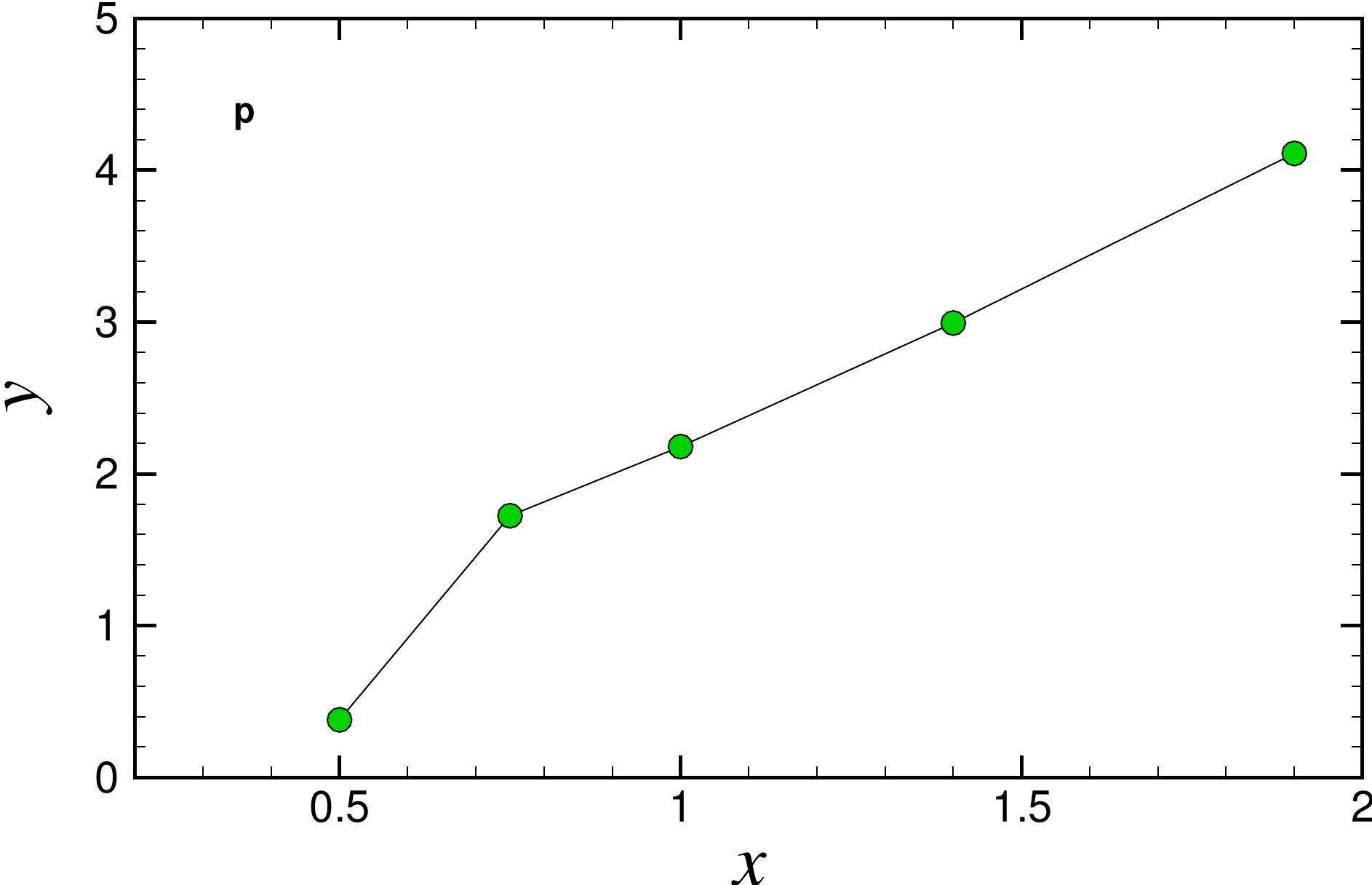} \hskip 1em
 \psfrag{y}[b][][1.2]{$x_{sep}, x_r$}
 \psfrag{p}[][][0.8]{$(b)$}
 \includegraphics[width=4.0cm,angle=270,clip]{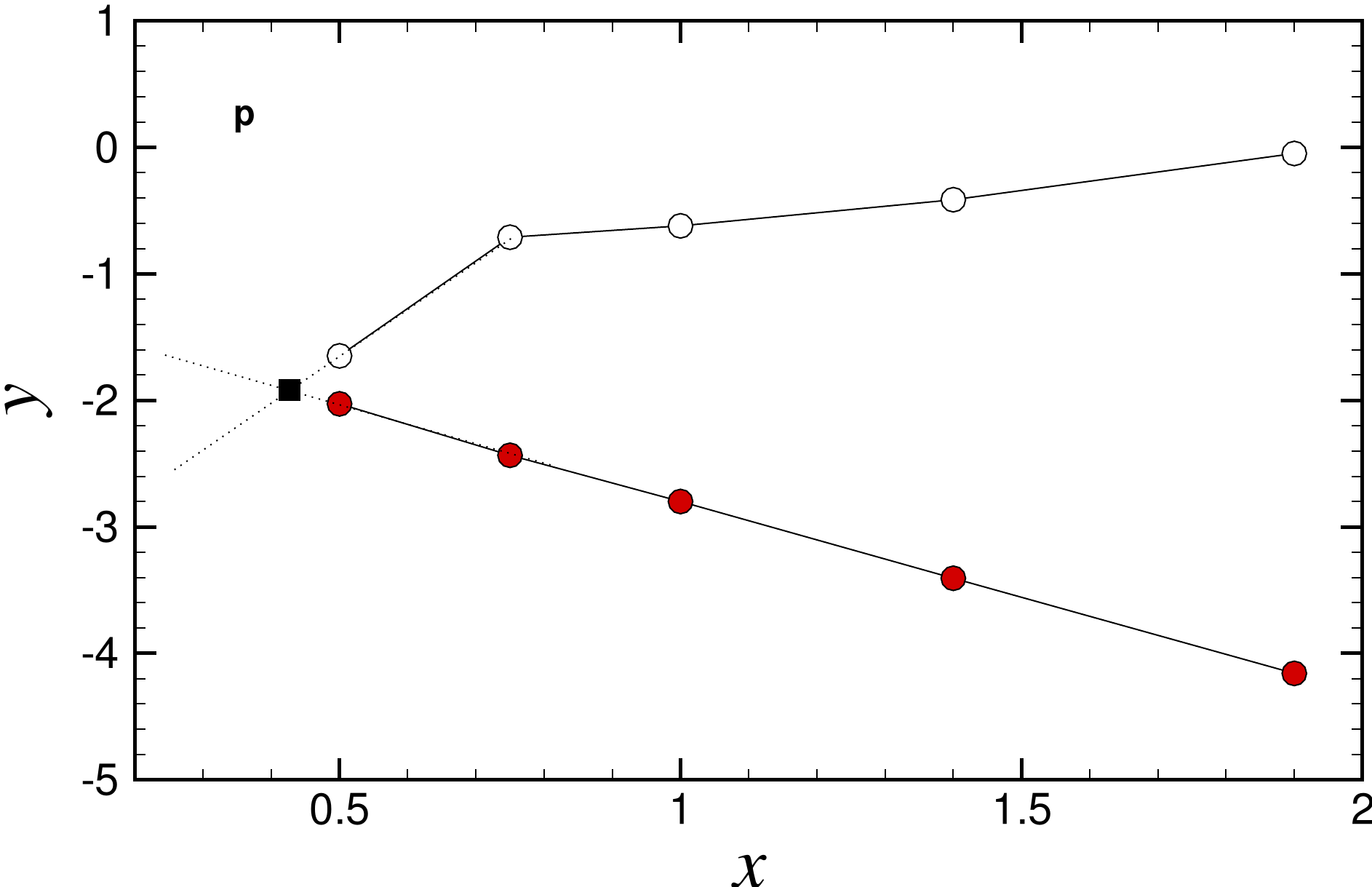} \vskip 0.5em
 \caption{Distribution of (a) mean separation length $L_{sep}$ and (b) location of the separation ($x_{sep}$, solid circles)
          and reattachment ($x_r$, open circles) points as a function of the wall-to-recovery temperature ratio.
          The black square is the extrapolated point corresponding to the condition of incipient separation.}
 \label{fig:lsep}
\end{figure}
 
The spatial distribution of the mean skin friction coefficient $C_f = 2 \tau_w / \rho_{\infty} u^2_{\infty}$
at various $s$ is depicted in figure~\ref{fig:cf_pmean} (a).
For reference purposes, we also report in the figure the skin friction distribution of
the cold and hot spatially evolving boundary layers BL-s0.5 and BL-s1.9 (dashed lines).
Upstream of the region of shock influence, a collapse of the curves for
the same temperature conditions is observed. The temperature step change produces an abrupt variation
of the skin friction, characterized by a maximum (minimum) when cooling (heating) the wall.
In the absence of the shock the skin friction distribution gradually relaxes to that of an
equilibrium boundary layer, and in agreement with previous studies~\citep{hadjadj15},
$C_f$ is increased by wall cooling and decreased by heating. 
In the presence of the impinging shock, the skin friction exhibits a sharp decrease
at the beginning of the interaction and for all cases mean flow separation is
observed. The extent of the recirculation region ($L_{\mathrm{sep}}$) is reported
in table~\ref{tab:testcases} and plotted in figure~\ref{fig:lsep}, where the location of
the separation and reattachment points is also shown.
Compared to the adiabatic case, wall cooling results in a significant reduction of $L_{\mathrm{sep}}$
($-74 \%$ for SBLI-s0.5), whereas heating the wall leads to the opposite effect ($+79.8 \%$ for SBLI-s1.9).
The location of the separation point is most affected by the wall temperature change, whereas the
boundary layer reattachment is less influenced by $s$, being
mainly controlled by the nominal (fixed) impinging shock location.
A simple extrapolation of the available data leads to a value of $s = 0.427$ to obtain the condition of incipient separation. 
We observe that, for all cases, the skin friction in the interaction region
exhibits the typical W-shape previously observed in both laminar and (adiabatic) turbulent
shock boundary layer interactions~\citep{katzer89,pirozzoli_11_3}, characterized
by two minima, which are both affected by $s$. In particular, an increase of the wall-to-recovery-temperature
ratio produces an upstream displacement of the first minimum, associated with the upstream shift of the
separation shock. The location of the second minimum is relatively insensitive to $s$
but its magnitude decreases when the wall temperature is raised.

\begin{figure}
 \centering
 \psfrag{x}[t][][1.2]{$x^*$}
 \psfrag{y}[b][][1.2]{$C_h (\cdot 10^3)$}
 \psfrag{p}[][][0.8]{$(a)$}
 \includegraphics[width=4.0cm,angle=270,clip]{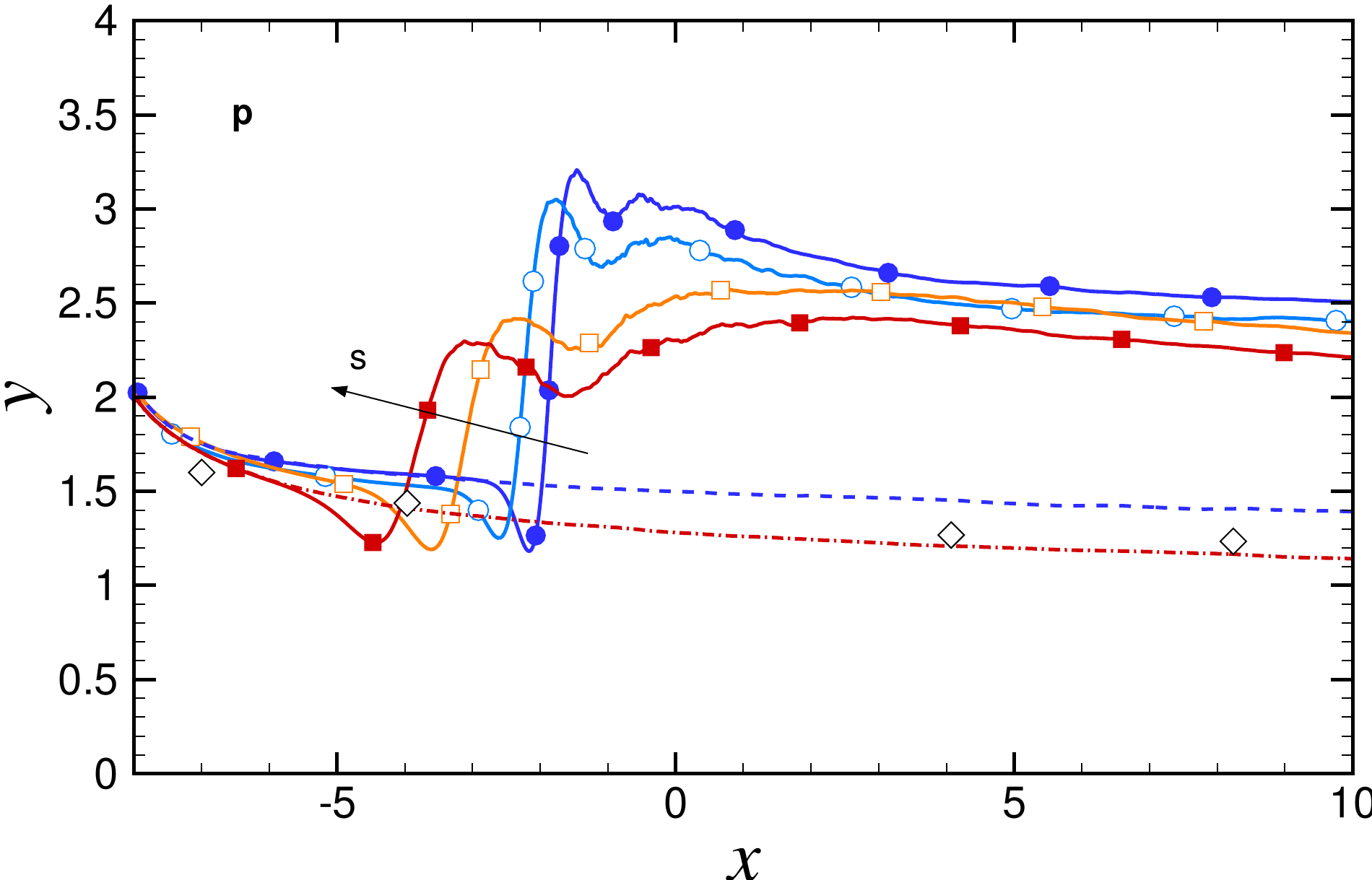} \hskip 1em
 \psfrag{y}[b][][1.2]{$q_w/(\rho_{\infty} u_{\infty} C_p T_r)$}
 \psfrag{p}[][][0.8]{$(b)$}
 \includegraphics[width=4.0cm,angle=270,clip]{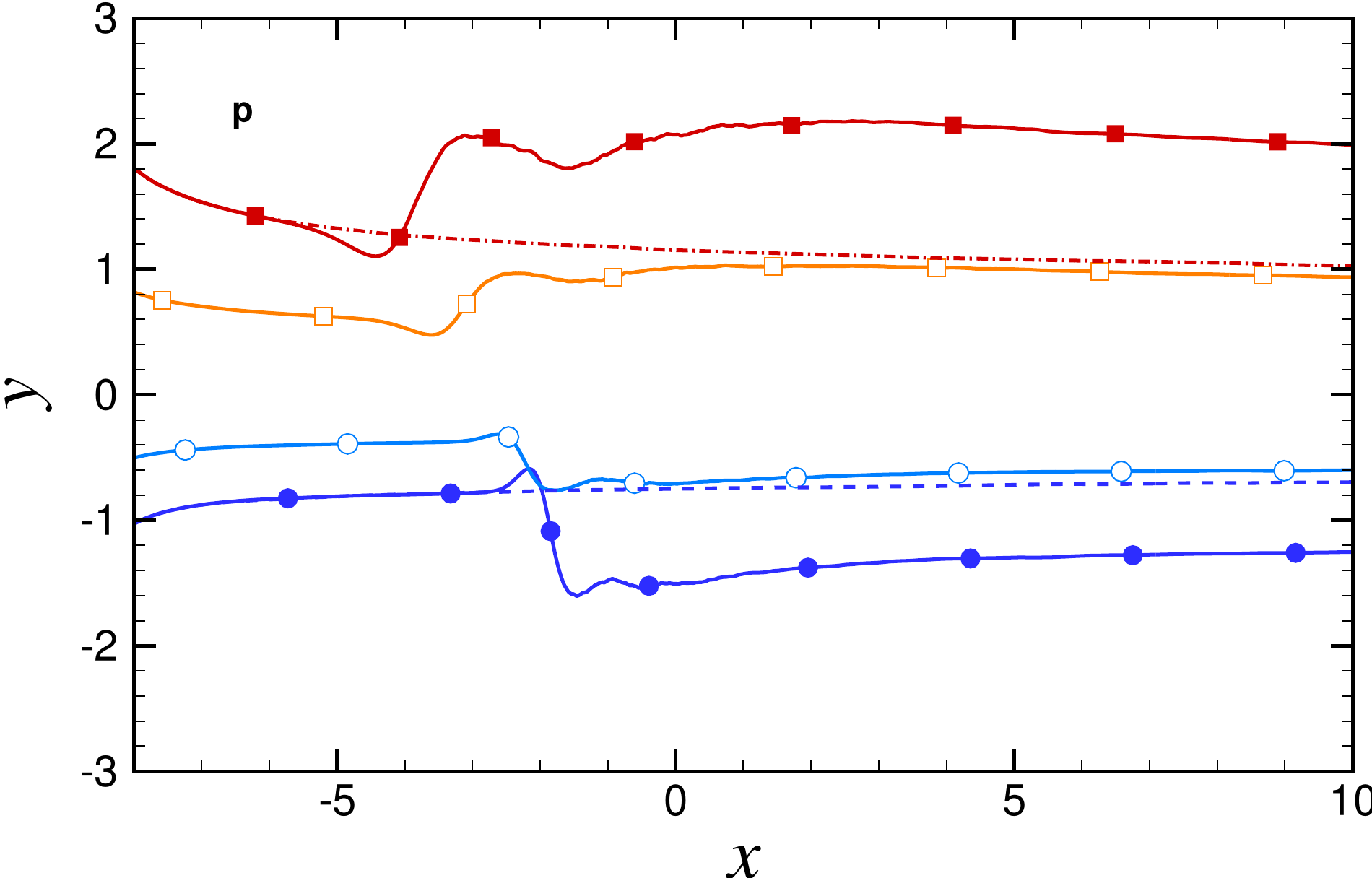} \vskip 0.5em
 \caption{Distribution of (a) Stanton number and (b) wall heat flux at various wall-to-recovery-temperature ratios.
          Refer to table~\ref{tab:testcases} for nomenclature of the DNS data.
          Open diamonds denote reference experiments with $s=2$ by~\citet{debieve97}.}
 \label{fig:stanton_qw}
\end{figure}
\begin{figure}
 \centering
 \psfrag{x}[t][][1.2]{$x^*$}
 \psfrag{y}[b][][1.2]{$p_{\textrm{rms}}/p_{\infty}$}
 \psfrag{p}[][][0.8]{$ $}
 \includegraphics[width=4.0cm,angle=270,clip]{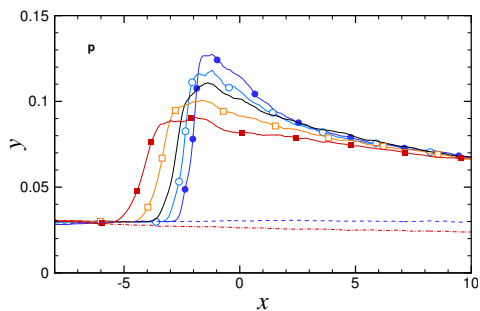}
 \caption{Distribution of root-mean-square wall pressure at various wall-to-recovery-temperature ratios.
          Refer to table~\ref{tab:testcases} for nomenclature of the DNS data.}
 \label{fig:prms}
\end{figure}
 
The major influence of cooling/heating is also apparent from the
mean wall pressure $p_w$, whose distribution is reported in figure~\ref{fig:cf_pmean} (b).
Heating the wall shifts upstream the beginning of the interaction, leading to a smoother pressure rise.
The opposite behavior occurs in the case of cooling, that produces a downstream shift of the upstream influence
and a steeper variation of $p_w$ within the interaction zone.
Interestingly, all the curves cross at the same point ($x^* = -1$) close to the nominal
impingment location, before gradually relaxing towards the value predicted by the inviscid theory.
In the downstream portion, contrary to some experimental observations~\citep{delery92},
our data do not show any overshoot with respect to the level of the inviscid fluid solution.

To characterize the heat transfer behavior across the interaction the spatial distribution of the Stanton number $C_h$ is reported
in figure~\ref{fig:stanton_qw} (a), for all flow cases (BL and SBLI) here investigated. As a reference purpose,
we also show in figure~\ref{fig:stanton_qw} (b) the wall heat flux $q_w$, that being normalized by the constant factor
$(\rho_{\infty} u_{\infty} C_p T_r)$, provides a perception of the direction and of the effective amount of heat exchanged at the wall in
the various cases.
A strong amplification of the heat transfer rate $C_h$ is found in the interaction region with respect
to the reference cooled/heated boundary layers, with a maximum increase of approximately a factor $2$
for the cooled and $1.7$ for the heated wall. A complex variation of the Stanton distribution is observed 
when varying the wall thermal condition, the curves being characterized by four local extrema.
First, $St$ decreases attaining a minimum in the proximity of the separation point, followed by a sharp increase in the interaction zone,
with the peak achieved at the same point where the skin friction features its local maximum.
In the case of heated wall, characterized by an extended separation, the Stanton number exhibits a curvature change with a second minimum around
the reattachment point and then increases again attaining a second broad maximum in the downstream relaxation region.
In the presence of cold wall, where the extent of the separation bubble
is strongly reduced, the curvature change is still observed but $St$ peaks immediately past the reattachment point.
These trends are very similar to those reported by~\citet{hayashi84}, who explored the effect of the
shock strength (by varying the shock generator wedge angle) under the same thermal condition (cold wall).
In particular, the Stanton number distribution found in the experiments for strong interactions is here recovered by
increasing the wall-to-recovery-temperature ratio.

We remark that, despite cooling the wall results in a weaker interaction (as far as the separation bubble size is concerned),
the reduction of the length-scales in the streamwise and wall normal direction produces stronger temperature gradients at the wall
thus leading to larger heating rates.
Similarly, since the shock penetrates deeper in the boundary layer and the pressure jump imparted by the shock must be sustained in a narrower region,
cooling the wall increases the root-mean-square wall pressure $p_{rms}$, as shown in figure~\ref{fig:prms}.
The location of the maximum values of $p_{rms}$
perfectly matches that of the first peak in the Stanton distribution, implying that the generation of
high thermal loads is likely to be associated with the turbulence amplification in the interaction region.

\begin{figure}
 \centering
 a)
 \includegraphics[width=1.9cm,angle=270,clip]{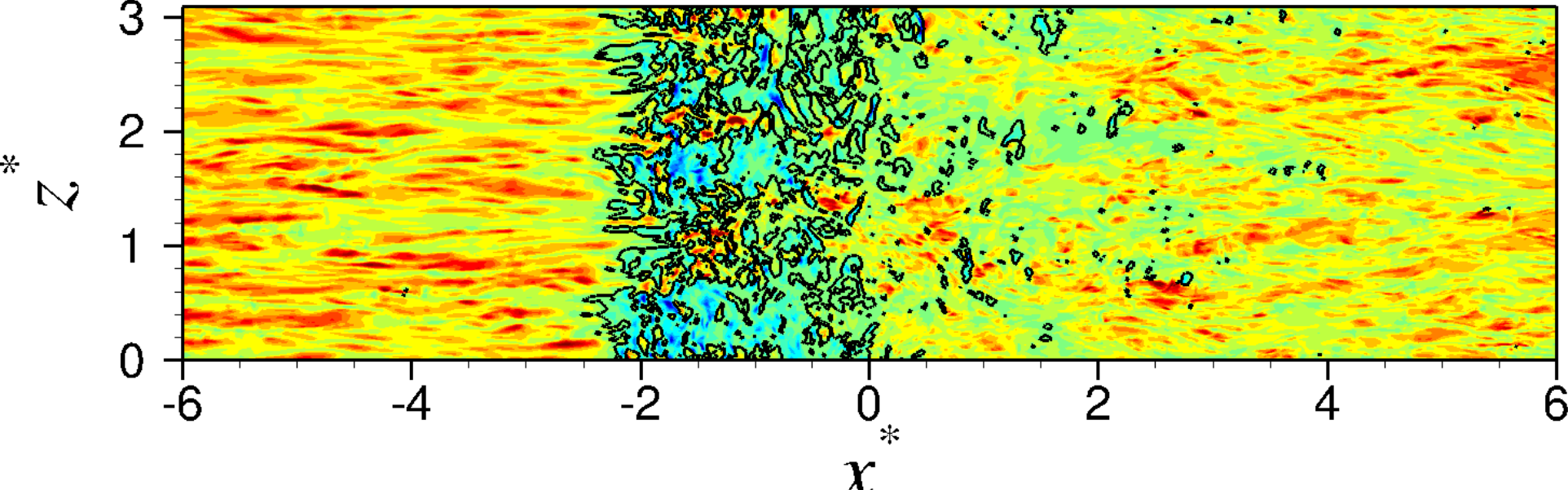} \hskip 0.5em
 b)
 \includegraphics[width=1.9cm,angle=270,clip]{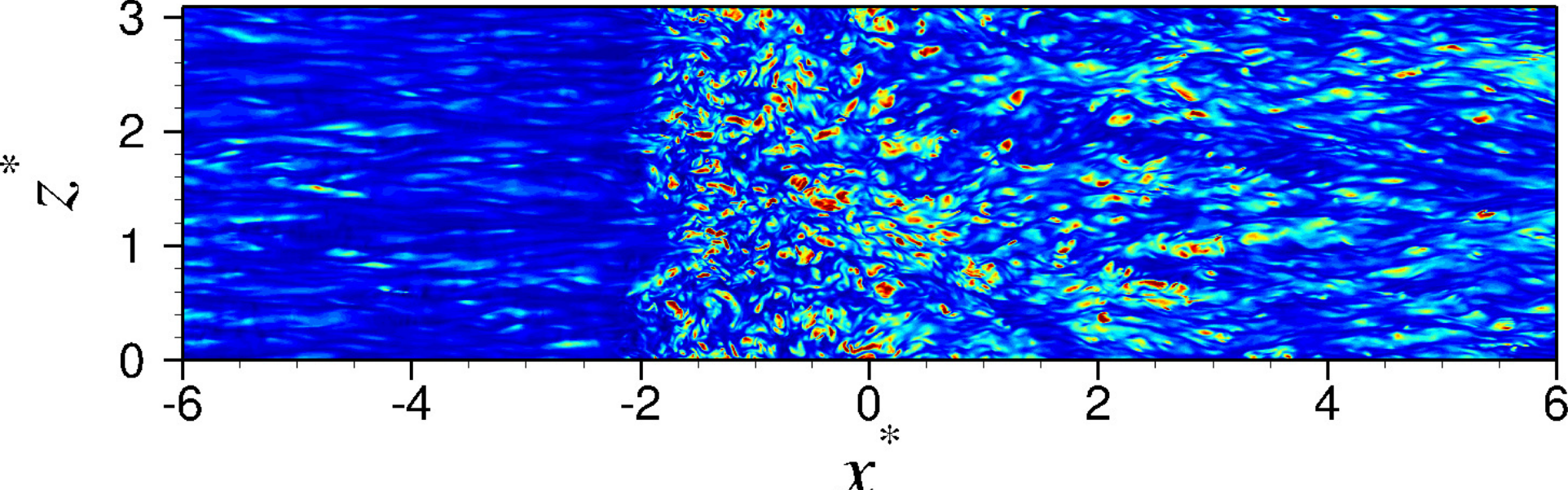} \vskip 0.5em
 c)
 \includegraphics[width=1.9cm,angle=270,clip]{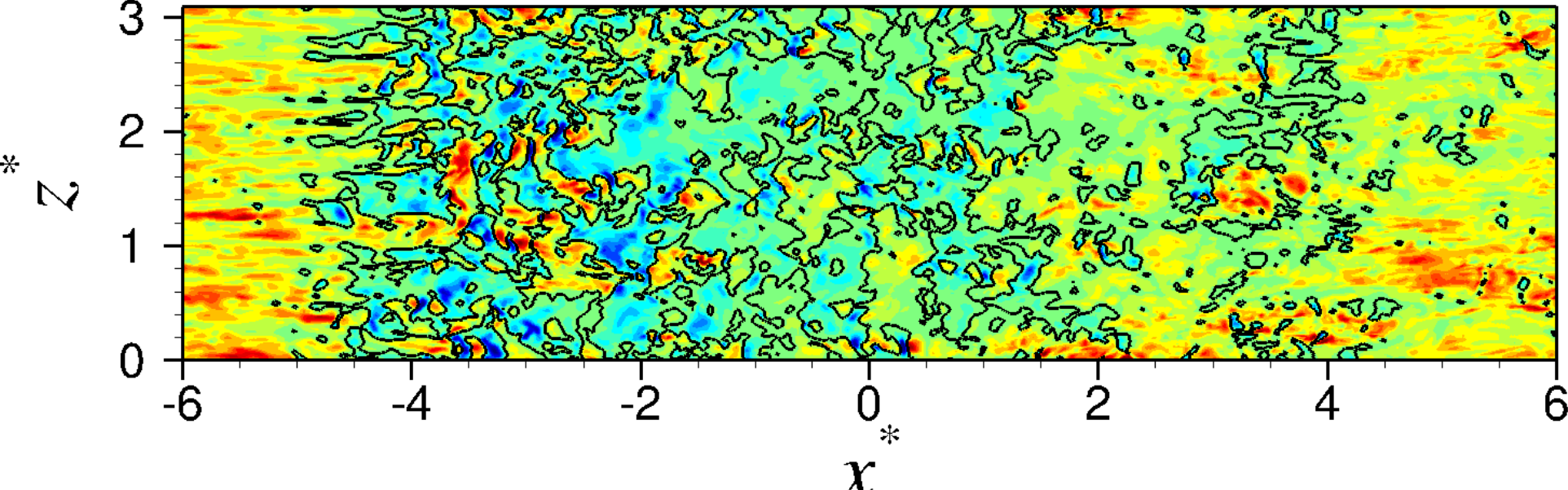} \hskip 0.5em
 d)
 \includegraphics[width=1.9cm,angle=270,clip]{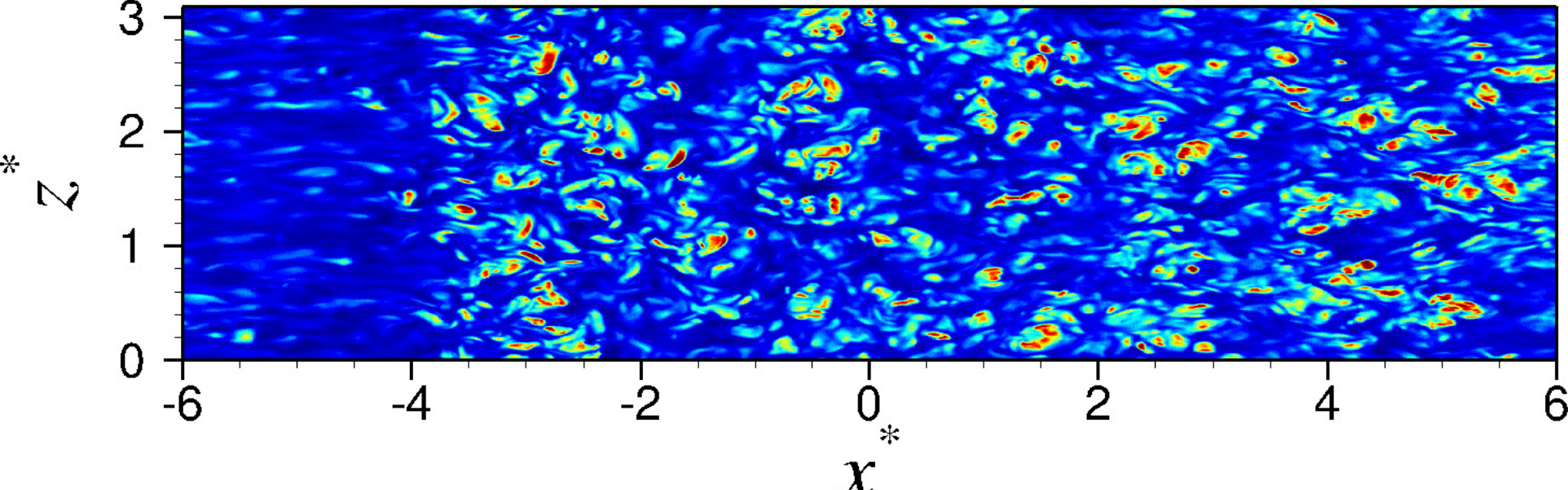} \vskip 0.5em
 \caption{Contours of (a-c) instantaneous skin friction and (b-d) Stanton number for flow cases SBLI-s0.5 (top panels) and SBLI-s1.9 (bottom panels).}
 \label{fig:cf_st}
\end{figure}
\begin{figure}
 \centering
 \psfrag{x}[t][][1.2]{$x^*$}
 \psfrag{y}[b][][1.2]{$R_{c_f c_h}$}
 \includegraphics[width=5.0cm,angle=270,clip]{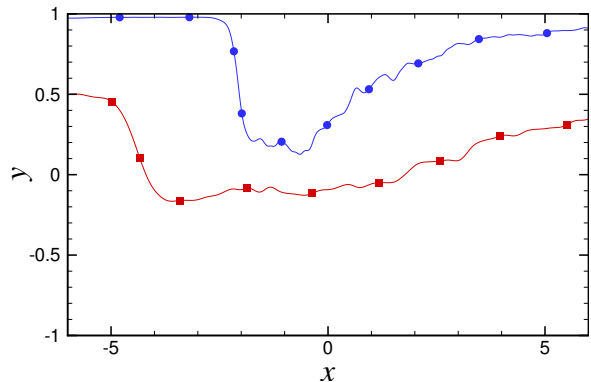} \hskip 0.5em
 \caption{Distribution of the correlation coefficient between $c_f$ and $c_h$ for flow cases SBLI-s0.5 abnd SBLI-s1.9.}
 \label{fig:rcfch}
\end{figure}
The results on the mean skin friction and the Stanton number reported in the previous figures confirm the experimental
observations of~\citet{schulein06}, who highlighted that the analogy between momentum and heat transfer, which is
well assessed in equilibrium flows and represents the basis of many simplified physical models is not valid in the interaction
region. This conclusion is not surprising, since even the most advanced and refined forms of the Reynolds analogy~\citep{zhang14}
are all based on the chief assumption/approximation of a quasi-one-dimensional flow, which clearly fails
in the presence of mean flow separation as in the present SBLI cases.

To examine in depth the relationship between momentum and heat transfer, and to better characterize the unsteady
behavior of the flow we show in figure~\ref{fig:cf_st} contours of the instantaneous skin friction
$c_f$
and instantaneous heat transfer coefficient
$c_h$
in the wall plane,
for the two extreme cases SBLIs-0.5 and SBLIs-1.9. We also provide more quantitative information in figure~\ref{fig:rcfch}
by reporting their correlation coefficient ($R_{c_f c_h}$), as a function of the streamiwse coordinate.
Upstream of the interaction, a streaky pattern typical of a zero-pressure-gradient boundary layer is found for $c_f$ and $c_h$
in both the cold and hot wall cases. This region is characterized by a positive correlation between the fluctuating friction
and heat transfer coefficients, especially in the case of cooling ($R_{c_f c_h} = 0.98$).
This scenario completely changes across the interaction, where flow patches of instantaneously reversed flow 
are found, starting from the beginning of the interaction and extending well into the recovery zone.
In this region the local Stanton number exhibits a strong intermittent behavior, characterized by scattered spots with extremely
high heat transfer rates and the correlation coefficient displays a rapid decay, attaining a nearly flat distribution
throughout the separation bubble. The relaxation region is characterized by a gradual recover of the upstream behavior which is not yet completed
at the end of the computational domain.

\begin{figure}
 \centering
 \includegraphics[width=3.0cm,angle=270,clip]{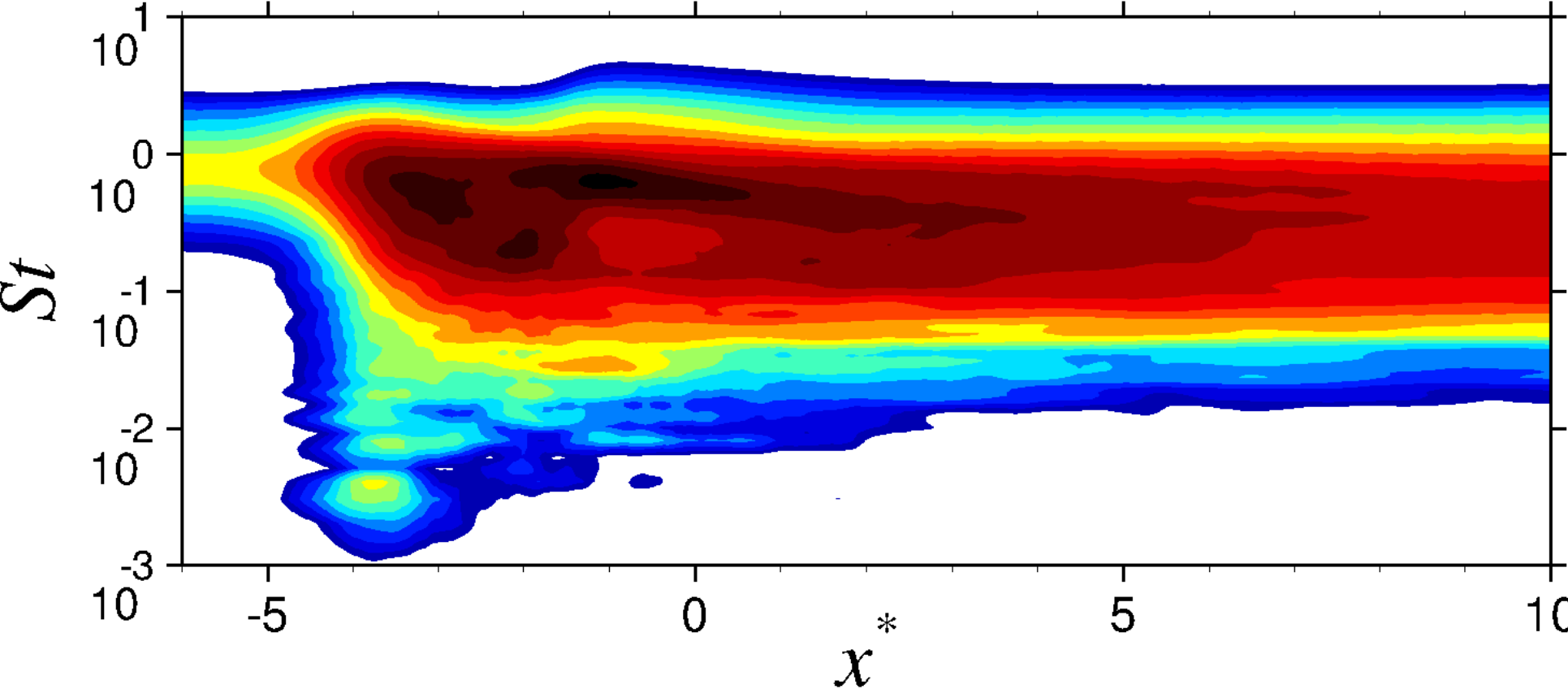} \vskip 1em
 \includegraphics[width=3.0cm,angle=270,clip]{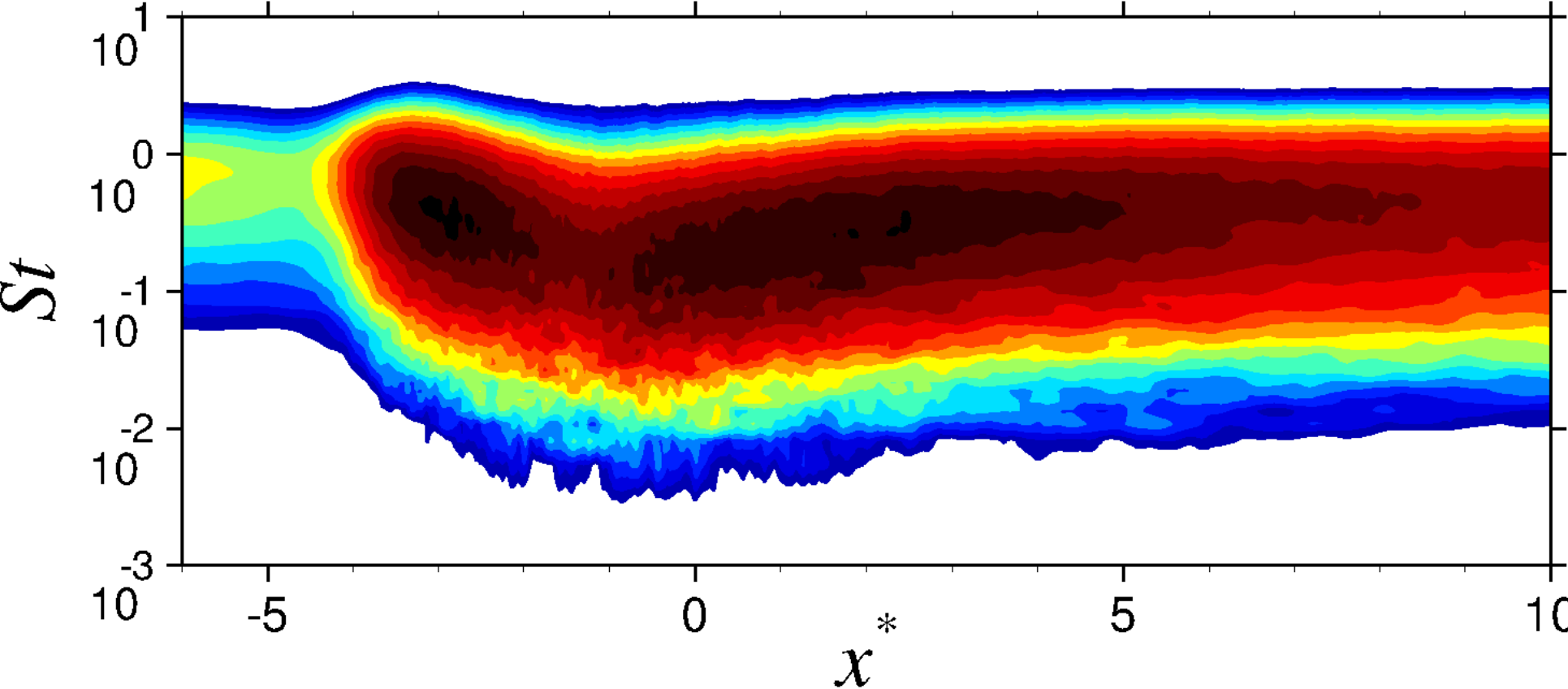} \vskip 1em
 \caption{Contours of pre-multiplied (a) pressure and (b) heat flux spectra ($f \, E(f)$) for flow case SBLI-s1.9
          as a function of streamwise location and Strouhal number.}
 \label{fig:mapspec}
\end{figure}
To further characterize the flow unsteadiness and to assess the possible influence of the reflected
shock motion on the wall heat flux, we report in figure~\ref{fig:mapspec} the pre-multiplied spectra
of both the wall pressure and the instantaneous heat flux as a function of Strouhal number
$St = f \, \delta_0 / u_{\infty}$ and streamwise position $x^*$.
The spectral maps refer to SBLIs1.9, which is characterized by extended separation and
correspond to the flow case for which the low-frequency shock motion is more evident.
The power spectral densities have been computed using the Welch method, subdividing the overall pressure
record into 4 segments with 50\% overlapping, which are individually Fourier-transformed. The frequency
spectra are then obtained by averaging the periodograms of the various segments, which allows to minimize
the variance of the PSD estimator, and by applying a Konno-Omachi smoothing filter~\citep{konno_98} that ensures a
constant bandwidth on a logarithmic scale.
The map of the wall pressure signal shows the typical features observed in previous studies~\citep{morgan13}.
Upstream of the interaction zone the spectra are bump-shaped as for canonical wall-bounded flows, with a
peak at $\St \sim O(1)$, associated with the energetic turbulent structures of the boundary layer. 
A similar shape is also found in the downstream relaxation region, although the spectral density is broadened and
the peak shifted to lower frequencies owing to the thicknening of the boundary layer.
A different behavior is observed at the beginning of the interaction region, close to the foot of the reflected shock,
where a broad peak appears in the map at low frequencies, centered at $St \approx 0.004$, corresponding to a Strouhal number
based on the separation length $\St_L = f L_{sep} / u_{\infty} \approx 0.025$.
This secondary peak is the signature of the broadband motion of the reflect shock, that in SBLI with massive
separation is known to be mainly driven by a donwstream mechanism associated with the
dynamics of the separation bubble~\citep{piponniau_09,clemens14}.

The power spectral density of the heat transfer coefficient brings to light a completely different picture.
In this case no evidence of any low frequency dynamics is apparent and most part of the energy is contained
at intermediate/high frequencies throughout the interaction. In particular a strong amplification of the heat transfer fluctuations
is found close to the separation and reattachement points, with a shift toward intermediate frequencies, classically associated with
the shedding of vortical structures in the shear layer that develops in the first part of the interaction~\citep{aubard13}.
This again suggests that in the flow cases here investigated, the primary mechanism responsible for the generation of peak heating
in the interaction zone is the turbulence amplification associated with the SBLI.

\section{Conclusions}

In the present work the influence of different wall thermal conditions on the properties of
impinging shock-wave/turbulent boundary layer interactions is investigated by means
of direct numerical simulations at $M_{\infty} = 2.28$ and shock angle $\varphi = 8^{\circ}$. 
Five different values of wall-to-recovery-temperature ratio are considered, corresponding to cold ($s = 0.5,0.75)$,
adiabatic ($s = 1$) and hot ($s = 1.4,1.9$) walls.
The characteristic features of SBLI are observed for all flow cases, but the
interaction properties are significantly affected by the wall temperature and our results
confirm the observations of the few experimental data available in literature.
Wall cooling has some beneficial effects on SBLI, leading to a considerable reduction of the interaction
scales and size of the separation bubble, whereas the opposite holds for wall heating.
A complex spatial variation of the Stanton number is found across the interaction,
whose structure strongly depends on the wall-to-recovery-temperature ratio.
The fluctuating heat flux exhibits a strong intermittent behavior, characterized by scattered spots with extremely high
values compared to the mean, and the analogy between momentum and heat transfer typical of equilirium boundary layers
is no longer valid in the interaction region.
The pre-multiplied spectra of the Stanton number do not show any evidence of the influence of the low-frequency
shock motion, and the primary mechanism for the generation of peak heating is found to be linked
with the turbulence amplification in the interaction region.

If the primary objective is to reduce flow separation, our results indicate
that wall cooling can be considered as an effective mean for flow control. However, since
the pressure jump imparted by the shock must be sustained by the boundary layer
in a narrower region, when the wall temperature decreases,
the maximum values of thermal (heat transfer rates) and dynamic loads (root-mean-square wall pressure)
are found in the case of cold wall. 

We expect that the DNS database developed in this work, whose statistics and raw data are available at
\texttt{http://newton.dima.uniroma1.it/osbli/},
would be useful for the high-speed turbulence modeling community, by fostering the development of advanced
models to improve the prediction of heat transfer in SBLI.
Future efforts will be made to extend our database to a wider range of flow conditions,
including different Mach numbers and shock strengths.

\begin{acknowledgments}
The simulations have been performed thanks to computational resources provided by the Italian
Computing center CINECA under the ISCRA initiative (grant jACOBI).
MB was supported by the SIR program 2014 (jACOBI project, grant RBSI14TKWU),
funded by MIUR (Ministero dell'Istruzione dell'Universit\`a e della Ricerca).
\end{acknowledgments}
\bibliographystyle{apsrev4-1}

\bibliography{references}

\begin{thebibliography}{41}%
\makeatletter
\providecommand \@ifxundefined [1]{%
 \@ifx{#1\undefined}
}%
\providecommand \@ifnum [1]{%
 \ifnum #1\expandafter \@firstoftwo
 \else \expandafter \@secondoftwo
 \fi
}%
\providecommand \@ifx [1]{%
 \ifx #1\expandafter \@firstoftwo
 \else \expandafter \@secondoftwo
 \fi
}%
\providecommand \natexlab [1]{#1}%
\providecommand \enquote  [1]{``#1''}%
\providecommand \bibnamefont  [1]{#1}%
\providecommand \bibfnamefont [1]{#1}%
\providecommand \citenamefont [1]{#1}%
\providecommand \href@noop [0]{\@secondoftwo}%
\providecommand \href [0]{\begingroup \@sanitize@url \@href}%
\providecommand \@href[1]{\@@startlink{#1}\@@href}%
\providecommand \@@href[1]{\endgroup#1\@@endlink}%
\providecommand \@sanitize@url [0]{\catcode `\\12\catcode `\$12\catcode
  `\&12\catcode `\#12\catcode `\^12\catcode `\_12\catcode `\%12\relax}%
\providecommand \@@startlink[1]{}%
\providecommand \@@endlink[0]{}%
\providecommand \url  [0]{\begingroup\@sanitize@url \@url }%
\providecommand \@url [1]{\endgroup\@href {#1}{\urlprefix }}%
\providecommand \urlprefix  [0]{URL }%
\providecommand \Eprint [0]{\href }%
\providecommand \doibase [0]{http://dx.doi.org/}%
\providecommand \selectlanguage [0]{\@gobble}%
\providecommand \bibinfo  [0]{\@secondoftwo}%
\providecommand \bibfield  [0]{\@secondoftwo}%
\providecommand \translation [1]{[#1]}%
\providecommand \BibitemOpen [0]{}%
\providecommand \bibitemStop [0]{}%
\providecommand \bibitemNoStop [0]{.\EOS\space}%
\providecommand \EOS [0]{\spacefactor3000\relax}%
\providecommand \BibitemShut  [1]{\csname bibitem#1\endcsname}%
\let\auto@bib@innerbib\@empty
\bibitem [{\citenamefont {Dolling}(2001)}]{dolling01}%
  \BibitemOpen
  \bibfield  {author} {\bibinfo {author} {\bibfnamefont {D.~S.}\ \bibnamefont
  {Dolling}},\ }\href@noop {} {\bibfield  {journal} {\bibinfo  {journal} {AIAA
  J.}\ }\textbf {\bibinfo {volume} {39}},\ \bibinfo {pages} {1517} (\bibinfo
  {year} {2001})}\BibitemShut {NoStop}%
\bibitem [{\citenamefont {Dolling}\ and\ \citenamefont
  {Murphy}(1983)}]{dolling83}%
  \BibitemOpen
  \bibfield  {author} {\bibinfo {author} {\bibfnamefont {D.}~\bibnamefont
  {Dolling}}\ and\ \bibinfo {author} {\bibfnamefont {M.}~\bibnamefont
  {Murphy}},\ }\href@noop {} {\bibfield  {journal} {\bibinfo  {journal} {AIAA
  J.}\ }\textbf {\bibinfo {volume} {21}},\ \bibinfo {pages} {1628} (\bibinfo
  {year} {1983})}\BibitemShut {NoStop}%
\bibitem [{\citenamefont {D{\'e}lery}\ and\ \citenamefont
  {Marvin}(1986)}]{delery86}%
  \BibitemOpen
  \bibfield  {author} {\bibinfo {author} {\bibfnamefont {J.}~\bibnamefont
  {D{\'e}lery}}\ and\ \bibinfo {author} {\bibfnamefont {J.}~\bibnamefont
  {Marvin}},\ }\href@noop {} {\emph {\bibinfo {title} {Shock-wave boundary
  layer interactions}}},\ \bibinfo {type} {AGARDograph}\ \bibinfo {number}
  {280}\ (\bibinfo  {institution} {DTIC Document},\ \bibinfo {year}
  {1986})\BibitemShut {NoStop}%
\bibitem [{\citenamefont {Dupont}\ \emph {et~al.}(2006)\citenamefont {Dupont},
  \citenamefont {Haddad},\ and\ \citenamefont {Debi{\`e}ve}}]{dupont_06}%
  \BibitemOpen
  \bibfield  {author} {\bibinfo {author} {\bibfnamefont {P.}~\bibnamefont
  {Dupont}}, \bibinfo {author} {\bibfnamefont {C.}~\bibnamefont {Haddad}}, \
  and\ \bibinfo {author} {\bibfnamefont {J.}~\bibnamefont {Debi{\`e}ve}},\
  }\href@noop {} {\bibfield  {journal} {\bibinfo  {journal} {J. Fluid Mech.}\
  }\textbf {\bibinfo {volume} {{559}}},\ \bibinfo {pages} {255} (\bibinfo
  {year} {2006})}\BibitemShut {NoStop}%
\bibitem [{\citenamefont {Piponniau}\ \emph {et~al.}(2009)\citenamefont
  {Piponniau}, \citenamefont {Dussauge}, \citenamefont {Debi{\`e}ve},\ and\
  \citenamefont {Dupont}}]{piponniau_09}%
  \BibitemOpen
  \bibfield  {author} {\bibinfo {author} {\bibfnamefont {S.}~\bibnamefont
  {Piponniau}}, \bibinfo {author} {\bibfnamefont {J.}~\bibnamefont {Dussauge}},
  \bibinfo {author} {\bibfnamefont {J.}~\bibnamefont {Debi{\`e}ve}}, \ and\
  \bibinfo {author} {\bibfnamefont {P.}~\bibnamefont {Dupont}},\ }\href@noop {}
  {\bibfield  {journal} {\bibinfo  {journal} {J.\ Fluid\ Mech.}\ }\textbf
  {\bibinfo {volume} {629}},\ \bibinfo {pages} {87} (\bibinfo {year}
  {2009})}\BibitemShut {NoStop}%
\bibitem [{\citenamefont {Humble}\ \emph {et~al.}(2009)\citenamefont {Humble},
  \citenamefont {Elsinga}, \citenamefont {Scarano},\ and\ \citenamefont {van
  Oudheusden}}]{humble09}%
  \BibitemOpen
  \bibfield  {author} {\bibinfo {author} {\bibfnamefont {R.}~\bibnamefont
  {Humble}}, \bibinfo {author} {\bibfnamefont {G.}~\bibnamefont {Elsinga}},
  \bibinfo {author} {\bibfnamefont {F.}~\bibnamefont {Scarano}}, \ and\
  \bibinfo {author} {\bibfnamefont {B.}~\bibnamefont {van Oudheusden}},\
  }\href@noop {} {\bibfield  {journal} {\bibinfo  {journal} {J. Fluid Mech.}\
  }\textbf {\bibinfo {volume} {635}},\ \bibinfo {pages} {47} (\bibinfo {year}
  {2009})}\BibitemShut {NoStop}%
\bibitem [{\citenamefont {Souverein}\ \emph {et~al.}(2010)\citenamefont
  {Souverein}, \citenamefont {Dupont}, \citenamefont {Debi{\`e}ve},
  \citenamefont {Dussauge}, \citenamefont {{van Oudheusden}},\ and\
  \citenamefont {Scarano}}]{souverein_10_b}%
  \BibitemOpen
  \bibfield  {author} {\bibinfo {author} {\bibfnamefont {L.}~\bibnamefont
  {Souverein}}, \bibinfo {author} {\bibfnamefont {P.}~\bibnamefont {Dupont}},
  \bibinfo {author} {\bibfnamefont {J.~F.}\ \bibnamefont {Debi{\`e}ve}},
  \bibinfo {author} {\bibfnamefont {J.~P.}\ \bibnamefont {Dussauge}}, \bibinfo
  {author} {\bibfnamefont {B.~W.}\ \bibnamefont {{van Oudheusden}}}, \ and\
  \bibinfo {author} {\bibfnamefont {F.}~\bibnamefont {Scarano}},\ }\href@noop
  {} {\bibfield  {journal} {\bibinfo  {journal} {AIAA J.}\ }\textbf {\bibinfo
  {volume} {48}},\ \bibinfo {pages} {1480} (\bibinfo {year}
  {2010})}\BibitemShut {NoStop}%
\bibitem [{\citenamefont {Adams}(2000)}]{Adams2000}%
  \BibitemOpen
  \bibfield  {author} {\bibinfo {author} {\bibfnamefont {N.}~\bibnamefont
  {Adams}},\ }\href@noop {} {\bibfield  {journal} {\bibinfo  {journal} {J.
  Fluid Mech.}\ }\textbf {\bibinfo {volume} {420}},\ \bibinfo {pages} {47}
  (\bibinfo {year} {2000})}\BibitemShut {NoStop}%
\bibitem [{\citenamefont {Wu}\ and\ \citenamefont
  {Martin}(2007)}]{wu_martin07}%
  \BibitemOpen
  \bibfield  {author} {\bibinfo {author} {\bibfnamefont {M.}~\bibnamefont
  {Wu}}\ and\ \bibinfo {author} {\bibfnamefont {M.}~\bibnamefont {Martin}},\
  }\href@noop {} {\bibfield  {journal} {\bibinfo  {journal} {AIAA J.}\ }\textbf
  {\bibinfo {volume} {45}},\ \bibinfo {pages} {879} (\bibinfo {year}
  {2007})}\BibitemShut {NoStop}%
\bibitem [{\citenamefont {Touber}\ and\ \citenamefont
  {Sandham}(2009)}]{touber_09}%
  \BibitemOpen
  \bibfield  {author} {\bibinfo {author} {\bibfnamefont {E.}~\bibnamefont
  {Touber}}\ and\ \bibinfo {author} {\bibfnamefont {N.}~\bibnamefont
  {Sandham}},\ }\href@noop {} {\bibfield  {journal} {\bibinfo  {journal}
  {Theor. Comput. Fluid Dyn.}\ }\textbf {\bibinfo {volume} {23}},\ \bibinfo
  {pages} {79} (\bibinfo {year} {2009})}\BibitemShut {NoStop}%
\bibitem [{\citenamefont {Pirozzoli}\ and\ \citenamefont
  {Bernardini}(2011{\natexlab{a}})}]{pirozzoli_11_3}%
  \BibitemOpen
  \bibfield  {author} {\bibinfo {author} {\bibfnamefont {S.}~\bibnamefont
  {Pirozzoli}}\ and\ \bibinfo {author} {\bibfnamefont {M.}~\bibnamefont
  {Bernardini}},\ }\href@noop {} {\bibfield  {journal} {\bibinfo  {journal}
  {AIAA J.}\ }\textbf {\bibinfo {volume} {49}},\ \bibinfo {pages} {1307}
  (\bibinfo {year} {2011}{\natexlab{a}})}\BibitemShut {NoStop}%
\bibitem [{\citenamefont {Grilli}\ \emph {et~al.}(2012)\citenamefont {Grilli},
  \citenamefont {Schmid}, \citenamefont {Hickel},\ and\ \citenamefont
  {Adams}}]{grilli12}%
  \BibitemOpen
  \bibfield  {author} {\bibinfo {author} {\bibfnamefont {M.}~\bibnamefont
  {Grilli}}, \bibinfo {author} {\bibfnamefont {P.}~\bibnamefont {Schmid}},
  \bibinfo {author} {\bibfnamefont {S.}~\bibnamefont {Hickel}}, \ and\ \bibinfo
  {author} {\bibfnamefont {N.}~\bibnamefont {Adams}},\ }\href@noop {}
  {\bibfield  {journal} {\bibinfo  {journal} {J. Fluid Mech.}\ }\textbf
  {\bibinfo {volume} {700}},\ \bibinfo {pages} {16} (\bibinfo {year}
  {2012})}\BibitemShut {NoStop}%
\bibitem [{\citenamefont {Aubard}\ \emph {et~al.}(2013)\citenamefont {Aubard},
  \citenamefont {Gloerfelt},\ and\ \citenamefont {Robinet}}]{aubard13}%
  \BibitemOpen
  \bibfield  {author} {\bibinfo {author} {\bibfnamefont {G.}~\bibnamefont
  {Aubard}}, \bibinfo {author} {\bibfnamefont {X.}~\bibnamefont {Gloerfelt}}, \
  and\ \bibinfo {author} {\bibfnamefont {J.}~\bibnamefont {Robinet}},\
  }\href@noop {} {\bibfield  {journal} {\bibinfo  {journal} {AIAA J.}\ }\textbf
  {\bibinfo {volume} {51}},\ \bibinfo {pages} {2395} (\bibinfo {year}
  {2013})}\BibitemShut {NoStop}%
\bibitem [{\citenamefont {Morgan}\ \emph {et~al.}(2013)\citenamefont {Morgan},
  \citenamefont {Duraisamy}, \citenamefont {Nguyen}, \citenamefont {Kawai},\
  and\ \citenamefont {Lele}}]{morgan13}%
  \BibitemOpen
  \bibfield  {author} {\bibinfo {author} {\bibfnamefont {B.}~\bibnamefont
  {Morgan}}, \bibinfo {author} {\bibfnamefont {K.}~\bibnamefont {Duraisamy}},
  \bibinfo {author} {\bibfnamefont {N.}~\bibnamefont {Nguyen}}, \bibinfo
  {author} {\bibfnamefont {S.}~\bibnamefont {Kawai}}, \ and\ \bibinfo {author}
  {\bibfnamefont {S.}~\bibnamefont {Lele}},\ }\href@noop {} {\bibfield
  {journal} {\bibinfo  {journal} {J. Fluid Mech.}\ }\textbf {\bibinfo {volume}
  {729}},\ \bibinfo {pages} {231} (\bibinfo {year} {2013})}\BibitemShut
  {NoStop}%
\bibitem [{\citenamefont {Clemens}\ and\ \citenamefont
  {Narayanaswamy}(2014)}]{clemens14}%
  \BibitemOpen
  \bibfield  {author} {\bibinfo {author} {\bibfnamefont {N.}~\bibnamefont
  {Clemens}}\ and\ \bibinfo {author} {\bibfnamefont {V.}~\bibnamefont
  {Narayanaswamy}},\ }\href@noop {} {\bibfield  {journal} {\bibinfo  {journal}
  {Annu. Rev. Fluid Mech.}\ }\textbf {\bibinfo {volume} {{46}}},\ \bibinfo
  {pages} {469} (\bibinfo {year} {2014})}\BibitemShut {NoStop}%
\bibitem [{\citenamefont {Delery}(1985)}]{delery85}%
  \BibitemOpen
  \bibfield  {author} {\bibinfo {author} {\bibfnamefont {J.}~\bibnamefont
  {Delery}},\ }\href@noop {} {\bibfield  {journal} {\bibinfo  {journal} {Prog.\
  Aerosp.\ Sci.}\ }\textbf {\bibinfo {volume} {22}},\ \bibinfo {pages} {209}
  (\bibinfo {year} {1985})}\BibitemShut {NoStop}%
\bibitem [{\citenamefont {Spaid}\ and\ \citenamefont
  {Frishett}(1972)}]{spaid72}%
  \BibitemOpen
  \bibfield  {author} {\bibinfo {author} {\bibfnamefont {F.}~\bibnamefont
  {Spaid}}\ and\ \bibinfo {author} {\bibfnamefont {J.}~\bibnamefont
  {Frishett}},\ }\href@noop {} {\bibfield  {journal} {\bibinfo  {journal} {AIAA
  J.}\ }\textbf {\bibinfo {volume} {10}},\ \bibinfo {pages} {915} (\bibinfo
  {year} {1972})}\BibitemShut {NoStop}%
\bibitem [{\citenamefont {Back}\ and\ \citenamefont {Cuffel}(1976)}]{back76}%
  \BibitemOpen
  \bibfield  {author} {\bibinfo {author} {\bibfnamefont {L.}~\bibnamefont
  {Back}}\ and\ \bibinfo {author} {\bibfnamefont {R.}~\bibnamefont {Cuffel}},\
  }\href@noop {} {\bibfield  {journal} {\bibinfo  {journal} {AIAA J.}\ }\textbf
  {\bibinfo {volume} {14}},\ \bibinfo {pages} {526} (\bibinfo {year}
  {1976})}\BibitemShut {NoStop}%
\bibitem [{\citenamefont {Delery}(1992)}]{delery92}%
  \BibitemOpen
  \bibfield  {author} {\bibinfo {author} {\bibfnamefont {J.}~\bibnamefont
  {Delery}},\ }\href@noop {} {\bibfield  {journal} {\bibinfo  {journal} {La
  Recherche Aerospatiale}\ }\textbf {\bibinfo {volume} {1}},\ \bibinfo {pages}
  {1} (\bibinfo {year} {1992})}\BibitemShut {NoStop}%
\bibitem [{\citenamefont {Jaunet}\ \emph {et~al.}(2014)\citenamefont {Jaunet},
  \citenamefont {Debi\'eve},\ and\ \citenamefont {Dupont}}]{jaunet14}%
  \BibitemOpen
  \bibfield  {author} {\bibinfo {author} {\bibfnamefont {V.}~\bibnamefont
  {Jaunet}}, \bibinfo {author} {\bibfnamefont {J.}~\bibnamefont {Debi\'eve}}, \
  and\ \bibinfo {author} {\bibfnamefont {P.}~\bibnamefont {Dupont}},\
  }\href@noop {} {\bibfield  {journal} {\bibinfo  {journal} {AIAA J.}\ }\textbf
  {\bibinfo {volume} {{52}}},\ \bibinfo {pages} {2524} (\bibinfo {year}
  {2014})}\BibitemShut {NoStop}%
\bibitem [{\citenamefont {Hayashi}\ \emph {et~al.}(1986)\citenamefont
  {Hayashi}, \citenamefont {Sakurai},\ and\ \citenamefont {Aso}}]{hayashi84}%
  \BibitemOpen
  \bibfield  {author} {\bibinfo {author} {\bibfnamefont {M.}~\bibnamefont
  {Hayashi}}, \bibinfo {author} {\bibfnamefont {A.}~\bibnamefont {Sakurai}}, \
  and\ \bibinfo {author} {\bibfnamefont {S.}~\bibnamefont {Aso}},\ }\href@noop
  {} {\bibfield  {journal} {\bibinfo  {journal} {NASA TM-77958}\ }\textbf
  {\bibinfo {volume} {57}},\ \bibinfo {pages} {455} (\bibinfo {year}
  {1986})}\BibitemShut {NoStop}%
\bibitem [{\citenamefont {Sch{\"u}lein}(2006)}]{schulein06}%
  \BibitemOpen
  \bibfield  {author} {\bibinfo {author} {\bibfnamefont {E.}~\bibnamefont
  {Sch{\"u}lein}},\ }\href@noop {} {\bibfield  {journal} {\bibinfo  {journal}
  {AIAA journal}\ }\textbf {\bibinfo {volume} {44}},\ \bibinfo {pages} {1732}
  (\bibinfo {year} {2006})}\BibitemShut {NoStop}%
\bibitem [{\citenamefont {Hadjadj}(2012)}]{hadjadj12}%
  \BibitemOpen
  \bibfield  {author} {\bibinfo {author} {\bibfnamefont {A.}~\bibnamefont
  {Hadjadj}},\ }\href@noop {} {\bibfield  {journal} {\bibinfo  {journal} {AIAA
  J}\ }\textbf {\bibinfo {volume} {50}},\ \bibinfo {pages} {2919} (\bibinfo
  {year} {2012})}\BibitemShut {NoStop}%
\bibitem [{\citenamefont {Nichols}\ \emph {et~al.}(2016)\citenamefont
  {Nichols}, \citenamefont {Larsson}, \citenamefont {Bernardini},\ and\
  \citenamefont {Pirozzoli}}]{Nichols2016}%
  \BibitemOpen
  \bibfield  {author} {\bibinfo {author} {\bibfnamefont {J.}~\bibnamefont
  {Nichols}}, \bibinfo {author} {\bibfnamefont {J.}~\bibnamefont {Larsson}},
  \bibinfo {author} {\bibfnamefont {M.}~\bibnamefont {Bernardini}}, \ and\
  \bibinfo {author} {\bibfnamefont {S.}~\bibnamefont {Pirozzoli}},\ }\href@noop
  {} {\bibfield  {journal} {\bibinfo  {journal} {Theoretical and Computational
  Fluid Dynamics}\ ,\ \bibinfo {pages} {1}} (\bibinfo {year}
  {2016})}\BibitemShut {NoStop}%
\bibitem [{\citenamefont {Fedorova}\ \emph {et~al.}(2001)\citenamefont
  {Fedorova}, \citenamefont {Fedorchenko},\ and\ \citenamefont
  {Sch{\"u}lein}}]{fedorova2001experimental}%
  \BibitemOpen
  \bibfield  {author} {\bibinfo {author} {\bibfnamefont {N.}~\bibnamefont
  {Fedorova}}, \bibinfo {author} {\bibfnamefont {I.}~\bibnamefont
  {Fedorchenko}}, \ and\ \bibinfo {author} {\bibfnamefont {E.}~\bibnamefont
  {Sch{\"u}lein}},\ }\href@noop {} {\bibfield  {journal} {\bibinfo  {journal}
  {Computational Fluid Dynamics Journal}\ }\textbf {\bibinfo {volume} {10}},\
  \bibinfo {pages} {390} (\bibinfo {year} {2001})}\BibitemShut {NoStop}%
\bibitem [{\citenamefont {Knight}\ \emph {et~al.}(2003)\citenamefont {Knight},
  \citenamefont {Yan}, \citenamefont {Panaras},\ and\ \citenamefont
  {Zheltovodov}}]{knight2003advances}%
  \BibitemOpen
  \bibfield  {author} {\bibinfo {author} {\bibfnamefont {D.}~\bibnamefont
  {Knight}}, \bibinfo {author} {\bibfnamefont {H.}~\bibnamefont {Yan}},
  \bibinfo {author} {\bibfnamefont {A.~G.}\ \bibnamefont {Panaras}}, \ and\
  \bibinfo {author} {\bibfnamefont {A.}~\bibnamefont {Zheltovodov}},\
  }\href@noop {} {\bibfield  {journal} {\bibinfo  {journal} {Progress in
  Aerospace Sciences}\ }\textbf {\bibinfo {volume} {39}},\ \bibinfo {pages}
  {121} (\bibinfo {year} {2003})}\BibitemShut {NoStop}%
\bibitem [{\citenamefont {Pirozzoli}\ \emph {et~al.}(2010)\citenamefont
  {Pirozzoli}, \citenamefont {Bernardini},\ and\ \citenamefont
  {Grasso}}]{pirozzoli_10_2}%
  \BibitemOpen
  \bibfield  {author} {\bibinfo {author} {\bibfnamefont {S.}~\bibnamefont
  {Pirozzoli}}, \bibinfo {author} {\bibfnamefont {M.}~\bibnamefont
  {Bernardini}}, \ and\ \bibinfo {author} {\bibfnamefont {F.}~\bibnamefont
  {Grasso}},\ }\href@noop {} {\bibfield  {journal} {\bibinfo  {journal} {J.
  Fluid Mech.}\ }\textbf {\bibinfo {volume} {657}},\ \bibinfo {pages} {361}
  (\bibinfo {year} {2010})}\BibitemShut {NoStop}%
\bibitem [{\citenamefont {Pirozzoli}\ and\ \citenamefont
  {Bernardini}(2011{\natexlab{b}})}]{PirozzoliJFM2011}%
  \BibitemOpen
  \bibfield  {author} {\bibinfo {author} {\bibfnamefont {S.}~\bibnamefont
  {Pirozzoli}}\ and\ \bibinfo {author} {\bibfnamefont {M.}~\bibnamefont
  {Bernardini}},\ }\href@noop {} {\bibfield  {journal} {\bibinfo  {journal} {J.
  Fluid Mech.}\ }\textbf {\bibinfo {volume} {688}},\ \bibinfo {pages} {120}
  (\bibinfo {year} {2011}{\natexlab{b}})}\BibitemShut {NoStop}%
\bibitem [{\citenamefont {Ducros}\ \emph {et~al.}(1999)\citenamefont {Ducros},
  \citenamefont {Ferrand}, \citenamefont {Nicoud}, \citenamefont {Darracq},
  \citenamefont {Gacherieu},\ and\ \citenamefont {Poinsot}}]{ducrosetal99}%
  \BibitemOpen
  \bibfield  {author} {\bibinfo {author} {\bibfnamefont {F.}~\bibnamefont
  {Ducros}}, \bibinfo {author} {\bibfnamefont {V.}~\bibnamefont {Ferrand}},
  \bibinfo {author} {\bibfnamefont {F.}~\bibnamefont {Nicoud}}, \bibinfo
  {author} {\bibfnamefont {D.}~\bibnamefont {Darracq}}, \bibinfo {author}
  {\bibfnamefont {C.}~\bibnamefont {Gacherieu}}, \ and\ \bibinfo {author}
  {\bibfnamefont {T.}~\bibnamefont {Poinsot}},\ }\href@noop {} {\bibfield
  {journal} {\bibinfo  {journal} {J. Comput. Phys.}\ }\textbf {\bibinfo
  {volume} {152}},\ \bibinfo {pages} {517–549} (\bibinfo {year}
  {1999})}\BibitemShut {NoStop}%
\bibitem [{\citenamefont {Kennedy}\ and\ \citenamefont
  {Gruber}(2008)}]{kennedy_08}%
  \BibitemOpen
  \bibfield  {author} {\bibinfo {author} {\bibfnamefont {C.}~\bibnamefont
  {Kennedy}}\ and\ \bibinfo {author} {\bibfnamefont {A.}~\bibnamefont
  {Gruber}},\ }\href@noop {} {\bibfield  {journal} {\bibinfo  {journal} {J.
  Comput. Phys.}\ }\textbf {\bibinfo {volume} {227}},\ \bibinfo {pages} {1676}
  (\bibinfo {year} {2008})}\BibitemShut {NoStop}%
\bibitem [{\citenamefont {Pirozzoli}(2010)}]{pirozzoli_10}%
  \BibitemOpen
  \bibfield  {author} {\bibinfo {author} {\bibfnamefont {S.}~\bibnamefont
  {Pirozzoli}},\ }\href@noop {} {\bibfield  {journal} {\bibinfo  {journal} {J.\
  Comput.\ Phys.}\ }\textbf {\bibinfo {volume} {229}},\ \bibinfo {pages} {7180}
  (\bibinfo {year} {2010})}\BibitemShut {NoStop}%
\bibitem [{\citenamefont {Bernardini}\ and\ \citenamefont
  {Pirozzoli}(2009)}]{bernardini09}%
  \BibitemOpen
  \bibfield  {author} {\bibinfo {author} {\bibfnamefont {M.}~\bibnamefont
  {Bernardini}}\ and\ \bibinfo {author} {\bibfnamefont {S.}~\bibnamefont
  {Pirozzoli}},\ }\href@noop {} {\bibfield  {journal} {\bibinfo  {journal} {J.
  Comput. Phys.}\ }\textbf {\bibinfo {volume} {228}},\ \bibinfo {pages} {4182}
  (\bibinfo {year} {2009})}\BibitemShut {NoStop}%
\bibitem [{\citenamefont {Simens}\ \emph {et~al.}(2009)\citenamefont {Simens},
  \citenamefont {Jim\'enez}, \citenamefont {Hoyas},\ and\ \citenamefont
  {Mizuno}}]{simens_09}%
  \BibitemOpen
  \bibfield  {author} {\bibinfo {author} {\bibfnamefont {M.}~\bibnamefont
  {Simens}}, \bibinfo {author} {\bibfnamefont {J.}~\bibnamefont {Jim\'enez}},
  \bibinfo {author} {\bibfnamefont {S.}~\bibnamefont {Hoyas}}, \ and\ \bibinfo
  {author} {\bibfnamefont {Y.}~\bibnamefont {Mizuno}},\ }\href@noop {}
  {\bibfield  {journal} {\bibinfo  {journal} {J. Comput. Phys.}\ }\textbf
  {\bibinfo {volume} {228}},\ \bibinfo {pages} {4218} (\bibinfo {year}
  {2009})}\BibitemShut {NoStop}%
\bibitem [{\citenamefont {Bermejo-Moreno}\ \emph {et~al.}(2014)\citenamefont
  {Bermejo-Moreno}, \citenamefont {Campo}, \citenamefont {Larsson},
  \citenamefont {Bodart}, \citenamefont {Helmer},\ and\ \citenamefont
  {Eaton}}]{bermejo14}%
  \BibitemOpen
  \bibfield  {author} {\bibinfo {author} {\bibfnamefont {I.}~\bibnamefont
  {Bermejo-Moreno}}, \bibinfo {author} {\bibfnamefont {L.}~\bibnamefont
  {Campo}}, \bibinfo {author} {\bibfnamefont {J.}~\bibnamefont {Larsson}},
  \bibinfo {author} {\bibfnamefont {J.}~\bibnamefont {Bodart}}, \bibinfo
  {author} {\bibfnamefont {D.}~\bibnamefont {Helmer}}, \ and\ \bibinfo {author}
  {\bibfnamefont {J.}~\bibnamefont {Eaton}},\ }\href@noop {} {\bibfield
  {journal} {\bibinfo  {journal} {J. Fluid Mech.}\ }\textbf {\bibinfo {volume}
  {758}},\ \bibinfo {pages} {5} (\bibinfo {year} {2014})}\BibitemShut {NoStop}%
\bibitem [{\citenamefont {El\'ena}\ and\ \citenamefont
  {Lacharme}(1988)}]{elena_88}%
  \BibitemOpen
  \bibfield  {author} {\bibinfo {author} {\bibfnamefont {M.}~\bibnamefont
  {El\'ena}}\ and\ \bibinfo {author} {\bibfnamefont {J.}~\bibnamefont
  {Lacharme}},\ }\href@noop {} {\bibfield  {journal} {\bibinfo  {journal} {J.
  M\'ec. Th\'eor. Appl.}\ }\textbf {\bibinfo {volume} {7}},\ \bibinfo {pages}
  {175} (\bibinfo {year} {1988})}\BibitemShut {NoStop}%
\bibitem [{\citenamefont {Schlatter}\ and\ \citenamefont
  {{\"O}rl{\"u}}(2010)}]{schlatter_10b}%
  \BibitemOpen
  \bibfield  {author} {\bibinfo {author} {\bibfnamefont {P.}~\bibnamefont
  {Schlatter}}\ and\ \bibinfo {author} {\bibfnamefont {R.}~\bibnamefont
  {{\"O}rl{\"u}}},\ }\href@noop {} {\bibfield  {journal} {\bibinfo  {journal}
  {J. Fluid Mech.}\ }\textbf {\bibinfo {volume} {659}},\ \bibinfo {pages} {116}
  (\bibinfo {year} {2010})}\BibitemShut {NoStop}%
\bibitem [{\citenamefont {Debi{\`e}ve}\ \emph {et~al.}(1997)\citenamefont
  {Debi{\`e}ve}, \citenamefont {Dupont}, \citenamefont {Smith},\ and\
  \citenamefont {Smits}}]{debieve97}%
  \BibitemOpen
  \bibfield  {author} {\bibinfo {author} {\bibfnamefont {J.}~\bibnamefont
  {Debi{\`e}ve}}, \bibinfo {author} {\bibfnamefont {P.}~\bibnamefont {Dupont}},
  \bibinfo {author} {\bibfnamefont {D.}~\bibnamefont {Smith}}, \ and\ \bibinfo
  {author} {\bibfnamefont {A.}~\bibnamefont {Smits}},\ }\href@noop {}
  {\bibfield  {journal} {\bibinfo  {journal} {AIAA J.}\ }\textbf {\bibinfo
  {volume} {35}},\ \bibinfo {pages} {51} (\bibinfo {year} {1997})}\BibitemShut
  {NoStop}%
\bibitem [{\citenamefont {Hadjadj}\ \emph {et~al.}(2015)\citenamefont
  {Hadjadj}, \citenamefont {Ben-Nasr}, \citenamefont {Shadloo},\ and\
  \citenamefont {Chaudhuri}}]{hadjadj15}%
  \BibitemOpen
  \bibfield  {author} {\bibinfo {author} {\bibfnamefont {A.}~\bibnamefont
  {Hadjadj}}, \bibinfo {author} {\bibfnamefont {O.}~\bibnamefont {Ben-Nasr}},
  \bibinfo {author} {\bibfnamefont {M.}~\bibnamefont {Shadloo}}, \ and\
  \bibinfo {author} {\bibfnamefont {A.}~\bibnamefont {Chaudhuri}},\ }\href@noop
  {} {\bibfield  {journal} {\bibinfo  {journal} {Int. J. Heat Mass Transfer}\
  }\textbf {\bibinfo {volume} {81}},\ \bibinfo {pages} {426} (\bibinfo {year}
  {2015})}\BibitemShut {NoStop}%
\bibitem [{\citenamefont {{Katzer, E.}}(1989)}]{katzer89}%
  \BibitemOpen
  \bibfield  {author} {\bibinfo {author} {\bibnamefont {{Katzer, E.}}},\
  }\href@noop {} {\bibfield  {journal} {\bibinfo  {journal} {J. Fluid Mech.}\
  }\textbf {\bibinfo {volume} {{206}}},\ \bibinfo {pages} {477} (\bibinfo
  {year} {1989})}\BibitemShut {NoStop}%
\bibitem [{\citenamefont {Zhang}\ \emph {et~al.}(2014)\citenamefont {Zhang},
  \citenamefont {Bi}, \citenamefont {Hussain},\ and\ \citenamefont
  {She}}]{zhang14}%
  \BibitemOpen
  \bibfield  {author} {\bibinfo {author} {\bibfnamefont {Y.}~\bibnamefont
  {Zhang}}, \bibinfo {author} {\bibfnamefont {W.}~\bibnamefont {Bi}}, \bibinfo
  {author} {\bibfnamefont {F.}~\bibnamefont {Hussain}}, \ and\ \bibinfo
  {author} {\bibfnamefont {Z.}~\bibnamefont {She}},\ }\href@noop {} {\bibfield
  {journal} {\bibinfo  {journal} {J. Fluid Mech.}\ }\textbf {\bibinfo {volume}
  {739}},\ \bibinfo {pages} {392} (\bibinfo {year} {2014})}\BibitemShut
  {NoStop}%
\bibitem [{\citenamefont {Konno}\ and\ \citenamefont
  {Ohmachi}(1998)}]{konno_98}%
  \BibitemOpen
  \bibfield  {author} {\bibinfo {author} {\bibfnamefont {K.}~\bibnamefont
  {Konno}}\ and\ \bibinfo {author} {\bibfnamefont {T.}~\bibnamefont
  {Ohmachi}},\ }\href@noop {} {\bibfield  {journal} {\bibinfo  {journal}
  {Bulletin of the Seismological Society of America}\ }\textbf {\bibinfo
  {volume} {88}},\ \bibinfo {pages} {228} (\bibinfo {year} {1998})}\BibitemShut
  {NoStop}%
\end{thebibliography}%

\end{document}